\documentclass[english,12pt,aps,prd,a4paper,preprintnumbers,floatfix,nofootinbib,superscriptaddress, notitlepage]{revtex4} 
 \pdfoutput=1
\usepackage[usenames,dvipsnames]{color}  
\usepackage{graphicx}
\usepackage{setspace}
\usepackage{caption}
\usepackage{amsmath}
\usepackage{slashed}
\usepackage{amssymb}
\usepackage[colorlinks=true,citecolor=darkred,urlcolor=darkred, pdfborder={0 0 0}]{hyperref}
\usepackage[normalem]{ulem}
\usepackage{xcolor}

\makeatletter
\def\p@subsection{}
\makeatother

\definecolor{darkred}{rgb}{0.6,0,0}

\definecolor{linkcolor}{rgb}{0,0,0.5}

\def\gsim{\raise0.3ex\hbox{$\;>$\kern-0.75em\raise-1.1ex\hbox{$\sim\;$}}}
\def\lsim{\raise0.3ex\hbox{$\;<$\kern-0.75em\raise-1.1ex\hbox{$\sim\;$}}}

\def\beqn#1{\begin{equation}\label{#1}}
\def\eeqn{\end{equation}}

\def\beqa#1{\begin{eqnarray}\label{#1}}
\def\eeqa{\end{eqnarray}}

\def\321{$\mathrm{SU(3) \otimes SU(2) \otimes U(1)}$ }
\def\lr{$\mathrm{SU(3) \otimes SU(2)_L \otimes SU(2)_R \otimes U(1)_{B-L}}$}

\def\Z2{$\mathcal{Z_2}$}

\newcommand {\ignore}[1]{}

\def\lnv{lepton number violation }
 \def\one{\ensuremath{\mathbf{1}}}
 \def\two{\ensuremath{\mathbf{2}}}
 \def\three{\ensuremath{\mathbf{3}}}
 \def\threeS{\ensuremath{\mathbf{3^*}}}

\def\SM{$\mathrm{SU(3)_c \otimes SU(2)_L \otimes U(1)_Y}$ }

\newcommand{\no}{\nonumber\\}

\newcommand{\AddrAHEP}{
  AHEP Group, Institut de F\'{i}sica Corpuscular --
  CSIC/Universitat de Val\`{e}ncia, Parc Cient\'ific de Paterna.\\
 C/ Catedr\'atico Jos\'e Beltr\'an, 2 E-46980 Paterna (Valencia) - SPAIN}
 
\newcommand{\AddrIISERB}{Department of Physics, Indian Institute of Science Education and Research - Bhopal, \\ 
Bhopal Bypass Road, Bhauri, Bhopal 462066, India}

\newcommand{\AddrCFTP}{Departamento de F\'{\i}sica and CFTP, Instituto Superior T\'ecnico, Universidade de Lisboa, Av. Rovisco Pais 1, 1049-001 Lisboa, Portugal}

\begin{document}
\title{\color{BrickRed} Phenomenology of the simplest linear seesaw mechanism}
\author{Aditya Batra}\email{aditya.batra@tecnico.ulisboa.pt}
\affiliation{\AddrIISERB}
\affiliation{\AddrCFTP}
\author{Praveen Bharadwaj}\email{praveen20@iiserb.ac.in}
\affiliation{\AddrIISERB}
\author{Sanjoy Mandal}\email{smandal@kias.re.kr}
\affiliation{Korea Institute for Advanced Study, Seoul 02455, Korea}
\author{Rahul Srivastava}\email{rahul@iiserb.ac.in}
\affiliation{\AddrIISERB}
\author{Jos\'{e} W. F. Valle}\email{valle@ific.uv.es}
\affiliation{\AddrAHEP}

\begin{abstract}
  \vspace{1cm} 
  The linear seesaw mechanism provides a simple way to generate neutrino masses. In addition to Standard Model particles, it includes quasi-Dirac leptons as neutrino mass mediators, and a leptophilic scalar doublet seeding small neutrino masses. Here we review its associated physics, including restrictions from theory and phenomenology. The model yields potentially detectable $\mu\to e\gamma$ rates as well as distinctive signatures in the production and decay of heavy neutrinos~($N_i$) and the charged Higgs boson~($H^\pm$) arising from the second scalar doublet. We have found that production processes such as $e^+e^-\to NN$, $e^-\gamma\to NH^-$ and $e^+ e^-\to H^+ H^-$ followed by the decay chain $H^\pm\to\ell_i^\pm N$, $N\to \ell_j^{\pm}W^\mp$ leads to striking \lnv signatures at high energies which may probe the Majorana nature of neutrinos.
\end{abstract}
\maketitle
\section{Introduction}
\label{sec:introduction}

Non-zero neutrino masses~\cite{Kajita:2016cak,McDonald:2016ixn,KamLAND:2002uet,K2K:2002icj} constitute one of the most convincing proofs of new physics.
Underpinning their ultimate origin stands out as one of the biggest challenges in elementary particle physics. 
Despite many attempts, the issue remains wide open.
A simple way to generate Majorana neutrino masses is to introduce a non-renormalizable dimension-five operator into the Standard Model (SM)~\cite{Weinberg:1980bf}. 
The effective dimension-five operator characterizing lepton number non-conservation is given by 
\begin{align}
    -\mathcal{L}_\nu^{d=5}=\frac{1}{\Lambda}({L} \Phi)\, (\Phi L) + \text{H.c.},
\end{align} 
where the contractions involve the left-handed lepton doublet spinors $L$ and the SM Higgs scalar doublet $\Phi$.
Here $\Lambda$ is the effective mass scale, flavor indices are omitted, for brevity, and 2-dimensional conjugation matrices in Lorentz and isospin space are understood.  

Neutrino mass model-building requires a completion of this operator.  
The seesaw mechanism provides a specially interesting one, and is most generally realized within the simplest \SM gauge structure~\cite{Schechter:1980gr}. 
In its type-I realization, neutrinos get mass due to the exchange of heavy singlet fermion mediators.  
This leads to a Majorana mass for the left-handed neutrinos as $m_\nu\sim m_D^2/M_N$, with $m_D=Y_\nu v/\sqrt{2}$, so that $\Lambda=M_N/Y_{\nu}^2$.
Hence, for $m_\nu\sim\mathcal{O}(0.1\,\text{eV})$ and a relatively large Yukawa coupling $Y_\nu\sim\mathcal{O}(1)$, $M_N$ must be large, i.e., $M_N\gg\mathcal{O}(\text{TeV})$.
As a result, the conventional high-scale implementation of the seesaw mechanism has few phenomenological implications other than those directly related to neutrino masses. 

However, the seesaw paradigm can arise from low-scale physics.
The low-scale seesaw varieties are the inverse~\cite{Mohapatra:1986bd,Gonzalez-Garcia:1988okv} and the linear seesaw mechanisms~\cite{Akhmedov:1995ip,Akhmedov:1995vm,Malinsky:2005bi}. 
These share a common ``(3,6)'' template~\cite{Schechter:1980gr,Schechter:1981cv}~\footnote{These contain 6 singlets that make up 3 heavy Dirac leptons in the limit of lepton number conservation.},
 which instead of a single right-handed neutrino, requires a sequential pair of isosinglet leptons associated to each family.
The possibility that the heavy neutrinos could be produced at high energy colliders~\cite{Dittmar:1989yg,Gonzalez-Garcia:1990sbd,AguilarSaavedra:2012fu,Das:2012ii,Deppisch:2013cya,Drewes:2019fou,Cottin:2022nwp}
was taken up by experiments, such as ATLAS and CMS at the LHC~\cite{ATLAS:2019kpx,CMS:2018iaf,CMS:2022nty} and also future proposals~\cite{FCC:2018evy,FCC:2018byv,Feng:2022inv}.  

Interestingly enough, leptonic flavour and CP can be violated even in the limit of massless neutrinos~\cite{Bernabeu:1987gr,Langacker:1988up,Branco:1989bn,Rius:1989gk}.
This implies that such processes need not be suppressed by the small neutrino masses, and can in fact have observable
rates~\cite{Bernabeu:1987gr,Langacker:1988up,Rius:1989gk,Gonzalez-Garcia:1991brm,Deppisch:2004fa,Deppisch:2005zm}~
\footnote{For generic references on cLFV in seesaw schemes see, for example~\cite{Ilakovac:1994kj,Arganda:2007jw,Abada:2014cca,Abada:2015oba}.}.
Detailed charged lepton flavour violation~(cLFV) predictions depend on whether one has an inverse or linear seesaw realization, and also on details of Yukawa coupling matrices~\cite{Boucenna:2014zba}. 

\par In this work, we examine the simplest SM-based variant of the linear seesaw mechanism. 
In contrast to most previous formulations~\cite{Akhmedov:1995ip,Akhmedov:1995vm,Malinsky:2005bi}, here we do not impose left-right symmetry.  
The linear seesaw mechanism is realized within the \SM gauge structure itself, in which lepton number symmetry is ungauged~\cite{Fontes:2019uld}. 
We assume at least two pairs of isosinglet leptons. In addition, the scalar sector contains a second Higgs doublet, carrying two units of lepton number.

Even for TeV-scale mediators, $M_N\sim\mathcal{O}(\text{TeV})$, neutrino masses are naturally small due to the small vacuum expectation value~(VEV) of this second Higgs doublet. 
For simplicity, we assume that lepton number symmetry is broken explicitly, but softly, in the scalar potential, thereby avoiding a Nambu-Goldstone boson and the associated
stringent astrophysical restrictions~\cite{Fontes:2019uld}. 
Such a ``neutrino-motivated'' version of the two-doublet model~\cite{Branco:2011iw,Bhattacharyya:2015nca,Wang:2022yhm}
allows for direct experimental tests~\cite{Eriksson:2009ws}, as we will discuss later. 
The charged scalar also contribute to lepton flavor violating decays such as, $\mu\to e\gamma$, with rates that can lie within reach of current experiment~\cite{MEG:2013oxv},
providing extra sensitivity to model parameters.

Note that our pair production of charged scalars at $e^+e^-$ collider is in sharp contrast with the usual two Higgs doublet model~(THDM),
in which it proceeds via the neutral current Drell-Yan mechanism involving s-channel $\gamma/Z$ exchange. 
In our linear seesaw scheme the pair production of charged scalars at $e^+e^-$ collider can be dominated by a t-channel heavy-neutrino-mediated diagram. 
The decays of the new scalars are controlled by the underlying $U(1)$ lepton symmetry and we find that when the
charged Higgs mass $m_{H^\pm}>M_{N_i}$, the decay chain $H^\pm\to\ell^\pm N_i, N_i\to \ell_j^{\pm}W^\mp$ leads to striking signatures. 
Our proposal also leads to new production mechanisms for heavy neutrinos involving t-channel charged Higgs mediation, $e^+e^-\to N_i N_i$ and associated production through $e^-\gamma\to N_i H^-$.
In contrast to other type-I seesaw schemes, in our linear seesaw model $e^+e^-\to N_i N_i$ production is not suppressed by light-heavy neutrino mixing.

%
\par The paper is organized as follows. To make our presentation self-contained, in Sec.~\ref{sec:model} we briefly recap the model, giving details of its new fields and their interactions.
In Sec.~\ref{sec:STU}, we discuss constraint from electroweak precision parameters $S$, $T$ and $U$.
In Sec.~\ref{sec:collider}, we discuss the existing collider constraints on the new scalar masses.
In Sec.~\ref{sec:lfv}, we discuss various phenomenological implications for charged lepton flavour violation processes.
In Sec.~\ref{sec:prod-decay} and \ref{sec:decay}, we discuss various possible production mechanisms at $e^+e^-$~\cite{Barklow:2015tja,FCC:2018evy,CLICdp:2018cto,CEPCStudyGroup:2018ghi},
$e^-\gamma$~\cite{Telnov:1999tb,Bechtel:2006mr,Ginzburg:1982bs,Ginzburg:1982yr,Velasco:2001fsi,Telnov:1989sd} colliders, and also various decay channels of heavy neutrinos and new scalars.
In Sec.~\ref{sec:collider-signatures} we show how our linear seesaw model can lead to promising signatures at future lepton colliders, such as the ILC~\cite{Barklow:2015tja}, FCC-ee~\cite{FCC:2018evy}, CLIC~\cite{CLICdp:2018cto}, and the CEPC~\cite{CEPCStudyGroup:2018ghi}.
Finally, in Sec.~\ref{sec:conclusions} we conclude.

\section{Linear seesaw model}
\label{sec:model}
The linear seesaw is a low-scale variant of the seesaw mechanism first proposed within the \lr  gauge group~\cite{Akhmedov:1995ip,Akhmedov:1995vm},
and subsequently shown to arise also within the SO(10) framework~\cite{Malinsky:2005bi}. 

\begin{table}[ht]
\begin{tabular}{ ||c | c c c c c c c| c c || }
\hline 
\hspace{0.5cm} &\hspace{0.05cm} $Q$ \hspace{0.05cm}& \hspace{0.05cm}$u^c$\hspace{0.05cm} &\hspace{0.05cm} $d^c$ \hspace{0.05cm}&\hspace{0.05cm} $L$\hspace{0.05cm} & \hspace{0.05cm}$e^c$ \hspace{0.05cm}& \hspace{0.05cm}$\nu^c $\hspace{0.05cm} &\hspace{0.05cm} $S$ \hspace{0.05cm} & \hspace{0.1cm} $\Phi$\hspace{0.05cm} &\hspace{0.05cm}  $\chi_L$  \hspace{0.05cm} \\ 
\hline \hline
$\rm SU(3)_C$ & $\three$ & $\threeS$ & $\threeS$ & $\one$ & $\one$ & $\one$ & $\one$ & $\one$ & $\one$  \\
$\rm SU(2)_L$ & $\two$ & $\one$ & $\one$ & $\two$ & $\one$ & $\one$ & $\one$ & $\two$ & $\two$  \\
$\rm U(1)_Y$ & $\frac{1}{6}$ & $-\frac{2}{3}$ & $\frac{1}{3}$ & $-\frac{1}{2}$ & $1$ & $0$ & $0$ & $\frac{1}{2}$ & $\frac{1}{2}$ \\[1mm]
\hline
$\rm U(1)_L$ & $0$ & $0$ & $0$ & $1$ & $-1$ & $-1$ & $1$ & $0$ & $-2$  \\
\hline 
\hline
\end{tabular}
\caption{Linear-seesaw particle content and transformation properties under the SM gauge and global $U(1)_L$ lepton number symmetry.
    The subscript ``L'' in $\chi_L$ denotes its non-zero charge under the $U(1)_L$ symmetry.}
\label{tab:content}
\end{table}

In this work, we propose the simplest variant of the linear seesaw mechanism, realized within the simplest \SM gauge structure itself. Particle content and their representations under the SM gauge and global $U(1)_L$ lepton number symmetry are given in Table.~\ref{tab:content}.
Here, in addition to the SM Higgs scalar $\Phi$, we add one more doublet $\chi_L$, carrying lepton number $L[\chi_L]=-2$ in order to seed neutrino mass generation.  
In contrast with Ref.~\cite{Fontes:2019uld} we do not add a gauge singlet scalar to implement the spontaneous breaking of the lepton number symmetry.
The assumed breaking is explicit, thus avoiding the existence of a physical (nearly) massless Nambu-Goldstone boson, and the associated restrictions from LEP~\cite{Joshipura:1992hp}
as well as the stringent astrophysical limits from stellar cooling~\cite{Fontes:2019uld}.  
In addition to the new scalar doublet we add three lepton singlets $\nu_i^c$ with lepton number $L[\nu^c_i]=-1$ and three lepton singlets $S_i$ with lepton number $L[S_i]=1$. 
The global $U(1)$ lepton number symmetry is broken only in the scalar sector, explicitly but softly. 

\subsection{Neutrino mass generation}
\label{subsec:neutrino-mass}
Here we focus on the simplest linear seesaw setup, a very simple extension of the Standard Model. 
In its simplest form the relevant lepton-number-invariant Lagrangian for neutrino mass generation is written as
\begin{equation}
  \label{eq:Yukawa}
  - \mathcal{L}_{\rm Yuk}= Y_{\nu}^{ij} L_i^T C \nu^c_j \Phi   + M_R^{ij} \nu^c_i C S_j + Y_{S}^{ij} L_i^T C  S_j \chi_L+ \text{h.c.}
\end{equation}
where $Y_{\nu}$ and $Y_{S}$ are dimensionless Yukawa couplings, $M_{R}$ is an arbitrary bare mass term, $\Phi$ is the SM Higgs doublet, while $\chi_L$ is the other scalar doublet. 
This form gives an effective description of more complete realizations with spontaneous breaking of gauged~\cite{Akhmedov:1995ip,Akhmedov:1995vm,Malinsky:2005bi} or
global lepton number~\cite{Fontes:2019uld}.   
\begin{figure}[!htbp]
	\includegraphics{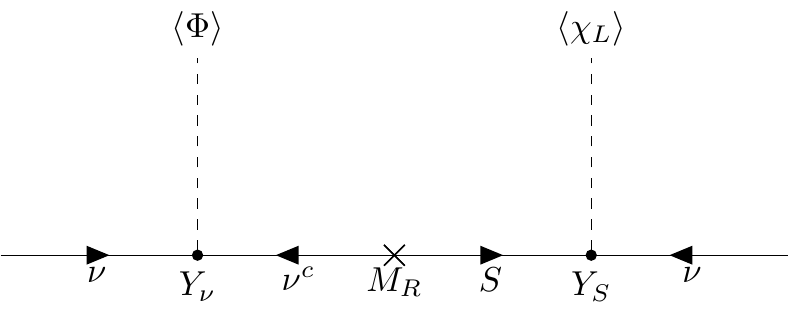}
	\caption{Neutrino mass generation in the linear seesaw mechanism. }
	\label{fig:neutrino}
\end{figure}

In the basis $\nu,\nu^c,S$ the resulting linear seesaw mass matrix obtained after \lnv is
\begin{align}
\mathcal{M}_{\nu}=
 \begin{pmatrix}
  0 & m_D  & M_L  \\
  m_D^T & 0 & M_R \\
  M_L^T &  M_R^T  &  0  \\
 \end{pmatrix}.
 \label{eq:neutrino-mass-matrix}
\end{align}
%
The mass entry $M_L$ is proportional to the VEV $v_\chi$ driving lepton number violation,
\begin{equation}\label{eq:ML}
  M_L=\frac{Y_S^{ij}v_{\chi}}{\sqrt{2}},
 \end{equation} 
while the other one is the conventional Dirac mass entry
\begin{equation}\label{eq:MD}
m_D=\frac{Y_\nu^{ij}v_{\Phi}}{\sqrt{2}},
 \end{equation} 
 given in terms of the SM Higgs VEV  $v_\Phi$.
 In order to generate VEV for $\chi_L$, we break the global $U(1)_L$ lepton symmetry explicitly in the scalar sector using the soft term $\mu_{12}^2(\Phi^\dagger\chi_L + h.c.)$. This gives the following induced VEV for $\chi_L$,
\begin{align}
v_\chi  \approx \frac{\mu_{12}^2 v_\Phi}{m_A^2},
\end{align}
where $m_A$ is the mass of pseudoscalar which we define later.
The full $9\times 9$ neutrino mass matrix in Eq.~\eqref{eq:neutrino-mass-matrix} can be diagonalized by a unitary matrix  
%
$\mathcal{U}^\dagger \mathcal{M}_{\nu} \mathcal{U^{*}}=\mathcal{M}_\nu^{\rm diag}$.
Here $\mathcal{U}$ is a product of a block-diagonalization followed by separate diagonalizations in the light and heavy sectors~\cite{Schechter:1981cv}.
The matrix $\mathcal{U}$ can be expressed as~\cite{Schechter:1981cv} 
\begin{align}
\mathcal{U}=\mathcal{U}_0 \mathcal{U}_1\,\,\text{ with }\,\, \mathcal{U}_0^\dagger \mathcal{M}_\nu \mathcal{U}_0^{*}=\begin{pmatrix} m_{\rm light} & 0 \\
0 & M_{\rm heavy}\\
\end{pmatrix}.
\end{align}
Hence, $\mathcal{U}_0$ first brings the full neutrino matrix to block diagonal form, while $\mathcal{U}_1 = \text{Diag}(U_{\rm lep}, \mathcal{U}_R)$
diagonalizes the mass matrices $m_{\rm light}$ and $M_{\rm heavy}$. This matrix $\mathcal{U}$ is expressed approximately as follows
\begin{eqnarray}
  \label{u-bdiag}
&
\mathcal{U}\approx
\left(
\begin{array}{ccc}
(1-\frac{1}{2}\epsilon)U_{\rm lep} & -\frac{i}{\sqrt{2}} V & \frac{1}{\sqrt{2}} V \\
0 & \frac{i}{\sqrt{2}}(1-\frac{1}{2}\epsilon') & \frac{1}{\sqrt{2}} (1-\frac{1}{2}\epsilon')\\
 -V^{\dagger} & -\frac{i}{\sqrt{2}}(1-\frac{1}{2}\epsilon') & \frac{1}{\sqrt{2}} (1-\frac{1}{2}\epsilon')
\end{array}
\right),  &
\end{eqnarray}
where $V$ is a $3\times 3$ matrix
\begin{equation} \label{eq:mixing}
 V = m_D \,(M_R^T)^{-1}. 
\end{equation} 
The parameters $\epsilon$ and $\epsilon'$ denote non-unitary corrections. Their explicit form is given as  
\begin{align}
& \epsilon \approx (m_{D}M_R^{-1T}) (M_R^{-1}m_D^{\dagger}) + \mathcal{O}(M_L^2/M_R^2),\\
& \epsilon'  \approx (M_R^{-1}m_D^{\dagger}) (m_{D}M_R^{-1T}) + \mathcal{O}(M_L^2/M_R^2).
\end{align}
The hierarchy $M_R\gg m_D\gg M_L$ implies an effective light neutrino mass matrix given as
\begin{equation}\label{lin}
m_{\rm light}=m_D(M_L M_R^{-1})^T+(M_L M_R^{-1}){m_D}^T.
\end{equation}
%
The limit $M_L\to 0$ leads to three massless neutrinos as in the Standard Model, plus three heavy Dirac neutrinos, $N$.   
Approximate analytical forms for the charged and neutral current weak interaction matrices in this limit are given in~\cite{Bernabeu:1987gr,Branco:1989bn,Rius:1989gk,Gonzalez-Garcia:1991brm}. 
The above form provides the template for the low-scale seesaw schemes, including the inverse~\cite{Mohapatra:1986bd,Gonzalez-Garcia:1988okv} and the
linear seesaw mechanisms~\cite{Akhmedov:1995ip,Akhmedov:1995vm,Malinsky:2005bi}. 
After \lnv, a non-zero $M_L$ is generated, leading to three light Majorana eigenstates $\nu_i$, with $i=1,2,3$ and six heavy neutrinos $N_j,j=4,..,9$ which form three pairs of quasi-Dirac
states~\cite{Valle:1982yw,Anamiati:2016uxp}. 
The neutrino mass generation mechanism is illustrated by the Feynman diagram shown in Fig. \ref{fig:neutrino}. 
One sees that, in contrast to the conventional type-I seesaw setups, the matrix $m_{\rm light}$ scales linearly with the Dirac Yukawa couplings contained in $m_D$,
thus the name linear seesaw mechanism. 
Note that neutrino masses will be suppressed by the small value of $M_L$ irrespective of how low the $M_R$ scale characterizing the heavy messengers is,
also allowing for non-negligible $Y_S$ values. 
This is achieved with a very small value of $v_\chi$. In the limit $\mu_{12} \to 0$ and hence $v_\chi \to 0$, lepton number is restored, so the construction is natural in t'Hooft's sense.

\par  For completeness we also specify the charged-lepton and quark Yukawa Lagrangian terms,
\begin{align}
 -\mathcal{L}_{\rm Yuk}=Y_e\bar{L}_L\Phi e_R + Y_u \bar{Q}_L \tilde{\Phi} u_R + Y_d \bar{Q}_L\Phi d_R + \text{h.c.} 
\end{align}
Notice that, due to the $U(1)_L$ lepton number symmetry, charged fermions acquire mass only through their Yukawa coupling with the SM Higgs doublet $\Phi$. 
Hence, the doublet $\chi_L$ is leptophilic. In this sense, as far as quarks are concerned, our simplest linear seesaw scheme resembles the Type I two-Higgs-doublet-Model~(THDM)~\cite{Branco:2011iw}.
It follows that the Higgs Yukawa Lagrangian can be written compactly as
\begin{align}
-\mathcal{L}_{\rm Yuk}=\frac{m_f}{v\sin\beta}\bar{\psi}_f \psi_f \Big(\cos\alpha\, h + \sin\alpha\, H \Big),
\end{align}
for all charged fermions.

\subsection{The scalar sector} 
\label{subsec:scalar-sector}
In addition to the SM Higgs doublet $\Phi$ we also have a second scalar doublet $\chi_L$, charged under lepton number.
The \SM gauge invariant scalar potential is given by 
\begin{align}
 V&=-\mu_\Phi^2 \Phi^{\dagger}\Phi - \mu_\chi^2 \chi_L^{\dagger}\chi_L+\lambda_1 (\Phi^{\dagger}\Phi)^2 + \lambda_2 (\chi_L^{\dagger}\chi_L)^2+\lambda_3 \chi_L^{\dagger}\chi_L \Phi^{\dagger}\Phi\nonumber \\
 & + \lambda_4 \chi_L^{\dagger}\Phi \Phi^{\dagger}\chi_L - \left(\mu_{12}^2 \Phi^{\dagger}\chi_L+ \text{H.c.} \right),
 \label{eq:potential}
\end{align}
For definiteness, we assume all parameters to be real.
In addition to breaking the electroweak gauge symmetry through the Higgs mechanism, this potential also breaks lepton number.
We choose to do this explicitly, but ``softly'', through the last bilinear term $\mu_{12}^2(\Phi^\dagger\chi_L + h.c.)$, which induces a non-zero VEV for $\chi_L$.

We now examine the consistency conditions of the potential. 
To ensure that the scalar potential is bounded from below and has a stable vacuum at any given energy scale, the following constraints must hold:  
\begin{align}
\lambda_1\geq 0,\,\, \lambda_2 \geq 0,\,\, \lambda_3 \geq -2\sqrt{\lambda_1 \lambda_2}\,\, \text{and}\,\, \lambda_3 + \lambda_4 \geq -2\sqrt{\lambda_1 \lambda_2}.
\label{vacstab}
\end{align}
To ensure perturbativity, we also restrict the scalar quartic couplings in Eq.~\ref{eq:potential} to the range $\lambda_i\leq 4\pi$.

\subsection*{Higgs boson mass spectrum}
After \SM and lepton-number symmetry breaking, we obtain the mass spectrum for the scalars by expanding the scalar fields $\Phi$ and $\chi_L$ as    
\begin{align}
 \Phi=
 \begin{pmatrix}
  \Phi^{+} \\
   \frac{1}{\sqrt{2}}(v_\Phi+h_\Phi+i\eta_\Phi)\\
 \end{pmatrix},\hspace{1cm}  \chi_L=
 \begin{pmatrix}
  \chi^{+} \\
   \frac{1}{\sqrt{2}}(v_\chi+h_\chi+i\eta_\chi)\\
 \end{pmatrix},
\end{align}
where $h_\Phi$, $h_\chi$ and $\eta_\Phi$, $\eta_\chi$ are CP even and CP odd neutral scalars, while $\chi^{\pm}$ and $\Phi^{\pm}$ are charged scalars.  
In order to get the physical states and describe the mixing between the two doublets $\Phi$ and $\chi_L$, we diagonalize the charged and neutral scalar mass matrix. 

In addition to the three unphysical Goldstone bosons $G^\pm,G^0$ which are ``eaten'' to become the longitudinal components of the SM $W^{\pm}$ and $Z$ gauge bosons,
there is one physical charged scalar $H^{\pm}$ and three neutral scalars $h$, $H$, $A$, making up the eight degrees of freedom of the two-doublet-Higgs boson system.

The mass matrix for the charged scalars in the basis $(\chi^+,\,\,\Phi^{+})$ is given by 
\begin{align}
 \mathcal{M}_{\chi}^2=
 \begin{pmatrix}
 \mu_{12}^2 \frac{v_\Phi}{v_\chi}-\lambda_4\frac{v_\Phi^2}{2}     &  -\mu_{12}^2+\lambda_4\frac{v_\Phi v_\chi}{2} \\
  -\mu_{12}^2+\lambda_4\frac{v_\Phi v_\chi}{2} &   \mu_{12}^2\frac{v_\chi}{v_\Phi}-\lambda_4\frac{v_\chi^2}{2} \\
 \end{pmatrix}.
\end{align}
This matrix has a zero eigenvalue corresponding to the charged Goldstone boson $G^{+}$, so the physical charged Higgs has a mass 
\begin{align}
 m_{H^{\pm}}^2=v^2\left(\frac{\mu_{12}^2}{v_\Phi v_\chi}-\frac{\lambda_4}{2}\right),
\end{align}
where $v=\sqrt{v_\Phi^2+v_\chi^2}$. The charged mass-eigenstates are obtained as
\begin{align}
\begin{pmatrix}
\chi^+ \\
\Phi^+
\end{pmatrix}=R(\beta)\begin{pmatrix}
G^+\\
H^+
\end{pmatrix}=\begin{pmatrix}
\cos\beta & -\sin\beta \\
\sin\beta &  \cos\beta 
\end{pmatrix}\begin{pmatrix}
G^+\\
H^+
\end{pmatrix}
\text{,~~~~with}~\tan\beta=\frac{v_\Phi}{v_\chi}.
\end{align}
%
Here, in defining the parameter $\tan\beta$, we have followed the Two-Higgs-Doublet-Model (2HDM) convention.
However, in contrast to standard 2HDM models, where $\tan\beta$ is usually constrained by perturbativity of Yukawa couplings,
in our case $\tan\beta$ can be naturally very large approaching $\tan\beta \to \infty$ as $v_\chi \to 0$.  

The mass matrix for CP even neutral scalars in the basis $(h_\chi\,\,h_\Phi)$ is given as 
\begin{align}
 \mathcal{M}_{h}^2=
\begin{pmatrix}
 A & C \\
 C  & B \\
\end{pmatrix}=
\begin{pmatrix}
 \mu_{12}^2\frac{v_\Phi}{v_\chi}+2\lambda_2 v_\chi^2  & -\mu_{12}^2+v_\Phi v_\chi \lambda_{34} \\
-\mu_{12}^2+v_\Phi v_\chi \lambda_{34} &  \mu_{12}^2\frac{v_\chi}{v_\Phi}+2\lambda_1 v_\Phi^2
\end{pmatrix},
\end{align}
where we defined $\lambda_{34} \equiv \lambda_3+\lambda_4$.   The masses of light and heavy eigenstates are given as
\begin{eqnarray}
m_{h}^2&=&\frac{1}{2}[A+B-\sqrt{(A-B)^2+4C^2}], \label{eq:mh0}\\
m_{H}^2&=&\frac{1}{2}[A+B+\sqrt{(A-B)^2+4C^2}]. \label{eq:mH0}
\end{eqnarray}
The lighter mass eigenstate $h$ is identified as the SM Higgs boson discovered at the LHC~\cite{ATLAS:2012yve,CMS:2012qbp}. 
Again following the 2HDM convention, the two mass eigenstates $h$ and $H$ are related with the $h_\chi$, $h_\Phi$ fields through the rotation matrix $R(\alpha)$ as, 
\begin{align}
\begin{pmatrix}
h_\chi \\
h_\Phi
\end{pmatrix}=R(\alpha)\begin{pmatrix}
H\\
h
\end{pmatrix}=\begin{pmatrix}
\cos\alpha & -\sin\alpha \\
\sin\alpha &  \cos\alpha 
\end{pmatrix} \begin{pmatrix}
H\\
h
\end{pmatrix},\,\,\text{with} \,\, \tan 2\alpha=\frac{2C}{A-B}.
\end{align}

The pseudoscalar mass matrix in the basis $(\eta_\chi,\,\, \eta_\Phi)$ is given by 
\begin{align}
 \mathcal{M}_{\eta}^2=
 \begin{pmatrix}
\mu_{12}^2 \frac{v_\Phi}{v_\chi}   & -\mu_{12}^2 \\
  -\mu_{12}^2 &     \mu_{12}^2\frac{v_\chi}{v_\Phi}  \\
 \end{pmatrix}.
\end{align}
One sees that this pseudoscalar mass matrix has a zero-mass eigenvalue, corresponding to the Goldstone boson $G^{0}$ eaten by the $Z$,
while the physical pseudoscalar Higgs has a mass
\begin{align}
m_{A}^{2}=\mu_{12}^{2}\frac{v^2}{v_\Phi v_\chi}.
 \label{eq:Amass}
\end{align}
The mass eigenstates are again obtained by rotating the component fields as 
\begin{align}
\begin{pmatrix}
\eta_\chi \\
\eta_\Phi
\end{pmatrix}=R(\beta)\begin{pmatrix}
G^0\\
A
\end{pmatrix}=\begin{pmatrix}
\cos\beta & -\sin\beta \\
\sin\beta &  \cos\beta 
\end{pmatrix}\begin{pmatrix}
G^0\\
A
\end{pmatrix}\,\,
\text{with} \,\, \tan\beta=\frac{v_\Phi}{v_\chi}.
\end{align}
From Eq.~\ref{eq:Amass} the pseudoscalar mass is proportional to $\mu_{12}$, which comes from the explicit lepton number soft breaking term $\mu_{12}^{2}\Phi^{\dagger}\chi_L$. 
Should this term not be present in the potential, this pseudoscalar would be an unwanted doublet ``majoron'',
ruled out by the measurements of the invisible decay width of the $Z$ boson at LEP~\cite{Joshipura:1992hp,ParticleDataGroup:2020ssz}.
Such a ``majoron" would also be copiously produced in stars, leading to an astrophysical disaster. The most straightforward way to avoid this is to give it a mass, through Eq.~\ref{eq:Amass}.

An alternative possibility to implement the spontaneous breaking of lepton number symmetry would be to ``invisibilize'' the majoron by adding another singlet scalar carrying lepton number.   
This possibility has already been examined and we refer the interested reader to~\cite{Fontes:2019uld}.
\par As we assume explicit lepton number violation, the Higgs potential is the minimal one. 
One can describe all its quartic couplings in terms of the just four physical masses, $m_h$, $m_H,m_A$ and $m_{H^\pm}$, and the angles $\beta$ and $\alpha$. 
Indeed, the quartic couplings $\lambda_1,\lambda_2,\lambda_3$ and $\lambda_4$ can be expressed as
\begin{align}
 \lambda_1&=\frac{1}{2v^2\sin^2\beta}\Big(m_H^2\sin^2\alpha+m_h^2\cos^2\alpha-m_A^2\cos^2\beta\Big),\label{eq:lam1}\\
 \lambda_2&=\frac{1}{2v^2\cos^2\beta}\Big(m_h^2\sin^2\alpha+m_H^2\cos^2\alpha-m_A^2\sin^2\beta\Big),\label{eq:lam2}\\
 \lambda_3&=\frac{1}{v^2}\Big(2m_{H^\pm}^2-m_A^2+\frac{(m_H^2-m_h^2)\sin (2\alpha)}{\sin (2\beta)}\Big),\label{eq:lam3}\\
 \lambda_4&=\frac{2}{v^2}\Big(m_A^2-m_{H^\pm}^2\Big),\label{eq:lam4}
\end{align}
where the VEV $v=246$~GeV. Note that since $v^2\sin^2\beta = v^2_\Phi$ and $v^2\cos^2\beta = v^2_\chi$,
  the quartic couplings $\lambda_1 \propto \frac{1}{v^2_\Phi}$ whereas $\lambda_2 \propto \frac{1}{v^2_\chi}$.
  This has important implications for the mass spectrum of the scalars as we discuss next.

\subsection*{Compressed spectrum} 
\begin{figure}[!htbp]
	\includegraphics[width=0.4\linewidth]{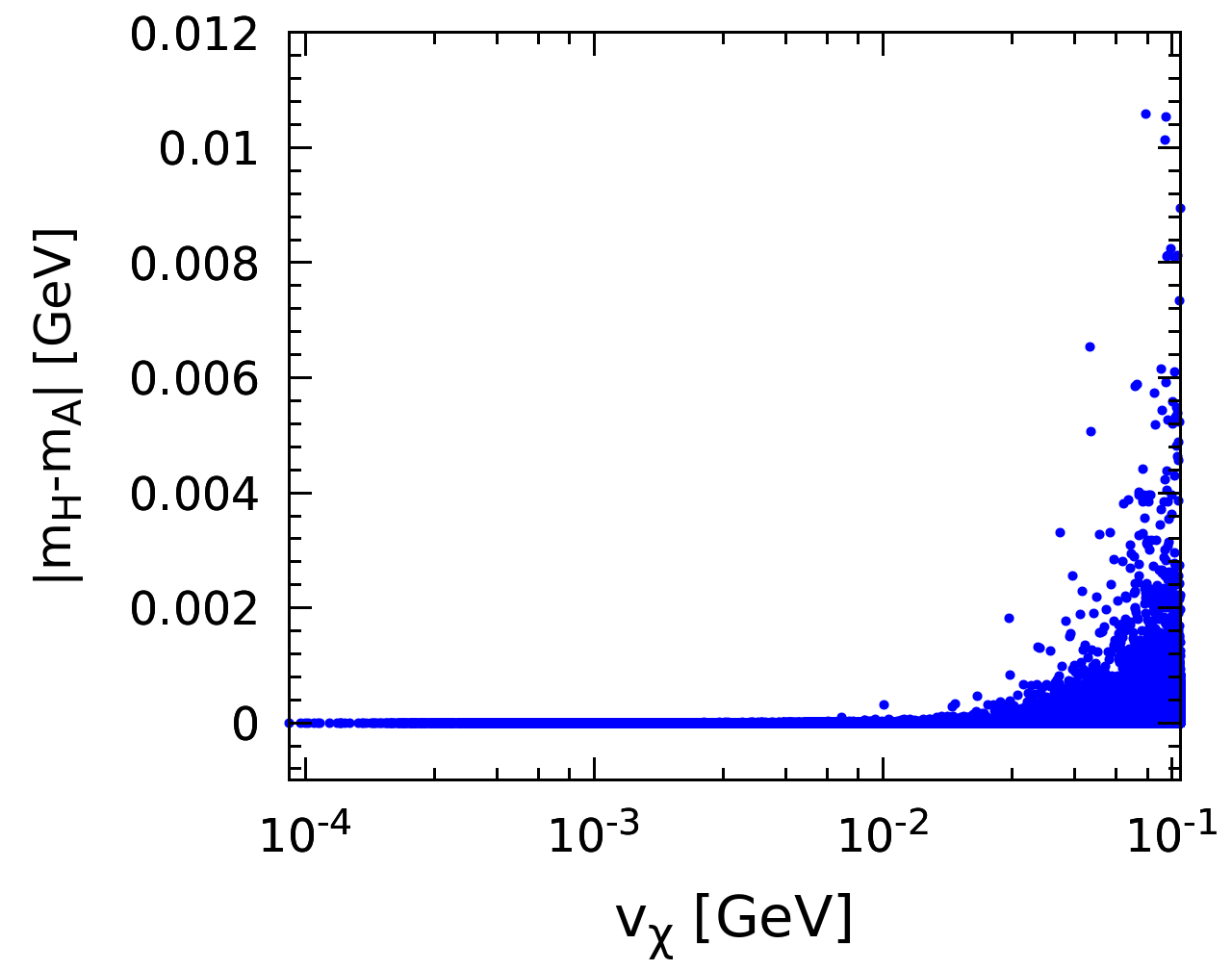}		
    	\includegraphics[width=0.4\linewidth]{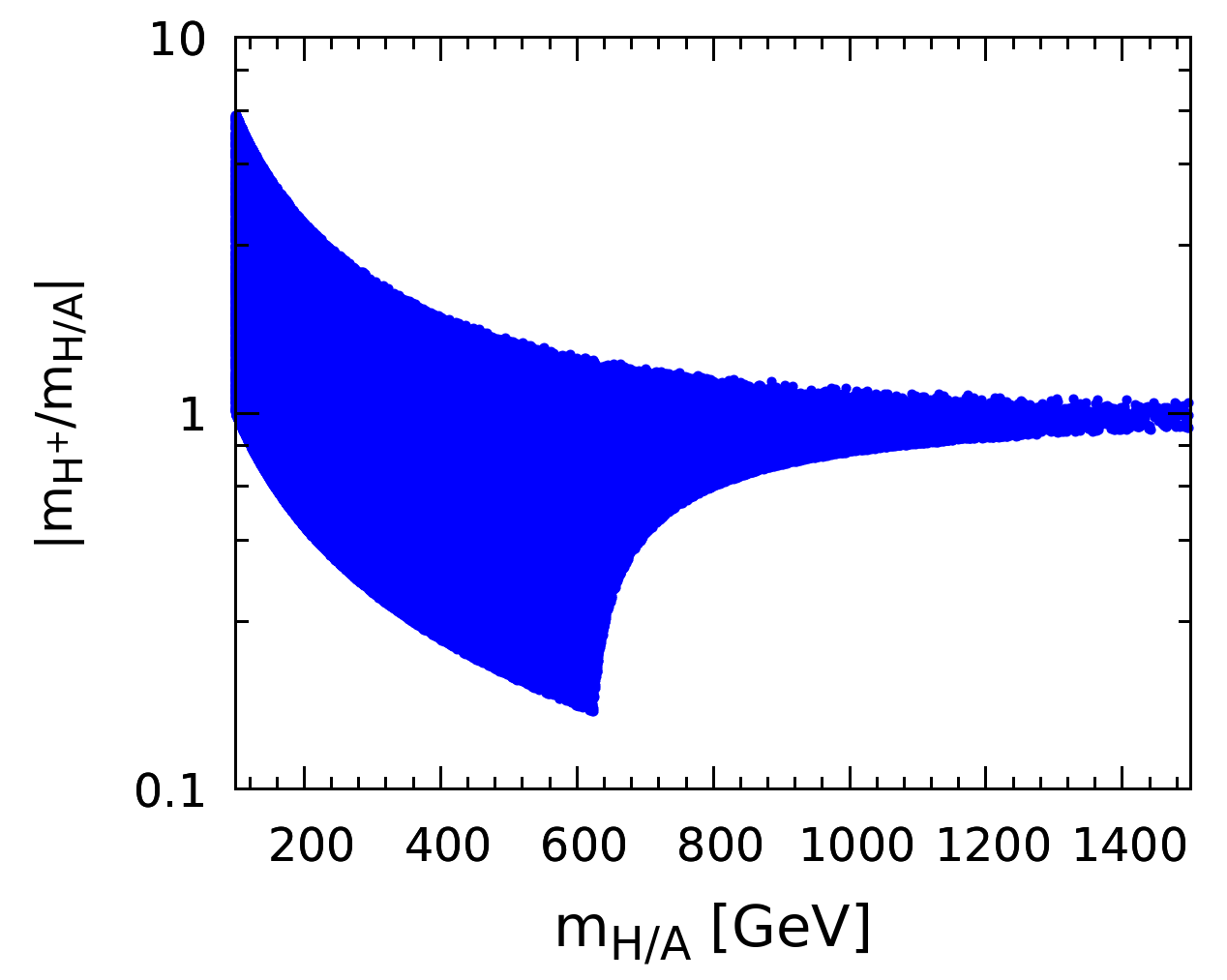}			
	\caption{
          Left panel: $|m_{H}-m_A|$ versus the lepton number breaking scale $v_\chi$.
          Right panel: $|m_{H^\pm}/m_{H/A}|$ versus $m_{H/A}$, for $v_\chi<10^{-1}$ GeV, so that $\alpha\approx 0$ and $m_{H}\approx m_A$.}
	\label{fig:compress}
\end{figure}

As we saw in Eq.~\ref{eq:ML}, the smallness of neutrino mass requires a very small value of the lepton number breaking scale $v_\chi$ but allows us to have a relatively large Yukawa coupling $Y_S$.
We note that scalar spectrum tends to be very compressed when $v_\chi$ is small.
This is required in order for $\lambda_2$ to be in the perturbative regime, as can be seen from the expression of $\lambda_2$ in Eq.~\ref{eq:lam2}.
  Indeed, as $\lambda_2$ is inversely proportional to $v_\chi^2$ and, for tiny $v_\chi$, a small numerator is achieved when $m_{H}\approx m_A$ and $\alpha\approx 0$.
The left panel of Fig.~\ref{fig:compress} shows the splitting $|m_H-m_A|$ with respect to $v_\chi$. One clearly sees that the splitting $|m_H-m_A|$ is very small as long as $v_\chi$ is small.
\par In the limit $\alpha\approx 0$, $m_H\approx m_A$ and $v_\chi\ll v_\Phi$, the quartic couplings $\lambda_3$, $\lambda_4$ simplify to $\lambda_3\approx (2m_{H^\pm}^2-m_A^2)/v^2$, $\lambda_4\approx 2(m_A^2-m_{H^\pm}^2)/v^2$.
  This suggests that the mass-splitting between $m_{H^\pm}$ and $m_{H/A}$ must be restricted in order to keep $\lambda_3$, $\lambda_4$ within the perturbative regime.
   This is shown in the right panel of Fig.~\ref{fig:compress} where we have plotted the ratio $|m_{H^\pm}/m_{H/A}|$ with respect to $m_{H/A}$.
    One sees that $H^\pm$ and $H/A$ can be non-degenerate in mass when $m_{H/A}$ is relatively small. However, for $m_{H/A}>1$~TeV, all scalars, $H^\pm$ and $H/A$, become very degenerate in mass.
    Note that all of this changes if $v_\chi$ is relatively large.
    In such a case, the mixing angle $\alpha$ can be large, and the mass degeneracy between $H, A$, and $H^\pm$ can be lifted.

\section{electroweak precision parameters $S$, $T$ and $U$}
\label{sec:STU}
The presence of the extra doublet $\chi_L$ in the linear seesaw model modifies the prediction for different radiative corrections, especially the oblique parameters $S$, $T$, $U$~\cite{Peskin:1991sw}.
The general form of these are given in Ref.~\cite{Batra:2022arl}. For the case of small $v_\chi$, the oblique parameters $S$, $T$ and $U$ take the following simple form: 
\begin{align}
T\approx \frac{1}{8\pi s_W^2 m_W^2} F(m_{H^\pm}^2,m_H^2),\,\, S\approx \frac{1}{12\pi}\text{log}\left(\frac{m_H^2}{m_{H^\pm}^2}\right),\,\, U\approx \frac{1}{12\pi} G(\frac{m_{H^\pm}^2}{m_W^2},\frac{m_H^2}{m_W^2}),
\label{eq:STU}
\end{align}
where the functions $F$ and $G$ are as shown in Ref.~\cite{Batra:2022arl}:
\begin{align}
F \left( x, y \right) &= \left\{
\begin{array}{l}
\displaystyle{\frac{x + y}{2} - \frac{x y}{x - y}\, \ln{\frac{x}{y}}}
\ \Leftarrow x \neq y,
\\
0 \ \Leftarrow x = y.
\end{array}
\right.
\\ \no
G \left( x, y \right) &= - \frac{16}{3}
+ 5 \left( x + y \right) - 2 \left( x - y \right)^2
+ 3 \left[ \frac{x^2 + y^2}{x - y}
- x^2 + y^2 + \frac{\left( x - y \right)^3}{3} \right] \ln{\frac{x}{y}}
\\ \nonumber
&+ \left[ 1 - 2 \left( x + y \right) + \left( x - y \right)^2 \right]
f \left( x + y - 1,
1 - 2 \left( x + y \right) + \left( x - y \right)^2 \right),
\label{rtycn} 
\end{align}
where
\begin{equation}
f \left( z, w \right) = \left\{
\begin{array}{l}
\displaystyle{\sqrt{w}\,
	\ln{\left| \frac{z - \sqrt{w}}{z + \sqrt{w}} \right|}}
\ \Leftarrow w > 0,
\\*[3mm]
0 \ \Leftarrow w = 0,
\\
\displaystyle{2 \sqrt{-w}\, \arctan{\frac{\sqrt{-w}}{z}}}
\
\Leftarrow w < 0.
\end{array} \right.
\end{equation}
Note that the function $G$ crucially depends on the mass splitting $m_{H^\pm}-m_{H/A}$ and goes to zero in the limit $m_{H^\pm}\approx m_{H/A}$.
Hence, within the linear seesaw model, the $U$ parameter is highly suppressed for small $v_\chi$. 
When $U$ is fixed at zero, the current global fit of electroweak precision data gives~\cite{ParticleDataGroup:2020ssz}:
\begin{align}
S=0.00\pm 0.07,\,\,\,   T= 0.05\pm 0.06.
\label{eq:STUfit}
\end{align}
Combining Eq.~\ref{eq:STU} with Eq.~\ref{eq:STUfit}, we obtain the following constraint on the mass-splitting:  
\begin{align}
|m_{H^\pm}-m_{H/A}|\leq 80\text{ GeV at }90\% \text{ C.L.}
\end{align}
On the other hand, a very recent measurement of $W$ boson mass at CDF shows about $7\sigma$ deviations from the SM predictions~\cite{CDF:2022hxs}.
If one takes this measurement seriously, the global electroweak fit will lead to~\cite{Lu:2022bgw}:
\begin{align}
S=0.15\pm 0.08,\,\, T=0.27\pm 0.06,
\end{align}
with the correlation $\rho_{ST}=0.93$. Hence, in order to accommodate this new $W$ mass measurement, one needs a sizable central value of the $T$ parameter.  
The latter is very sensitive to the mass-splitting $m_{H^\pm}-m_{H/A}$ and vanishes for $m_{H^\pm}=m_{H/A}$. 
As a result, in order to explain the CDF-II measurements the $H/A$ should not be exactly degenerate in mass with $H^\pm$.  
We have shown in our previous paper~\cite{Batra:2022arl} that the $W$ boson mass will be compatible with the CDF-II measurements (at 3-$\sigma$) only when this mass difference lies in the following region:
\begin{align}
50\text{ GeV } \leq |m_{H^\pm}-m_{H/A}| \leq 120 \text{ GeV at }95\% \text{ C.L.}
\end{align}
Note that although the absolute scale of the charged Higgs boson mass is not fixed, the CDF-II result suggests that it must lie below a few TeV.

\section{Charged lepton flavour violation} 
\label{sec:lfv}

\begin{figure}[!htbp]
	\centering
	\includegraphics[width=0.33\linewidth]{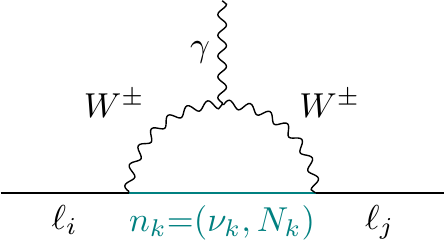}\quad\quad
	\includegraphics[width=0.33\linewidth]{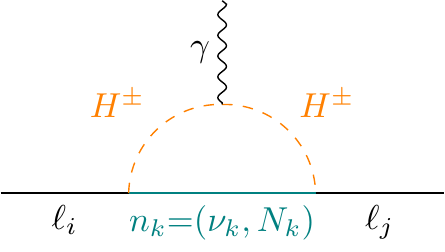}
	\caption{ Feynman diagrams for charged LFV processes e.g. $\ell_i\to \ell_j\gamma$ present in the model.}
	\label{fig:LFV0}
\end{figure}

The Yukawa interactions are not only responsible for neutrino mass generation, but they also give rise to charged lepton flavour violation~(cLFV). 
In this section we provide the theoretical formulas for the two-body decay amplitudes $\ell_i\to\ell_j\gamma$.
In Fig.~\ref{fig:LFV0} we show relevant Feynman diagrams for $\ell_i\to\ell_j\gamma$ in the mass basis. 

The first diagram is the conventional one involving light-heavy neutrino mixing in the effective charged current interactions.
Ref.~\cite{Schechter:1980gr} provides a thorough description of the effective lepton mixing matrix $K$ that characterizes the charged current weak interaction of
mass-eigenstate neutrinos in any kind of seesaw model. It can be expressed in rectangular form  
 \begin{align}
 -\mathcal{L}_{\rm CC}=\frac{g}{2\sqrt{2}}\sum_{\alpha=1}^3\sum_{\beta=1}^9 K_{\alpha\beta}\bar{\ell}_{\alpha}\gamma_\mu (1-\gamma_5) n_\beta W^\mu + \text{H.c.}, 
 \end{align}
 where $n_\beta=(\nu, N)$. In the diagonal basis for the charged lepton mass matrix, we can write the $K$ matrix as follows,
 \begin{align}
 K=\begin{pmatrix}
 K_L  &  K_H 
 \end{pmatrix},
 \end{align}
 where $K_L$ is a 3 by 3 matrix and $K_H$ is a 3 by 6 matrix. The submatrices $K_L$ and $K_H$ are not unitary. 
 From Eq.~\ref{u-bdiag}, we can express $K_L$ and $K_H$ as

\begin{align}
& K_L=\left(1-\frac{1}{2}m_D M_R^{-1T}M_R^{-1}m_D^\dagger\right)U_{\rm lep} =\left(1-\frac{1}{2}V V^\dagger\right)U_{\rm lep},\\
& K_H=\left(-\frac{i}{\sqrt{2}}V  \,\,\,\,\, \frac{1}{\sqrt{2}} V\right).
\end{align}
We parametrize the deviations from unitarity as follows:
\begin{align} K_L=(1-\eta) U_{\rm lep} \text{ with } \eta =
\frac{1}{2} V V^\dagger.
\end{align}

For the template scheme with massless neutrinos, these blocks can be
parametrized in a vey simple manner, see e.g. eqs.(11,12)
in~\cite{Dittmar:1989yg}, because of the high degree of symmetry.
It follows that the matrix $\eta$ characterizes unitarity deviation in
the light-active $3\times 3$ sub-block of the lepton mixing
matrix~\footnote{
  For high scale type-I seesaw, the deviations from unitarity are negligible, $V\sim 10^{-10}$, however they can lead to a rich phenomenology in low-scale seesaw.}.
As this non-unitarity parameter can be relatively large for low-scale seesaw schemes, such as the linear seesaw,
it will break the Glashow-Illiopoulos-Maiani~(GIM) cancellation mechanism for the light-neutrino contribution~\cite{Glashow:1970gm},
see the left panel of Fig.~\ref{fig:LFV0}.
As far as the charged-current interaction is concerned, besides the enhanced light-neutrino contribution,
the $\ell_i\to\ell_j\gamma$ decay also proceeds through the exchange of the six sub-dominantly coupled heavy states~\cite{Forero:2011pc}.

The radiative decay rate is given by~\cite{Minkowski:1977sc,Marciano:1977wx,Cheng:1980tp,Lim:1981kv,Langacker:1988up},
\begin{align} \text{BR}(\ell_i\to\ell_j\gamma)^{\rm
CC}_n=\frac{\alpha_w^3
s_w^2}{256\pi^2}\Big(\frac{m_{\ell_i}}{M_W}\Big)^4\Big(\frac{m_{\ell_i}}{\Gamma_{\ell_i}}\Big)\Bigg|\sum_{k=1}^9
K_{ik}^{*}K_{jk}G_{\gamma}^W\left(\frac{m_{n_k}^2}{M_W^2}\right)\Bigg|^2\label{eq:LFV-BR1},
\end{align}
where $\alpha_w=g_w^2/4\pi$, $s_w^2=\sin^2\theta_w$. The loop function
$G_\gamma^W(x)$ is given as:
\begin{align} & G_\gamma^W(x)=\frac{1}{12(1-x)^4}\left(10-43 x+78
x^2-49 x^3+18x^3\text{ln}x+4x^4\right).
\label{eq:GFunc}
\end{align}

In order to examine the variation of the cLFV rates in parameter space we perform a scan procedure using the approximate Casas-Ibarra-like expression~\cite{Casas:2001sr} of the
Dirac Yukawa couplings in terms of oscillation parameters given as~\cite{Forero:2011pc,Cordero-Carrion:2018xre,Cordero-Carrion:2019qtu},
 \begin{align} Y_\nu=\frac{\sqrt{2}}{v_\Phi}U_{\rm lep}
\text{diag}\{\sqrt{m_i}\} \, {\mathcal A^T} \text{diag}\{\sqrt{m_i}\}
\, U_{\rm lep}^T \left(M_L^T\right)^{-1} \,M_R^T,
 \label{eq:Ynu}
 \end{align}
 where $U_{\rm lep}$ is approximately the mixing matrix determined in oscillation experiments~\cite{deSalas:2020pgw},
 $m_i$ are the three light neutrino masses and ${\mathcal A}$ has the following general form:
\begin{equation}\label{A}
  {\mathcal A} = \left(
\begin{array}{ccc} \frac{1}{2} & a & b\\
-a & \frac{1}{2}& c\\ -b & -c & \frac{1}{2}
\end{array} \right),
\end{equation}
with $a,b,c$ are real numbers. Using this analytical parametrization optimizes the scan, ensuring that only viable solutions
consistent with oscillation data are included.
Having said that, we stress that in the numerical code, the exact expressions are used in order to ``extract'' the Yukawas
from the measured neutrino observables.

\begin{figure}[!htbp]
	\includegraphics[width=0.49\linewidth]{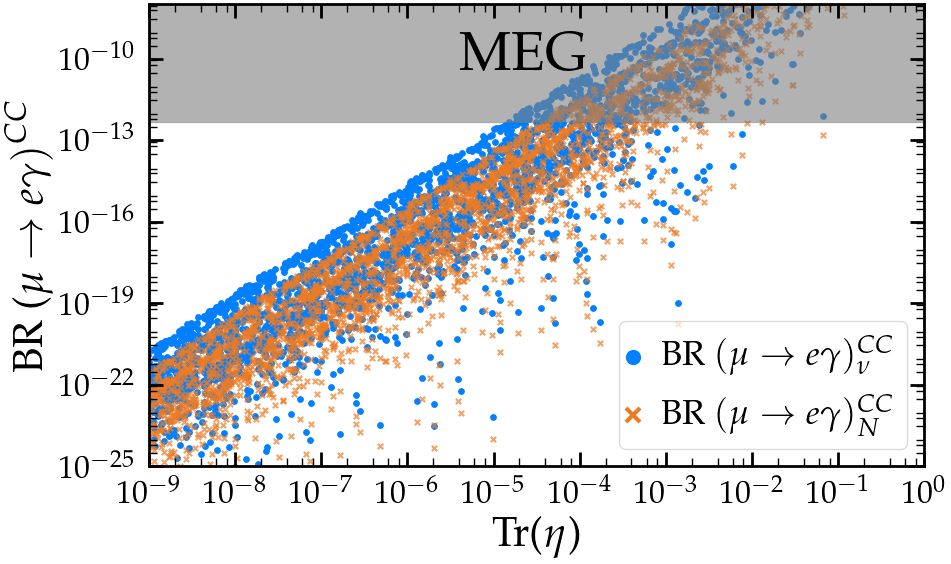}
	\caption{ Charged current contribution to $\mu\to
	e\gamma$ as a function of the unitarity violation parameter
	$\eta$.
          A full calculation is performed varying the parameters
          described in Table~\ref{tab:param} as explained in the text.
          The horizontal band is excluded by the MEG
          experiment~\cite{MEG:2013oxv}.}
	\label{fig:LFVplotsCC}
\end{figure}

\begin{table}[!htbp]
	\begin{center}
		\begin{tabular}{|c|c| }
			\hline Parameter & Range \\ \hline
			&\\[-12.5pt] $m_{H^\pm}$ & $ [100, 2000]
			\text{ GeV} $ \\[2.5pt] $M_{N_i}$ & $[1, 100]
			\text{ TeV} $ \\[2.5pt] $a,\,b,\,c$ &
			$[0,10^{-2}]$ \\[2.5pt] $ \alpha$ &
			$[-\frac{\pi}{2},\frac{\pi}{2}]$ \\ [2.5pt]
			$v_\chi$ & $ [10^{-9}, 1] \text{ GeV} $
			\\[2.5pt] $Y_S^{ii}$ & $[10^{-4},\sqrt{4\pi}]$
			\\[2.5pt] \hline
	\end{tabular}
	\end{center}
	\caption{ Parameter range used for the numerical scan
	of cLFV processes.}
	\label{tab:param}
\end{table}
In Fig.~\ref{fig:LFVplotsCC}, we show the charged current contribution
to the $\mu\to e\gamma$ rate involving light neutrinos~(blue) and heavy neutrinos~(orange)
as a function of the relevant unitarity violation parameter $\eta$ varying the parameter space according to Table.~\ref{tab:param}. As
$\eta$ is a $3\times 3$ matrix, we choose to show the results in terms
of the parameter $\text{Tr}(\eta)$.
From Fig.~\ref{fig:LFVplotsCC}, one sees that the light neutrino
contribution~(blue points) can exceed that coming from the heavy
neutrinos~(orange points).
  It is clear from this figure that the charged current contribution
  need not be suppressed by the small neutrino masses, and can lead to
  observable rates even in the limit of vanishing $v_\chi$ as
  neutrinos become
  massless~\cite{Bernabeu:1987gr,Branco:1989bn,Rius:1989gk,Gonzalez-Garcia:1991brm,Deppisch:2004fa,Deppisch:2005zm}.
 The GIM cancelation is broken in the light neutrino sector due to the
 non-unitarity of the leptonic mixing matrix.
 In particular, the contribution involving light neutrino exchange can
 be as large as that coming from heavy neutrinos.
 
 Notice that cLFV processes also receive contributions coming
from the Yukawa interactions. The relevant Yukawa interactions are
given as
 \begin{align} &-\mathcal{L}^{\rm Yuk}_{N}\approx  \frac{\sin\beta}{2\sqrt{2}}\sum_{\alpha,\beta=1}^3Y_S^{\alpha\beta}\bar{\ell}_{\alpha}(1+\gamma_5)(i N_{2\beta+2}+N_{2\beta+3})H^{-}+\text{H.c.}, \label{eq:LFV-Yuk2}\\ &-\mathcal{L}^{\rm
 Yuk}_{\nu}\approx\frac{v\sin^2\beta}{2\sqrt{2}}\sum_{\alpha,\beta=1}^3(Y_S M_R^{-1} Y_\nu^\dagger)_{\alpha\beta}\bar{\ell}_\alpha(1+\gamma_5)\nu_\beta H^{-} + \text{H.c.} 
 \end{align}
  The individual contributions to the rate for the radiative
 $\ell_i\to\ell_j\gamma$ decay coming from these Yukawa terms is given
 by~\cite{Lavoura:2003xp},
\begin{align} & \text{BR}(\ell_i\to\ell_j\gamma)^{\rm
Yuk}_N=\frac{\alpha_{\rm
em}}{4}\Bigg(\frac{m_{\ell_i}^5}{\Gamma_{\ell_i}}\Bigg)\Bigg|(Y_{p}
Y_{p}^\dagger)_{ji}I_p\left(M_{H^\pm},\frac{M_N^2}{M_{H^\pm}^2}\right)\Bigg|^2\label{eq:LFV-BR2},\\
& \text{BR}(\ell_i\to\ell_j\gamma)^{\rm Yuk}_\nu=\frac{1}{(192 \pi^2
M_{H^\pm}^2)^2}\frac{\alpha_{\rm
em}}{4}\Bigg(\frac{m_{\ell_i}^5}{\Gamma_{\ell_i}}\Bigg)\Bigg|\frac{v^2\sin^4\beta}{2M_N^2}(Y_S
Y_\nu^\dagger Y_\nu Y_S^\dagger)_{ji}\Bigg|^2,
\label{eq:LFV-BR3}
\end{align}
 where $\alpha_{\rm em}=e^2/4\pi$ and $Y_P=Y_S\sin\beta$. The
loop function $I_{p}(m_B,x)$ has the following form:
\begin{align} & I_p(m_B,x) = -\frac{1}{16\pi^2
m_B^2}\Big(\frac{(3x-1)}{4(x-1)^2}-x^2\frac{\text{log}\,x}{2(x-1)^3}\Big)+\frac{3}{32\pi^2
m_B^2}\Big(\frac{11x^2-7x+2}{18(x-1)^3}\nonumber \\ &
-x^3\frac{\text{log}\,x}{3(x-1)^4}\Big).
\end{align}

\begin{figure}[!htbp]
	\includegraphics[width=0.49\linewidth]{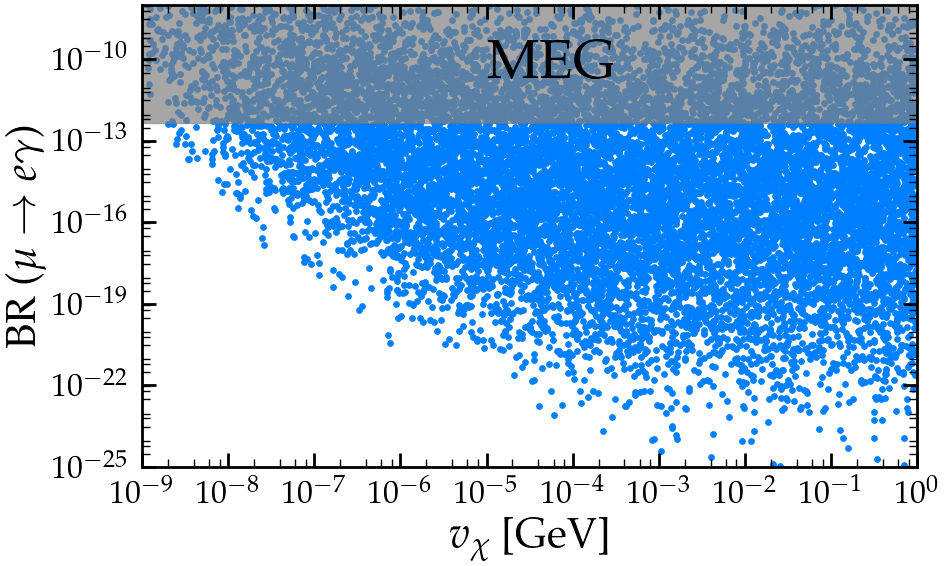}
	\caption{
          Total $\text{BR}(\mu\to e\gamma)$ as a function of $v_\chi$.
          We have varied the parameters according to Table~\ref{tab:param} as explained in the text.
          The upper band indicates the limit from the MEG experiment~\cite{MEG:2013oxv}.}
	\label{fig:LFV}
\end{figure}

\begin{figure}[!htbp]
	\includegraphics[width=0.7\linewidth]{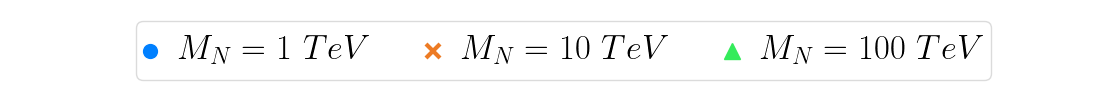}\vspace{-0.4em}
	\includegraphics[width=0.49\linewidth]{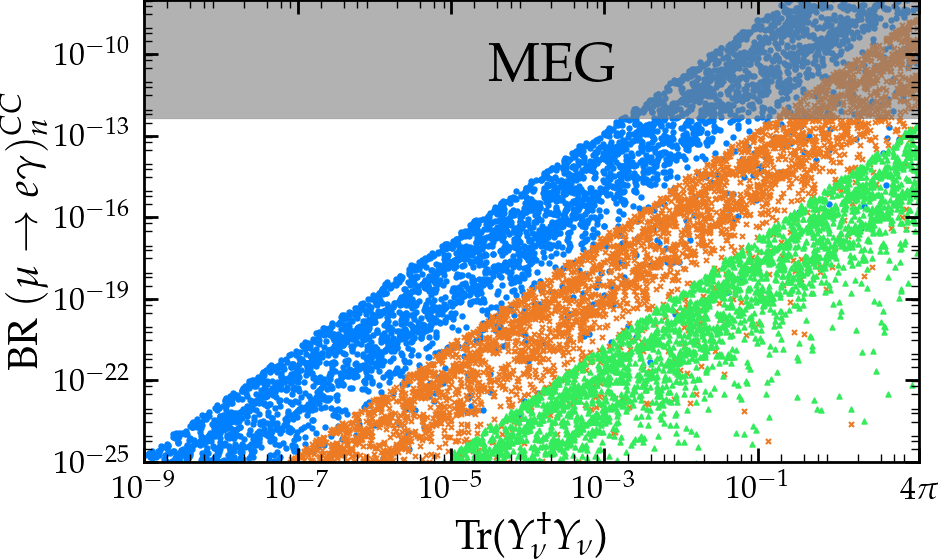}
	\includegraphics[width=0.49\linewidth]{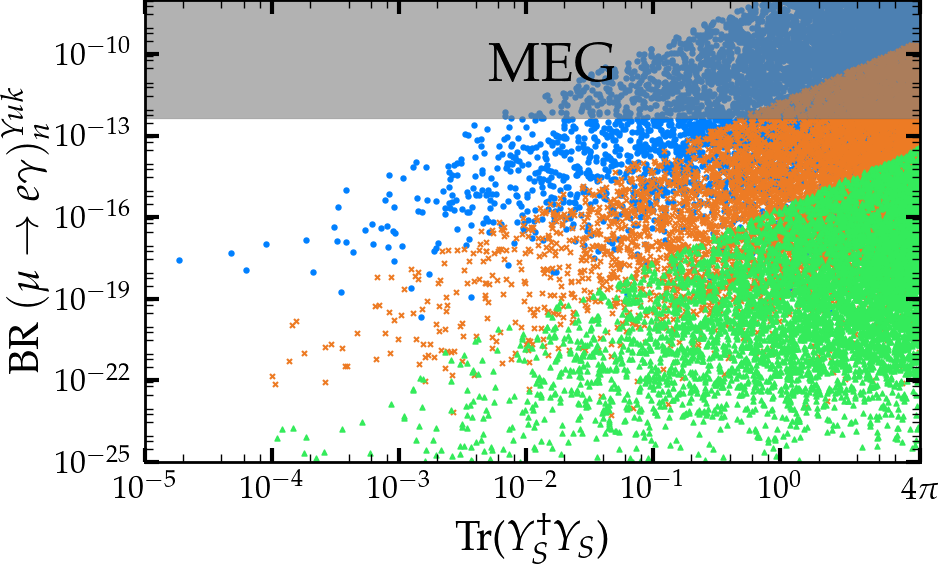}
	\caption{
           $\text{BR}(\mu\to e\gamma)$ for various mediator mass
           values $M_N$.  Left panel: Charged current contribution
           versus $\text{Tr}(Y_\nu^\dagger Y_\nu)$.  Right panel:
           Yukawa contribution versus $\text{Tr}(Y_S^\dagger Y_S)$. The
           horizontal band indicates the limit from the MEG
           experiment~\cite{MEG:2013oxv}. The charged Higgs boson mass is
           taken as $m_{H^\pm}=$ 100 GeV, 1 TeV and 2
           TeV (blue, orange and green points, respectively).  Other
           parameters are varied as in Table~\ref{tab:param}. }
	\label{fig:LFVplots}
\end{figure}

In Fig.~\ref{fig:LFV} and ~\ref{fig:LFVplots}, we show the resulting
cLFV decay rates as a function of the relevant parameters, varied
according to Table.~\ref{tab:param}.
We impose the additional constraint of perturbativity of the Yukawa
couplings $Y_\nu$ and $Y_S$.
Notice that in the limit $v_\chi \to 0$, $M_L\to 0$, neutrinos are
massless and the Yukawa coupling $Y_\nu$ becomes unrestricted by
neutrino mass limits.
The latter affects the magnitude of $M_L$, which can be dynamically
suppressed by $v_\chi $ even if the Yukawa couplings $Y_\nu$ and $Y_S$
are sizeable.
As a result one can have large contributions to cLFV from charged
current interaction.
This is indeed confirmed in Fig.~\ref{fig:LFV}, that shows the total
$\mu\to e\gamma$ rate as a function of $v_\chi$.\\
As noted long
ago~\cite{Bernabeu:1987gr,Branco:1989bn,Rius:1989gk,Gonzalez-Garcia:1991brm,Deppisch:2004fa,Deppisch:2005zm},
the fact that the charged-current contributions to cLFV can be
sizeable is a generic feature of low-scale seesaw mechanisms,
including also the inverse seesaw
mechanism~\cite{Mohapatra:1986bd,Gonzalez-Garcia:1988okv}.
A novel feature of the linear seesaw is the presence of a second
Yukawa interaction characterized by the coupling matrix $Y_S$ and
involving a second doublet Higgs scalar.
Indeed, this Yukawa coupling $Y_S$ can be sizeable for small $v_\chi$,
as it is not directly restricted by the neutrino mass constraint.
In the left and right panels of Fig.~\ref{fig:LFVplots}, we show
the total $\text{BR}(\mu\to e \gamma)$ as a function of
$\text{Tr}(Y_\nu^\dagger Y_\nu)$ and $\text{Tr}(Y_S^\dagger Y_S)$,
respectively.
One sees that the Yukawa contributions to the $\mu\to e\gamma$ decay
rate can exceed those of the charged current and also exceed the
present experimental bound from the MEG experiment~\cite{MEG:2013oxv}
for reasonable choices for the Yukawa couplings $Y_\nu$ and $Y_S$.

In summary, as the main message, we stress that the rates for cLFV
processes need not be ``neutrino-mass-suppressed'' so that cLFV
processes can be non-zero even in the massless neutrino limit.
Moreover, the linear seesaw framework brings in novel and potentially dominant cLFV contributions associated with the Yukawa sector and the charged scalar boson.\\[-.3cm]

Before closing, we comment on another class of relevant processes, involving (total) lepton number violation, such as neutrinoless double beta decay~($0\nu\beta\beta$).
In addition to light and heavy neutrino contributions, within the linear seesaw model there will be a contribution to $0\nu\beta\beta$ from the charged Higgs boson exchange.
However, very much like the charged-Higgs-boson contribution~\cite{Schechter:1981bd} present in the triplet seesaw mechanism~\cite{Schechter:1980gr}, 
the charged-Higgs-boson of the linear seesaw is also strongly ``leptophilic'' in the small $v_\chi$ limit, leading to a negligible contribution to $0\nu\beta\beta$.
See Ref.~\cite{Batra:2023ssq} for further discussion on lepton number violating processes.

\section{Collider Constraints}
\label{sec:collider}  
\par 
At the LHC, the  additional neutral Higgs scalars $H$ and $A$ are produced dominantly through gluon-gluon fusion, generated by top (t) and bottom (b) quark exchange in the loops~\cite{ATLAS:2017eiz,ATLAS:2020zms}.
The CMS and ATLAS collaborations have searched for such new scalars decaying to various SM channels. 
The official CMS and ATLAS searches~\cite{CMS:2015lsf,ATLAS:2018gfm,ATLAS:2018ntn} employ $gg\to tbH^\pm$ and $gb\to t H^\pm$ as the production channels for the singly charged Higgs. 
However, within our linear seesaw model, the relevant couplings involved in these processes are suppressed as $\mathcal{O}(v_\chi/v)$, hence these constraints are not directly applicable. 

The LEP experiments have looked for pair production of charged Higgs bosons through the process $e^+e^-\to\gamma/Z\to H^\pm H^\mp$. 
Although the couplings that appear in the production process are gauge couplings, the $H^\pm$ decay to hadronic states are again suppressed as $\mathcal{O}(v_\chi/v)$~\cite{ALEPH:2013htx}.
In conclusion, due to the suppressed $H, A$ and $H^\pm$ couplings to SM particles, all the constraints coming from searches for additional scalars at LHC and LEP can be easily satisfied.
Thus $H,A$ and $H^\pm$ are allowed to have broad mass ranges.

Moreover, the precise measurements of the $W$ and $Z$ widths at LEP require~\cite{Cao:2007rm,Gustafsson:2007pc}: 
\begin{equation}
m_H+m_A \text{, } 2m_{H^\pm}>m_Z \text{, and } m_{H/A}+m_{H^\pm}>m_W.
\end{equation}
In the limit $v_\chi\to 0$, $\alpha$ and $\beta$ can be approximated as $\alpha\approx 0$ and $\beta\approx \pi/2$.  
Therefore, $\Phi$ behaves almost identically to the SM Higgs doublet, so we do not anticipate any observable deviation from the Higgs couplings to the SM particles.
A possible exception are loop-induced couplings, such as $h\gamma\gamma$, to which we turn next.

\subsection*{Constraints from Higgs Physics} 
\begin{figure}[!htbp]
	\centering
	\includegraphics{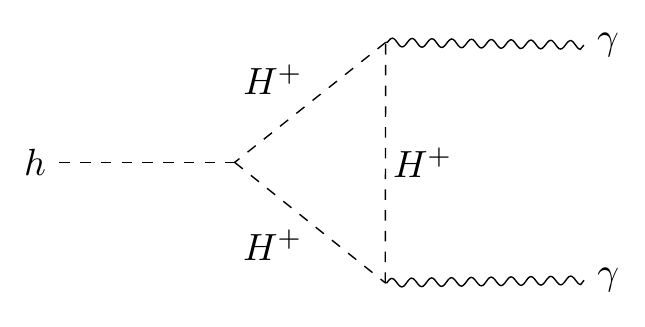}
	\caption{ $h \to \gamma \gamma$ process mediated by the new charged scalar $ H^+ $ running in the loop.}
	\label{fig:htogamgam}
\end{figure}

The $ h \to \gamma \gamma $ decay width is modified in the linear seesaw model due to the existence of a new physical charged scalar $ H^+ $ running in the loop as shown in Fig. \ref{fig:htogamgam}. 

The $h \to \gamma \gamma$ decay width including this new contribution can be written as~\cite{Posch:2010hx, Batra:2022wsd}
\begin{align}
\Gamma(h \to\gamma\gamma) = 
\frac{G_\mu \alpha^2 m_{h}^3}{128\sqrt{2}\pi^3}  
\bigg|\sum_{f} g_f N_c Q_f^2 A_{1/2}^h(\tau_f) 
+ g_W A_1^h(\tau_W) 
+ g_h A_0^h(\tau_{H^\pm})\bigg|^2,
\end{align}
where
\begin{equation}
g_W=\sin{(\beta-\alpha)}, \hspace{0.5cm}
g_f=\frac{\cos{\alpha}}{\sin{\beta}}, \hspace{0.5cm}
g_h=-\frac{m_W}{g m^2_{H^\pm}}\lambda_{hH^+H^+}.
\end{equation}
$ Q_f $ and $ N_c $ are the electric charge and colour of the fermion $f$, while $ g $ is the weak coupling constant and $ \lambda_{hH^+H^+} $ is the trilinear $ hH^+H^+ $ coupling,  
\begin{equation}
	\lambda_{hH^+H^+}=-\frac{1}{2v \sin(2\beta)}[(m_h^2-2 m_{H^\pm}^2)\cos(\alpha-3\beta)+(3m_h^2+2m_{H^\pm}^2-4 m_A^2)\cos(\alpha+\beta)].
\end{equation}
Notice that, among all SM fermions $f$, the dominant contribution comes from the top quark, followed by a small bottom-quark contribution.
The form factors $A_{1/2}^h$, $A_{1}^h$ and $A_{0}^h$ are given as
\begin{align} 
A_{1/2}^h(\tau) & = 2 [\tau +(\tau -1)f(\tau)]\, \tau^{-2} , \nonumber \\   
A_1^h(\tau) & = - [2\tau^2 +3\tau+3(2\tau -1)f(\tau)]\, \tau^{-2},\nonumber \\
A_{0}^h(\tau) & = - [\tau -f(\tau)]\, \tau^{-2} ,
\label{eq:Ascalar}
\end{align}
where $\tau_i = M^2_{h}/4M^2_i$; $i = f, W, H^{\pm}$ with $M_i$ denoting the mass of the particle running in the $h \to \gamma \gamma$ loop,
and the function $f(\tau)$ is defined as:
\begin{eqnarray}
f(\tau)=\left\{
\begin{array}{ll}  \displaystyle
\arcsin^2\sqrt{\tau} & \tau\leq 1 \\
\displaystyle -\frac{1}{4}\left[ \log\frac{1+\sqrt{1-\tau^{-1}}}
{1-\sqrt{1-\tau^{-1}}}-i\pi \right]^2 \hspace{0.5cm} & \tau>1
\end{array} \right.
\label{eq:ftau}
\end{eqnarray}
To quantify the deviation from the Standard Model prediction, we define the following parameter
\begin{equation}
	R_{\gamma\gamma}=\frac{\text{BR}(h \to\gamma\gamma)}{\text{BR}(h \to\gamma\gamma)_{\rm SM}}.
\end{equation}
The value we use for the Standard Model is $\text{BR}(h \to\gamma\gamma)_{\rm SM}\approx 2.27\times 10^{-3}$. 
This decay mode has been explored by the ATLAS and CMS collaborations, and their combined study of the 8 TeV data yields $R_{\gamma\gamma}^{\rm exp}=1.16_{-0.18}^{+0.20}$~\cite{ATLAS:2016neq}.
There is currently no combined final data for the 13 TeV Run-2, and the available data is separated by the production process~\cite{ATLAS:2019nkf}.  
In our analysis we use the 13~TeV ATLAS result which gives the global signal strength measurement of $R_{\gamma\gamma}^{\rm exp}=1.04_{-0.09}^{+0.10}$~\cite{ATLAS:2022tnm}.  
\begin{figure}[!htbp]
	\centering
	\includegraphics[width=0.5\linewidth]{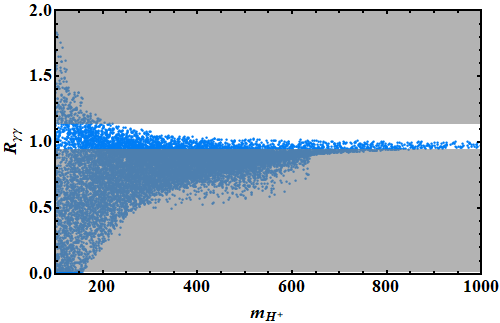}
	\caption{
          Dependence of the diphoton signal strength $ R_{\gamma\gamma} $ on the charged scalar mass.
          For $v_\chi\in [10^{-9}, 10]\,\text{GeV}$ we varied the other relevant parameters as 
          $m_{H^\pm}, m_{H/A}\in [100, 2000]$ and $\alpha\in [-\frac{\pi}{2}, \frac{\pi}{2}]$.
          The shaded regions are excluded from the experimental limits on $ R_{\gamma\gamma} $~\cite{ATLAS:2022tnm}.}
	\label{fig:Rgamgam}
\end{figure}

Fig.~\ref{fig:Rgamgam} shows a scatter plot of $ R_{\gamma\gamma} $ as a function of $ m_{H^\pm} $.
One sees that no charged-Higgs mass values can be ruled out from the experimental limits on $ R_{\gamma\gamma} $ from ATLAS~\cite{ATLAS:2022tnm}.

\section{Production of heavy neutrinos and new scalars}
\label{sec:prod-decay}

Within our linear seesaw scheme, the heavy neutrino mediators $N_i$ as
well as the new scalars $H,A,H^{\pm}$ can all naturally lie below the TeV
scale, hence accessible to direct experimental discovery at future particle colliders.
The possibility of producing the heavy neutrinos that mediate neutrino
mass generation at high energy
colliders~\cite{ATLAS:2019kpx,CMS:2018iaf,CMS:2022nty}
has a long history~\cite{Dittmar:1989yg,Gonzalez-Garcia:1990sbd,Atre:2009rg,AguilarSaavedra:2012fu,Das:2012ii,Deppisch:2013cya,Banerjee:2015gca,Das:2017rsu,Das:2018usr,Drewes:2019fou,Drewes:2022rsk,Cottin:2022nwp,Mekala:2022cmm,Das:2023tna,Abdullahi:2022jlv}.
\begin{figure}[h]
	\includegraphics[width=0.22\linewidth]{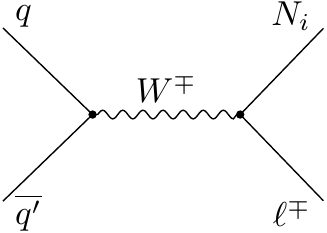}~~~
	\includegraphics[width=0.22\linewidth]{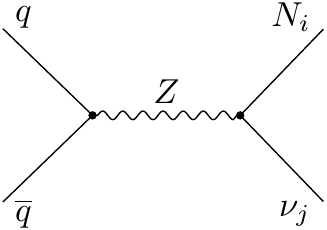}~~~
	\includegraphics[width=0.22\linewidth]{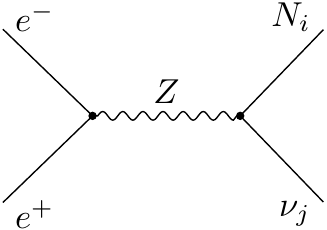}~~~
	\includegraphics[width=0.22\linewidth]{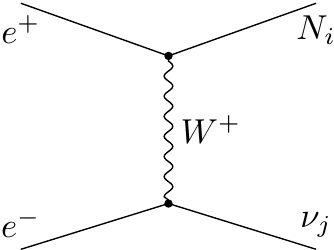}
	\caption{
          Feynman diagrams for single heavy neutrino production through the SM gauge portal at $pp$ and $e^+e^-$ colliders.
          However, these channels are suppressed by light-heavy neutrino mixing.  }
	\label{fig:N-production}
\end{figure}

Indeed, heavy neutrino mediators can be produced at proton-proton or $e^+e^-$ colliders in a variety of ways.
At $pp$ colliders, the most studied production mechanism is the
charged- and neutral-current-induced Drell-Yan process, $pp\to W^{\pm
*}\to\ell^{\pm} N$ and $pp\to Z^{*}\to \nu N$, see
Fig.~\ref{fig:N-production}.
At $e^+e^-$ colliders, the heavy neutrinos can be singly-produced as
$e^+e^-\to\nu N$ through gauge-mediated t and s-channel processes, see
Fig.~\ref{fig:N-production}.
However, barring resonant production~\cite{Dittmar:1989yg}, the heavy neutrino's production cross-section is very small due to light-heavy
neutrino mixing supression~($\mathcal{O}(m_DM_R^{-1})$ in the amplitude).

In what follows we discuss the new unsuppressed production mechanisms
for neutrino mass mediators within our linear seesaw scheme.  We show
how they could produce interesting distinctive signatures at various
collider setups.
%

\subsection{Direct heavy-neutrino pair production at $e^+e^-$  collider}
 \label{sec:direct-prod-pair}

\begin{figure}[h]
	\includegraphics[width=0.22\linewidth]{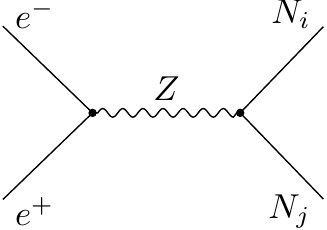}~~~~~
	\includegraphics[width=0.22\linewidth]{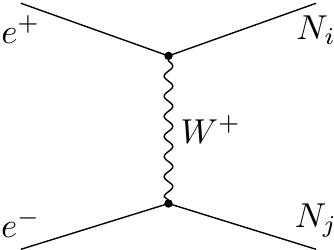}~~~~~
    \includegraphics[width=0.22\linewidth]{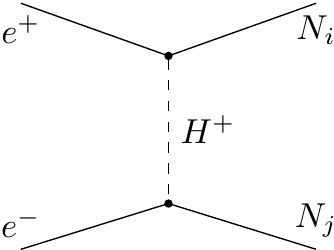}
    \caption{
      Feynman diagrams for heavy-neutrino pair production at an
      $e^+e^-$ collider. The gauge diagrams are suppressed by
      light-heavy neutrino mixing~(first two panels) while the
      t-channel scalar exchange diagram is unsuppressed~(last
      panel). }
	\label{fig:NN-production}
\end{figure}

Our linear seesaw scheme offers new unsuppressed heavy-neutrino
 production mechanisms.
For example, as illustrated in the right panel of
Fig.~\ref{fig:NN-production}, heavy neutrinos can be produced as
$e^+e^-\to NN$ through t-channel exchange of the charged Higgs boson.
In contrast to the first two diagrams in Fig.~\ref{fig:NN-production}
which are light-heavy neutrino mixing suppressed,
the contribution to the production cross section arising from the
Higgs exchange diagram in Fig.~\ref{fig:NN-production} is proportional
to $Y_S^4$.
As we already discussed, for small $v_\chi$, sizeable $Y_S$ values are consistent with small neutrino masses.
The analytical expression for this production cross-section is given explicitly as
\small
\begin{align}
\frac{d\sigma}{d\cos\theta}&=\frac{|Y_S^{1i}|^4}{256\pi\sqrt{s}}\frac{\big(s-4M_{N_i}^2\big)^{\frac{3}{2}}}{\Big(\big(2m_{H^\pm}^2-2M_{N_i}^2+s\big)^2-s\cos^2\theta\big(s-4M_{N_i}\big)^2\Big)^2}\nonumber\\
&
\Big[2\cos^2\theta\Big(2\big(m_{H^\pm}^2-M_{N_i}^2\big)-s^2-2s\big(m_{H^\pm}^2-2M_{N_i}^2\big)\Big)+\big(s+2m_{H^\pm}^2-2M_{N_i}^2\big)^2
\nonumber \\ &+ s\cos^4\theta \big(s-4M_{N_i}^2\big)\Big],
\label{eq:Sigma_eeNN}
\end{align}
\normalsize
\begin{figure}[h]
	\includegraphics[width=0.49\linewidth]{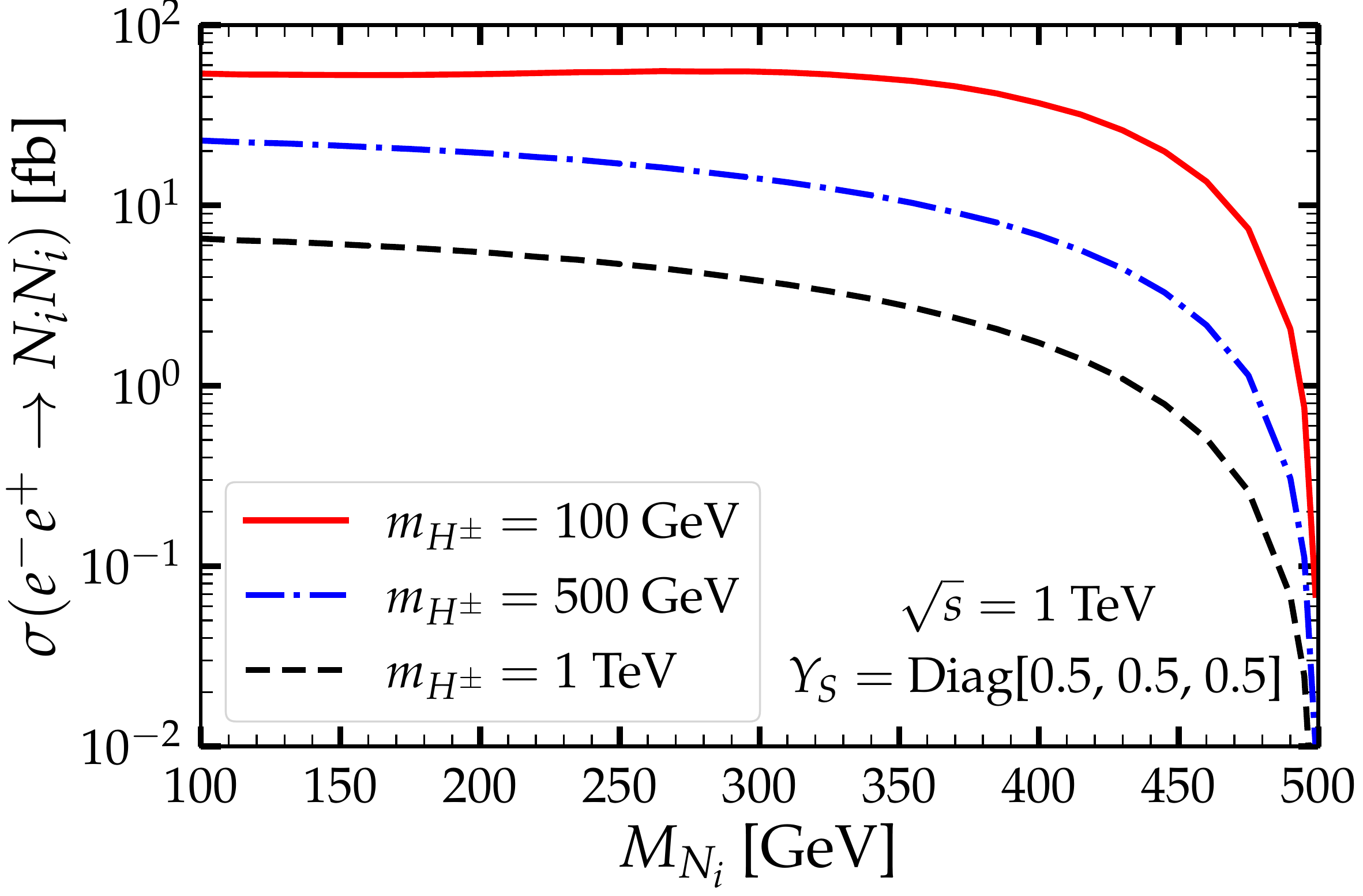}
	\includegraphics[width=0.49\linewidth]{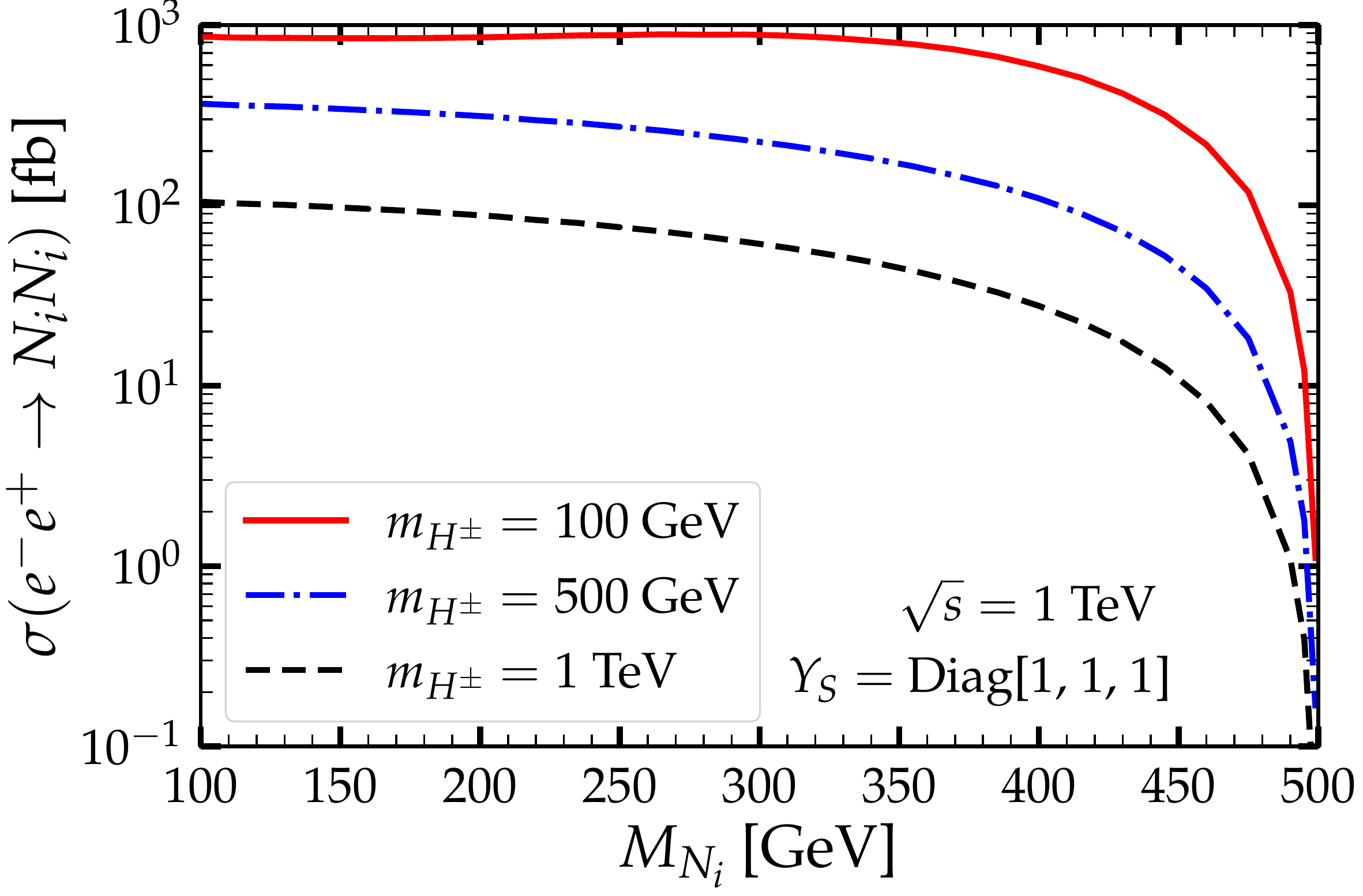}
	\includegraphics[width=0.49\linewidth]{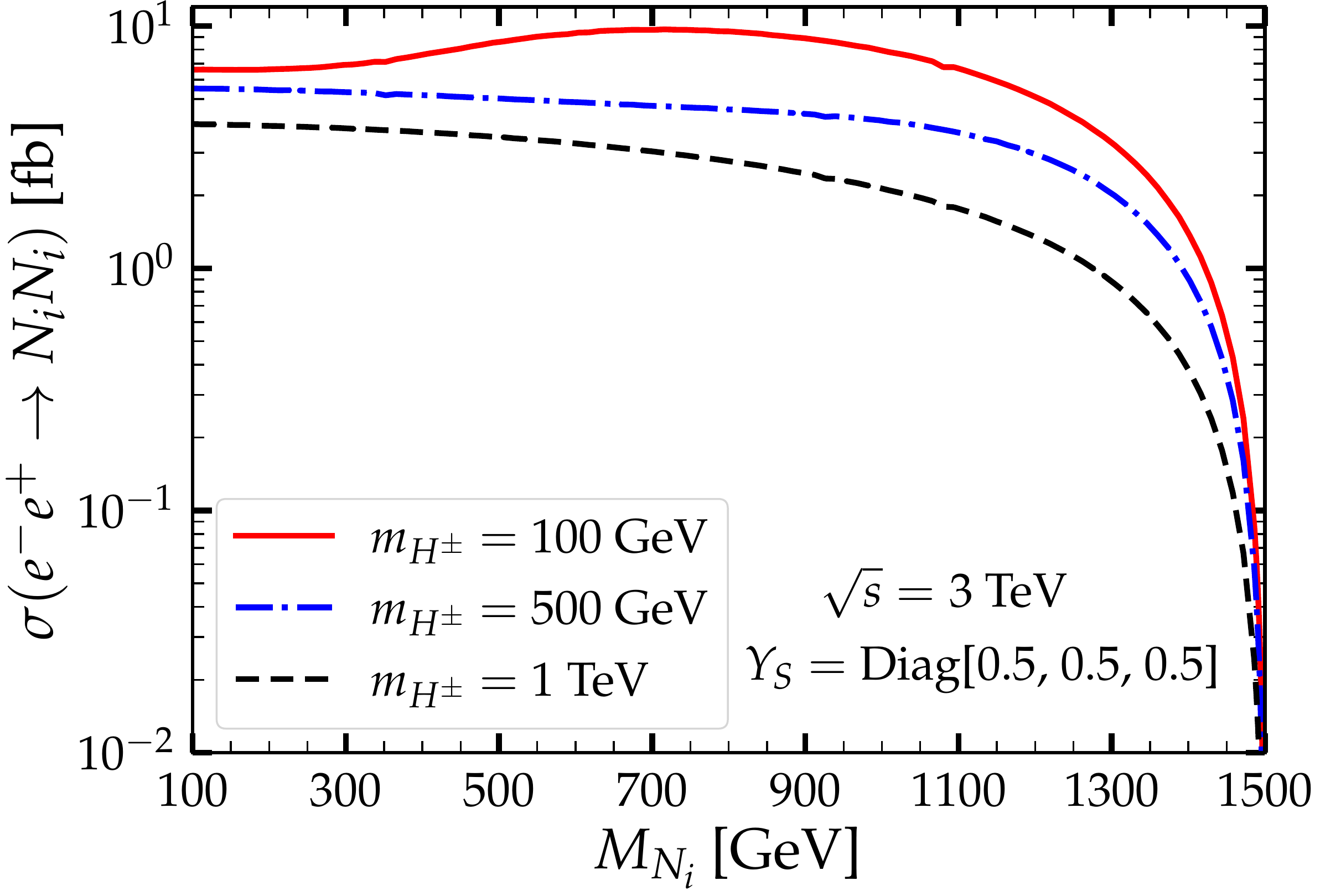}
    \includegraphics[width=0.49\linewidth]{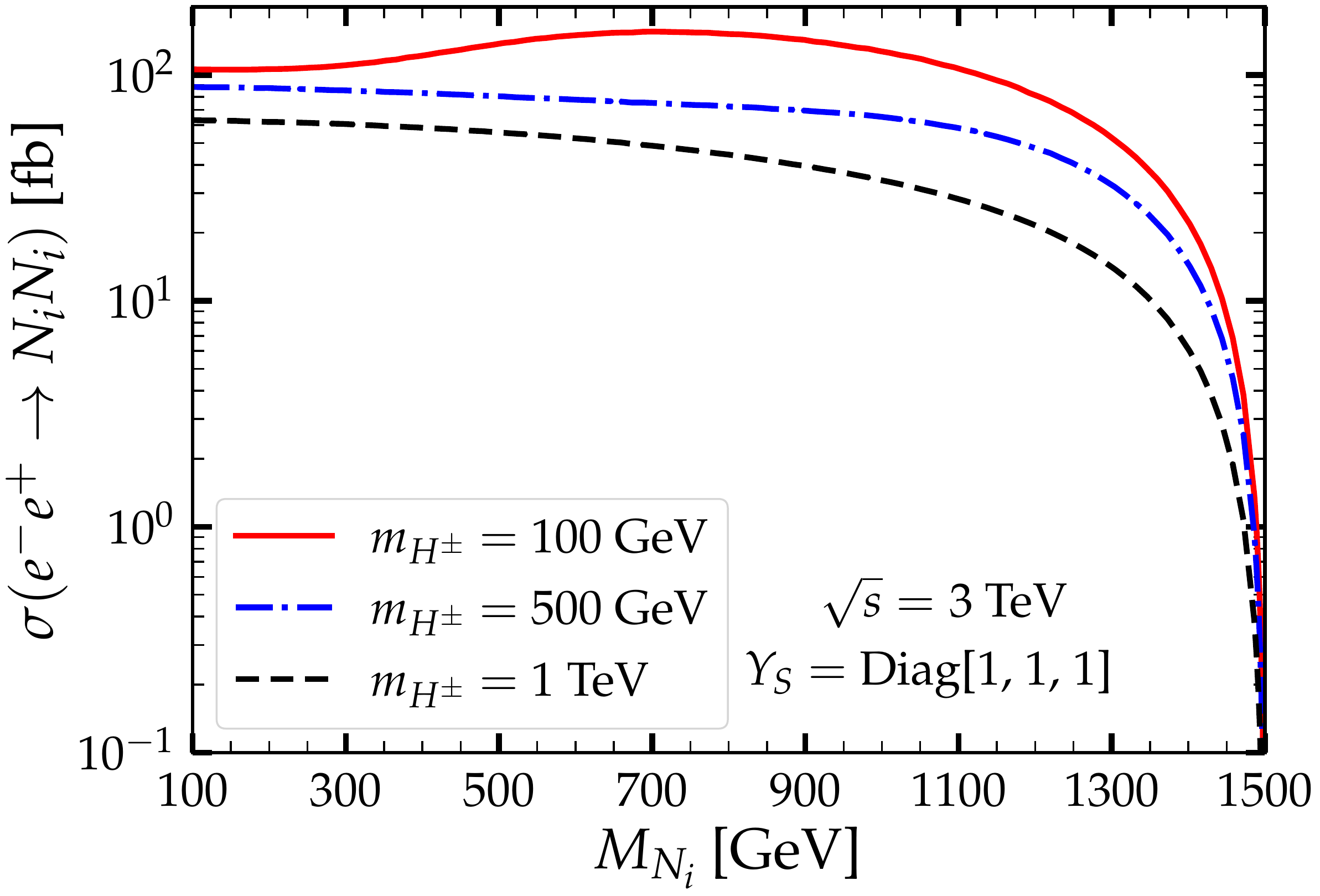}
	\caption{
          Cross-section for heavy-neutrino pair production versus its
          mass $M_{N_i}$ at an $e^+e^-$ collider with center of mass
          energy $\sqrt{s}=1$ TeV (top) and $\sqrt{s}=3$ TeV (bottom).  Results are shown for different
          $Y_S$ and charged-Higgs boson mass values, as indicated.}
	\label{fig:eetoNN}
\end{figure}
 where $\sqrt{s}$ and $\theta$ are the CM energy and scattering angle. One sees from Eq.~\ref{eq:Sigma_eeNN} that, for adequately
large but experimentally allowed values of the Yukawa coupling $Y_S$,
the process $e^+e^-\to NN$ provides a promising way to produce heavy
neutrinos.
In Fig.~\ref{fig:eetoNN} we display the heavy-neutrino pair-production
cross-section $\sigma(e^+ e^-\to N_i N_i)$ for two benchmark Yukawa
couplings $Y_S=\text{Diag}(0.5,0.5,0.5)$~(left panels) and
$Y_S=\text{Diag}(1,1,1)$~(right panels) for center of mass
energy $\sqrt{s}=1$ TeV (top panels) and $\sqrt{s}=3$ TeV (bottom panels).
The three lines in each panel correspond to three charged-Higgs masses
$m_{H^\pm}=100\,\text{GeV}$, 500 GeV and 1000 GeV.
One can clearly see that a relatively large Yukawa coupling $Y_S$ leads to a sizeable t-channel cross section.
Therefore, from this discussion it follows that heavy-neutrino pair-production at $e^+e^-$ colliders deserves further study to
ascertain the potential detectability of the associated signatures.

\subsection{Heavy neutrino production associated with charged Higgs at
$e^-\gamma$ collider}

\begin{figure}[h]
	\includegraphics[width=0.22\linewidth]{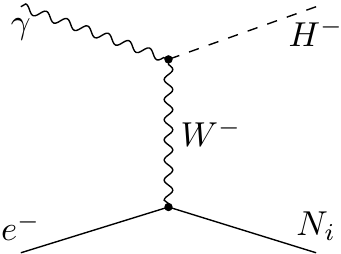}~~~~~
    	\includegraphics[width=0.22\linewidth]{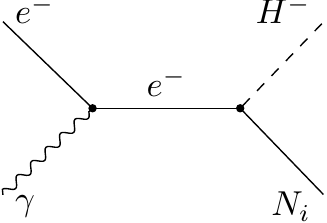}~~~~~
	\includegraphics[width=0.22\linewidth]{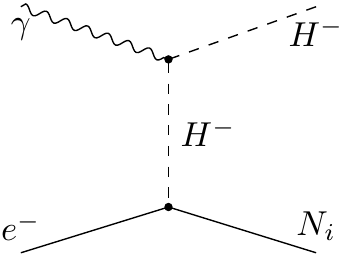}
	\caption{
          Feynman diagrams for $NH^{-}$ production at an $e^-\gamma$
          collider. The first diagram is light-heavy neutrino mixing
          suppressed.}
	\label{fig:HpN-production}
\end{figure}

In the context of $e^+e^-$ colliders, the $e^+$ beam can be replaced by a back-scattered photon, leading to an $e^-\gamma$ collider
which can have extremely rich physics potential~\cite{Telnov:1999tb,Bechtel:2006mr,Ginzburg:1982bs,Ginzburg:1982yr,Velasco:2001fsi,Telnov:1989sd}.
Within our linear seesaw setup an $e^-\gamma$ collider can also
produce the heavy neutrino mediator in association with the charged
Higgs boson.
The relevant Feynman diagrams for this process are shown in
Fig.~\ref{fig:HpN-production}. Notice that the first diagram is
suppressed by light-heavy neutrino mixing, whereas the last two
diagrams are proportional to $Y_S^2$.
Hence, for relatively large $Y_S$, we also expect to have large cross
section for this process.

The analytical expression for this production cross-section is given as
\small
\begin{align} \frac{d\sigma}{d\cos\theta}&=\frac{\alpha_{\rm em}
|Y_S^{1i}|^2}{16 s^2
\big(m_{H^\pm}^2-t\big)^2}\lambda^{\frac{1}{2}}\big(1,\frac{M_{N_i}^2}{s},\frac{m_{H^\pm}^2}{s}\big)\nonumber\\
&\Big[t\big(2M_{N_i}^4-2M_{N_i}^2(s+t)+t(s+t)\big)-m_{H^\pm}^2\big(2M_{N_i}^4+t^2\big)+m_{H^\pm}^4\big(2M_{N_i}^2+s+t\big)-m_{H^\pm}^6\Big],
\label{eq:Sigma_eaNH}
\end{align}
\normalsize
\begin{figure}[h]
\includegraphics[width=0.49\linewidth]{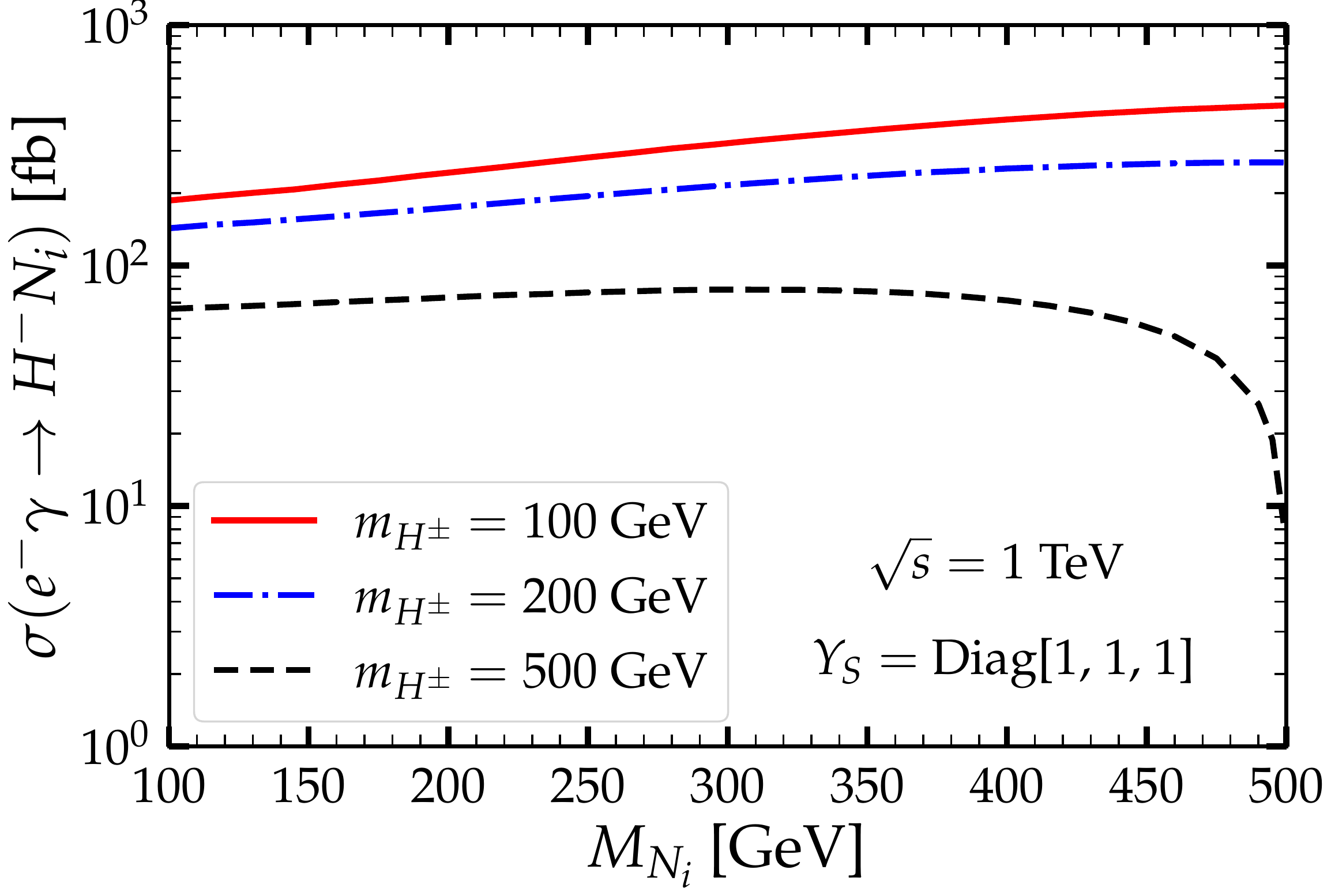}
\includegraphics[width=0.49\linewidth]{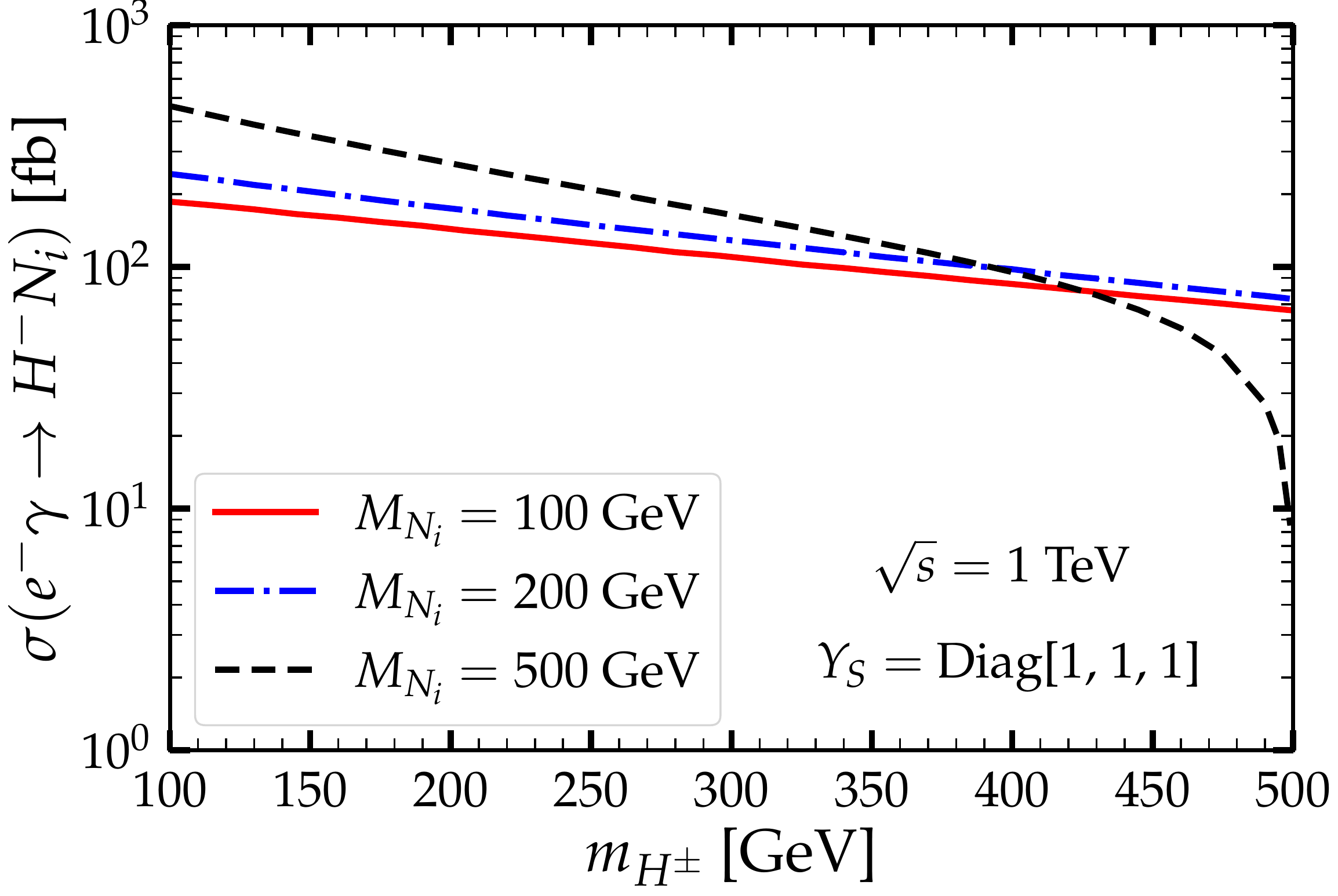}
\includegraphics[width=0.49\linewidth]{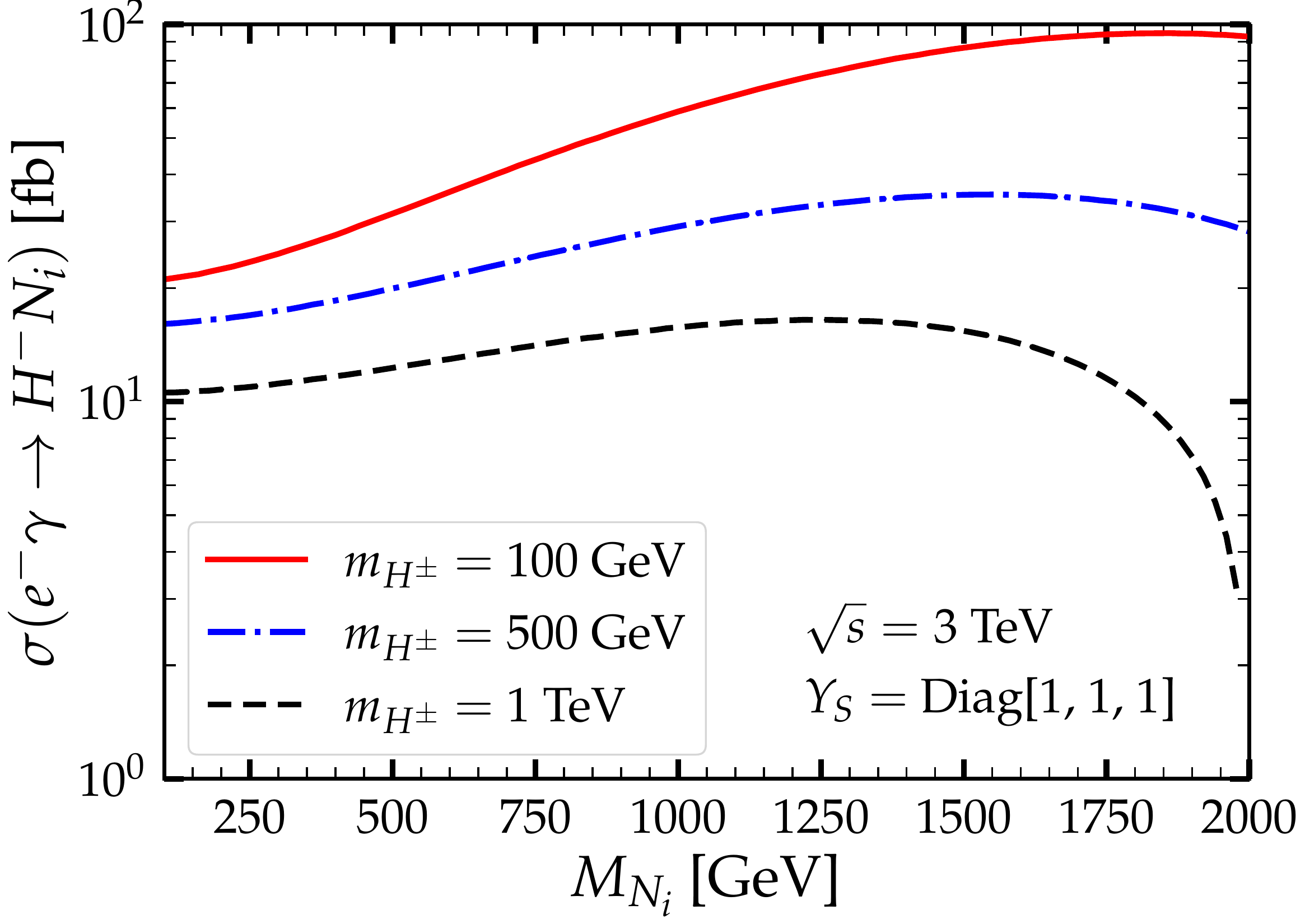}
\includegraphics[width=0.49\linewidth]{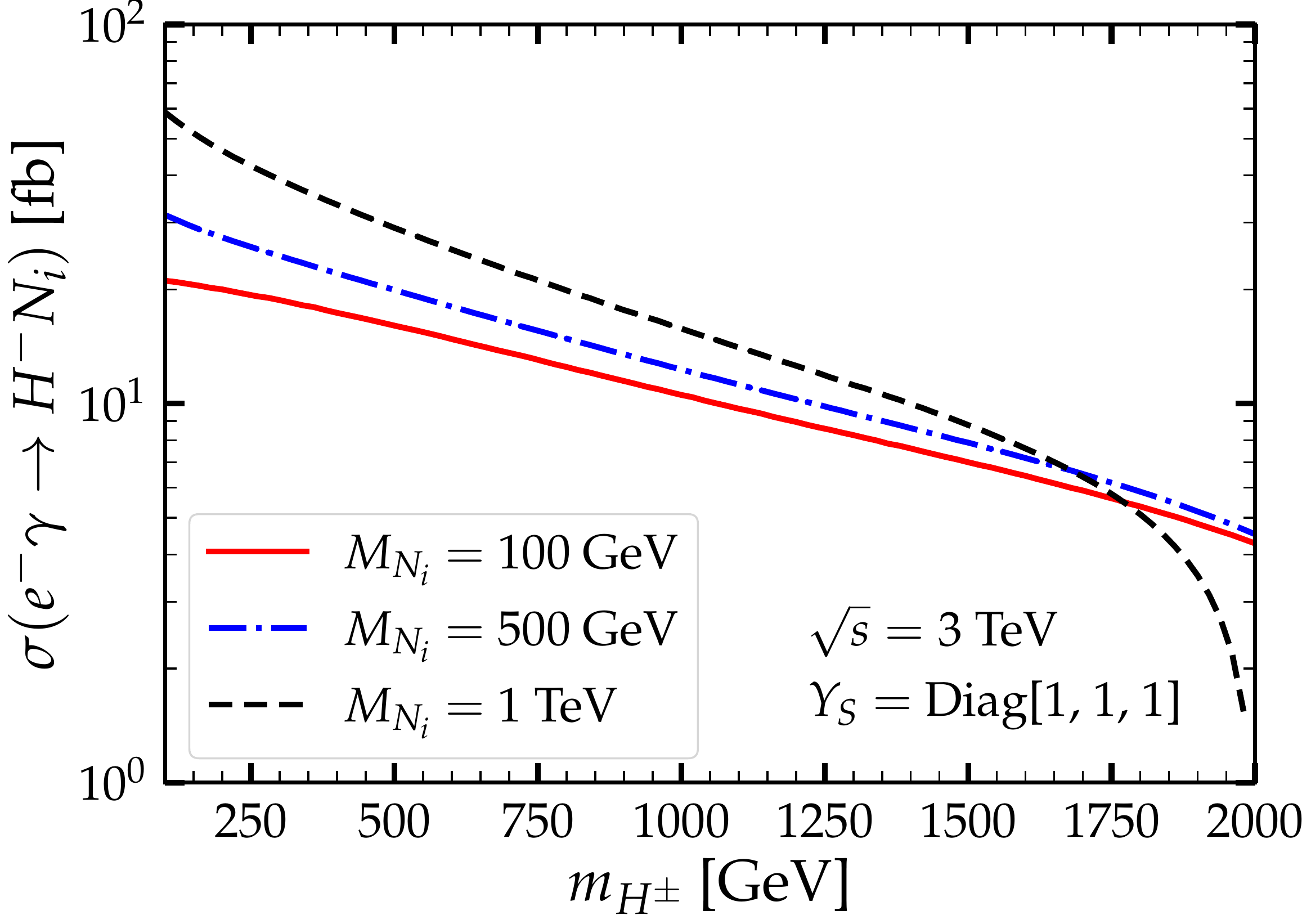}
\caption{$NH^-$ production cross-section versus the heavy neutrino mass $M_{N_i}$~(left panel) and versus the charged scalar mass
    $m_{H^\pm}$~(right panel) at an $e^-\gamma$ collider at center of mass energy $\sqrt{s}=1$ TeV (top) and $\sqrt{s}=3$ TeV (bottom).
    Left panels~(right panels) correspond to three values of the charged Higgs~(heavy neutrino) mass
    $m_{H^\pm}(M_{N_i})=100\,\text{GeV}$, 200 GeV and 500 GeV for the top panels, and $m_{H^\pm}(M_{N_i})=100\,\text{GeV}$, 500 GeV and 1000 GeV
    for the bottom panels.}
	\label{fig:eatohmN}
\end{figure} 
\begin{figure}[h] \includegraphics[width=0.5\linewidth]{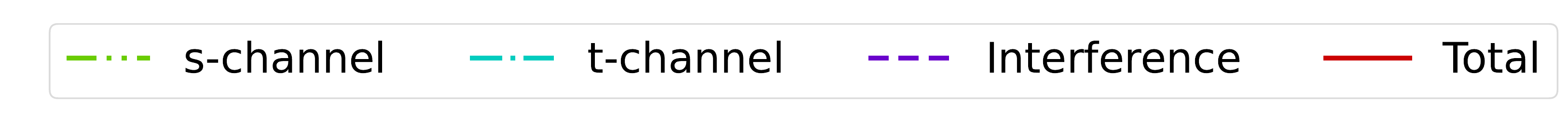}
\includegraphics[height=4.5cm,width=0.45\linewidth]{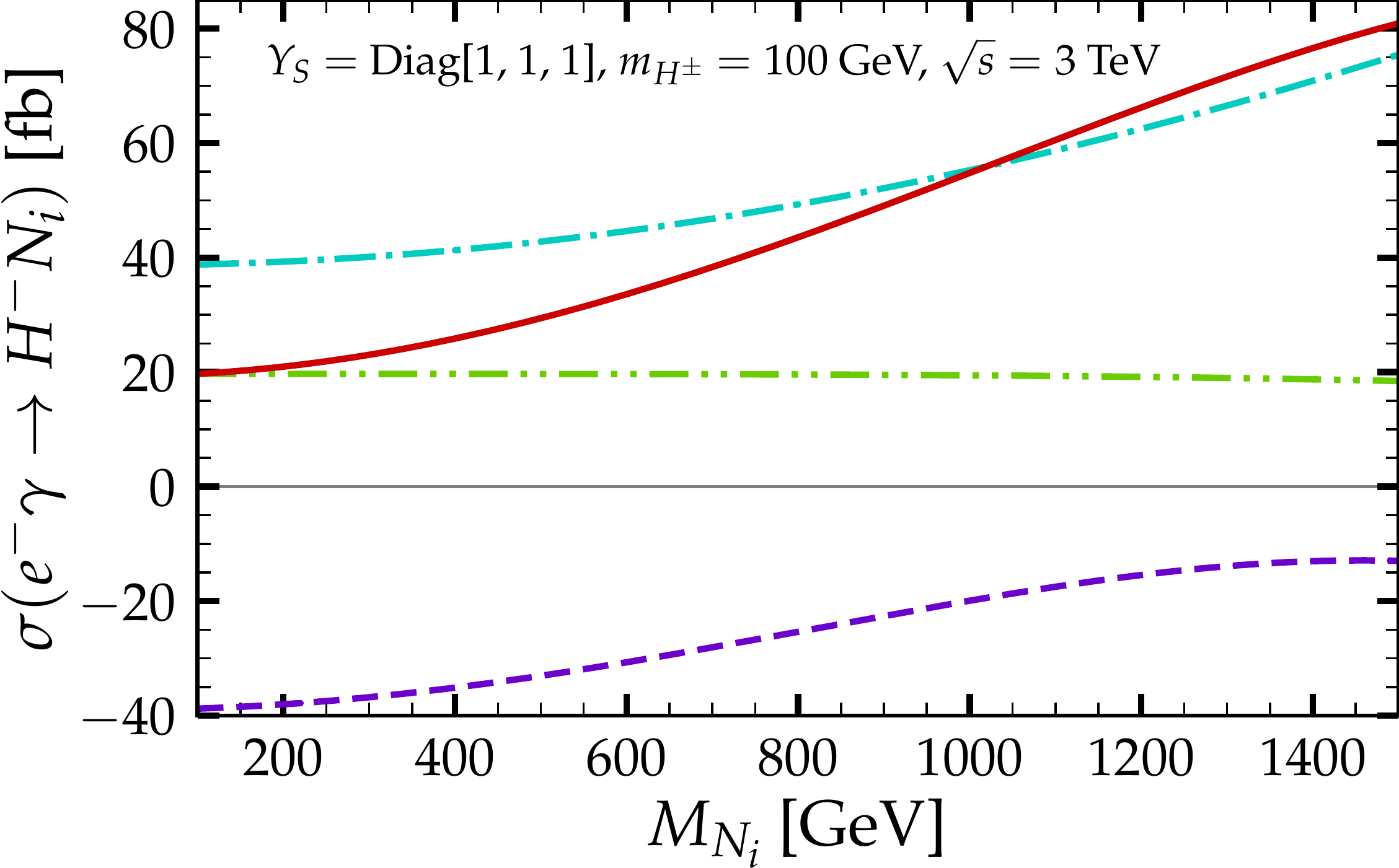}~~~
\includegraphics[height=4.5cm,width=0.45\linewidth]{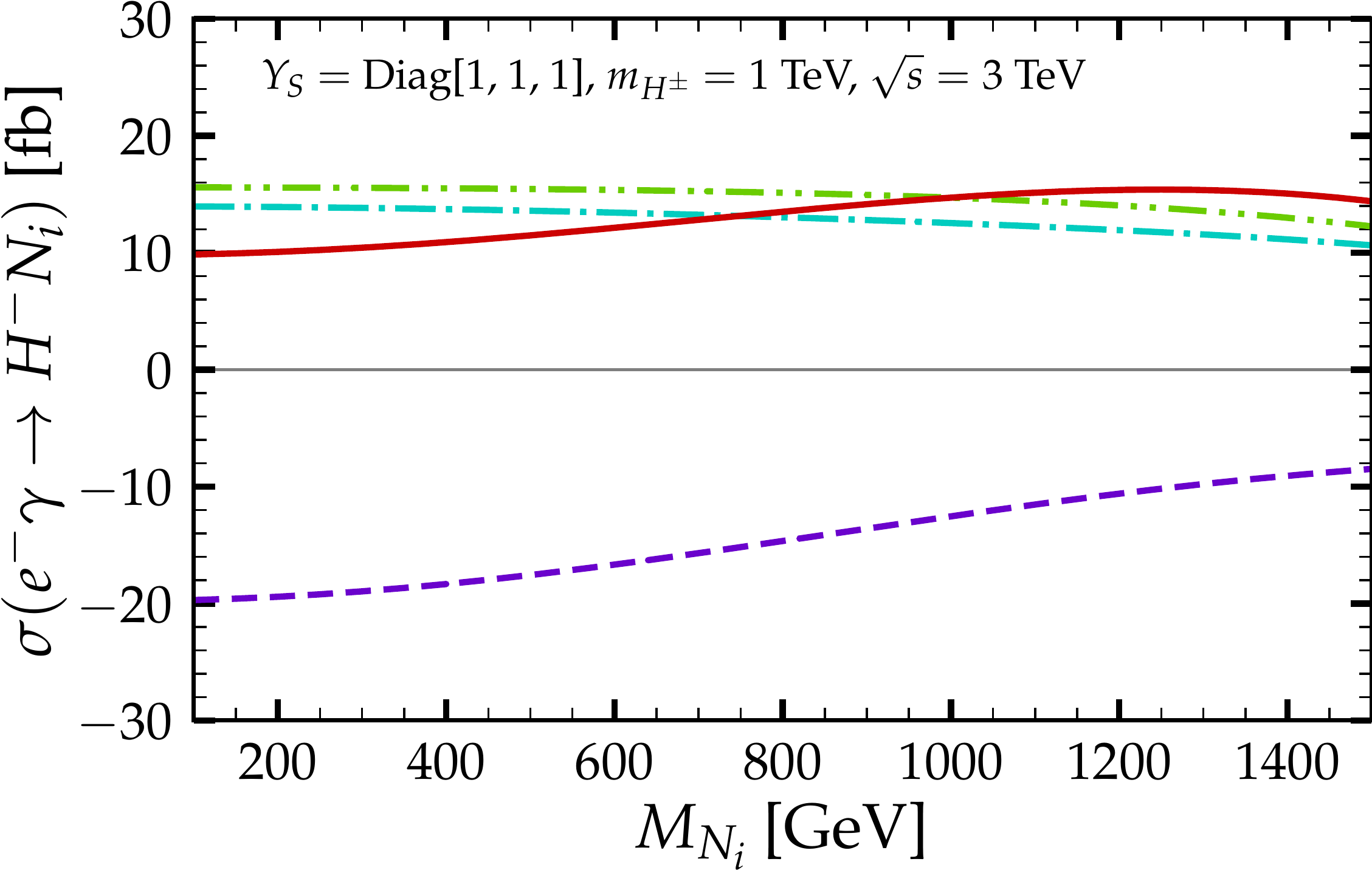}
\caption{
  Comparison of individual contributions to the cross-section for the process $e^- \gamma\to N_i H^-$ at center of mass energy
  $\sqrt{s}=3$~TeV.  Left~(right) panels correspond to charged Higgs masses $m_{H^\pm}=100$~GeV~(1 TeV). See text for details.}
\label{fig:eyNH}
\end{figure}
with $t=-\frac{1}{2}(s-M_{N_i}^2-m_{H^\pm}^2)+\frac{\cos\theta}{2}\lambda^{\frac{1}{2}}(s,M_{N_i}^2, m_{H^\pm}^2)$. Using Eq.~\ref{eq:Sigma_eaNH}
we obtain Fig.~\ref{fig:eatohmN}, which shows the $N_i H^-$ production cross section versus $M_{N_i}$~(left panels) and $m_{H^\pm}$~(right panels)
for center of mass energy $\sqrt{s}=1$ TeV (top panels) and $\sqrt{s}=3$ TeV (bottom panels).
For all panels we fix the Yukawa coupling to be $Y_S=\text{Diag}(1,1,1)$.
The three lines in the left panels~(right panels) correspond to three
values of charged Higgs~(heavy neutrino) mass
$m_{H^\pm}(M_{N_i})=100\,\text{GeV}$, 200 GeV and 500 GeV for the top panels and $m_{H^\pm}(M_{N_i})=100\,\text{GeV}$, 500 GeV and 1000 GeV for the bottom panels.
Comparing different lines in Fig.~\ref{fig:eatohmN}, we see that the cross-section tends to increase with the heavy neutrino mass $M_{N_i}$, provided the phase-space is sufficiently open (left panels).
This behavior can also be seen in Fig.~\ref{fig:eyNH}, where we compare the leading contributions coming from s- and t-channels in Fig.~\ref{fig:HpN-production}, as well as the interference between them. 

The left and right panels in Fig.~\ref{fig:eyNH} correspond to two different charged Higgs masses, $m_{H^\pm}=100$~GeV and $m_{H^\pm}=1$~TeV, respectively.
From the left panel, we see that for lighter charged Higgs mass $m_{H^\pm}=100$~GeV, the t-channel contribution dominates over
s-channel and grows with heavy neutrino mass $M_{N_i}$, leading to the rising trend of the total cross-section with $M_{N_i}$.
For larger charged Higgs mass $m_{H^\pm}=1$~TeV, the t-channel and s-channel contributions are of the same order, and the t-channel
contribution does not grow with heavy neutrino mass $M_{N_i}$ due to the phase space suppression.
This is why the cross-section in Fig.~\ref{fig:eatohmN} grows with mass $M_{N_i}$ for light charged Higgs but decreases when the charged
Higgs or heavy neutrino become heavy enough to suppress the phase space.
Note however that, even in the idealized limit of infinite energy (corresponding to a collider with infinite $\sqrt{s}$ energy)
  the cross-section will ultimately stop growing owing to the unitarity of the theory. 


\subsection{Heavy neutrinos from charged-scalar pair and associated
pair production}

%
In contrast to the inverse seesaw, the presence of a charged Higgs
boson in our linear seesaw is a key feature required in order to seed
the neutrino mass, see Eqs.~\ref{eq:ML} and \ref{lin}.
This brings in the possibility of producing the heavy neutrinos via
the charged and neutral Higgs boson decays.
\begin{figure}[h!]
    \includegraphics[width=0.22\linewidth]{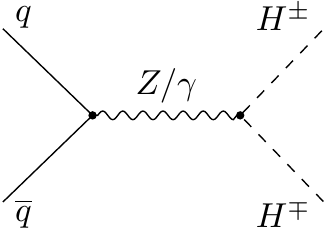}~~~~~
	\includegraphics[width=0.22\linewidth]{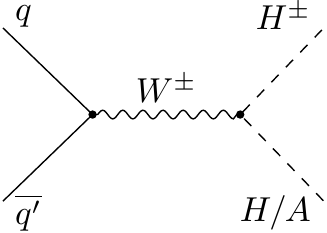}~~~~~
	\includegraphics[width=0.22\linewidth]{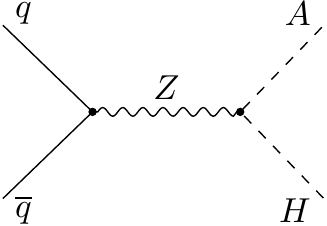}\\
      \includegraphics[width=0.22\linewidth]{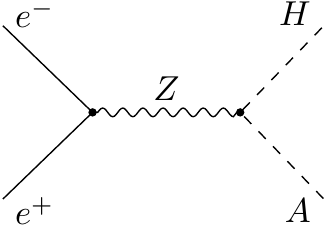}
      \caption{
        Feynman diagrams for Drell-Yan Higgs boson production at $pp$~(upper panel) and $e^+e^-$~(bottom panel) colliders.
        Charged Higgs production at an $e^+e^-$ collider is discussed separately below, see Fig.~\ref{fig:feyn-prod-hphm},~\ref{fig:HpHm-t} and the accompanying discussion.}
	\label{fig:feynman-production}
\end{figure}

\begin{figure}[t]
	\includegraphics[width=0.49\linewidth]{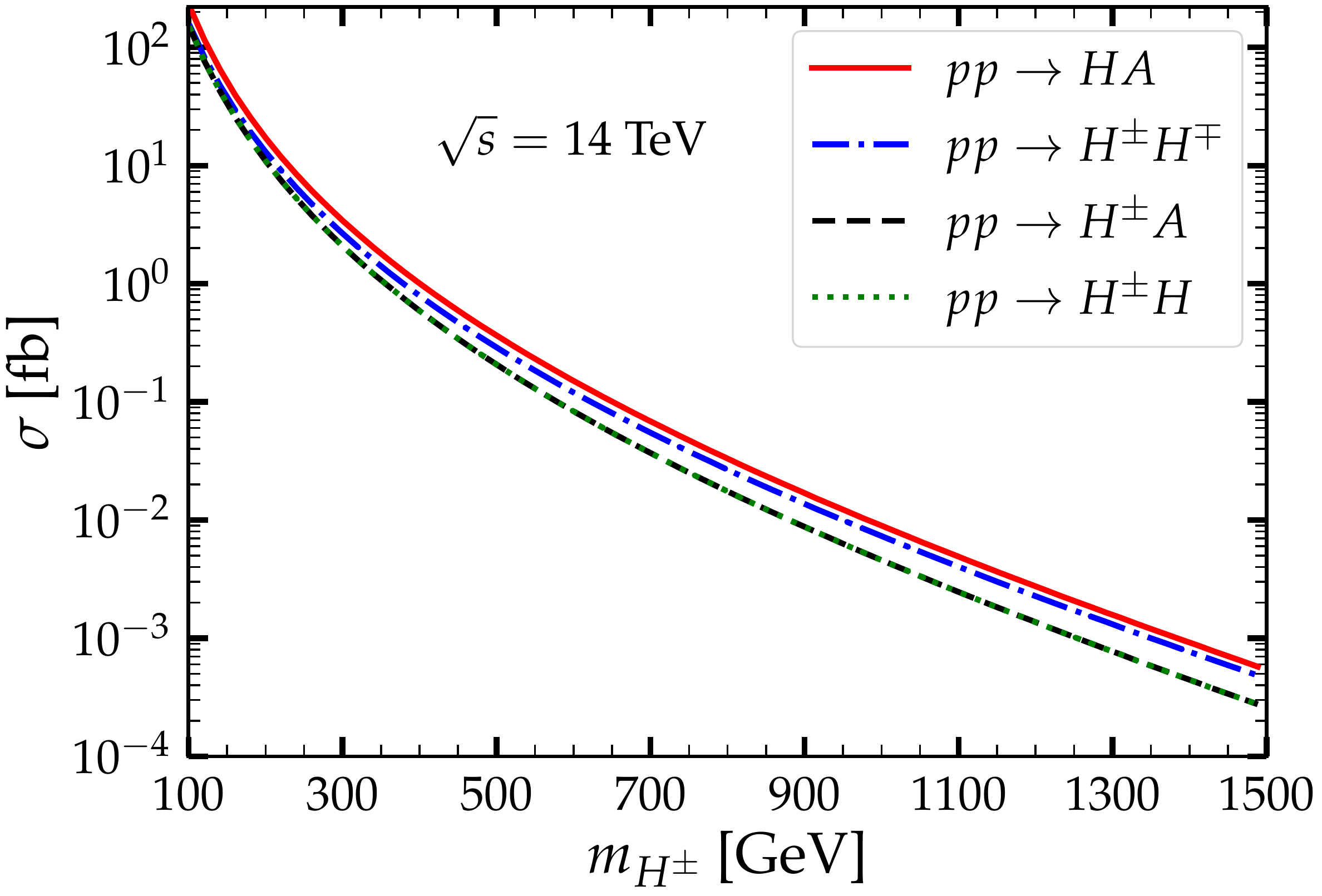}
	\includegraphics[width=0.49\linewidth]{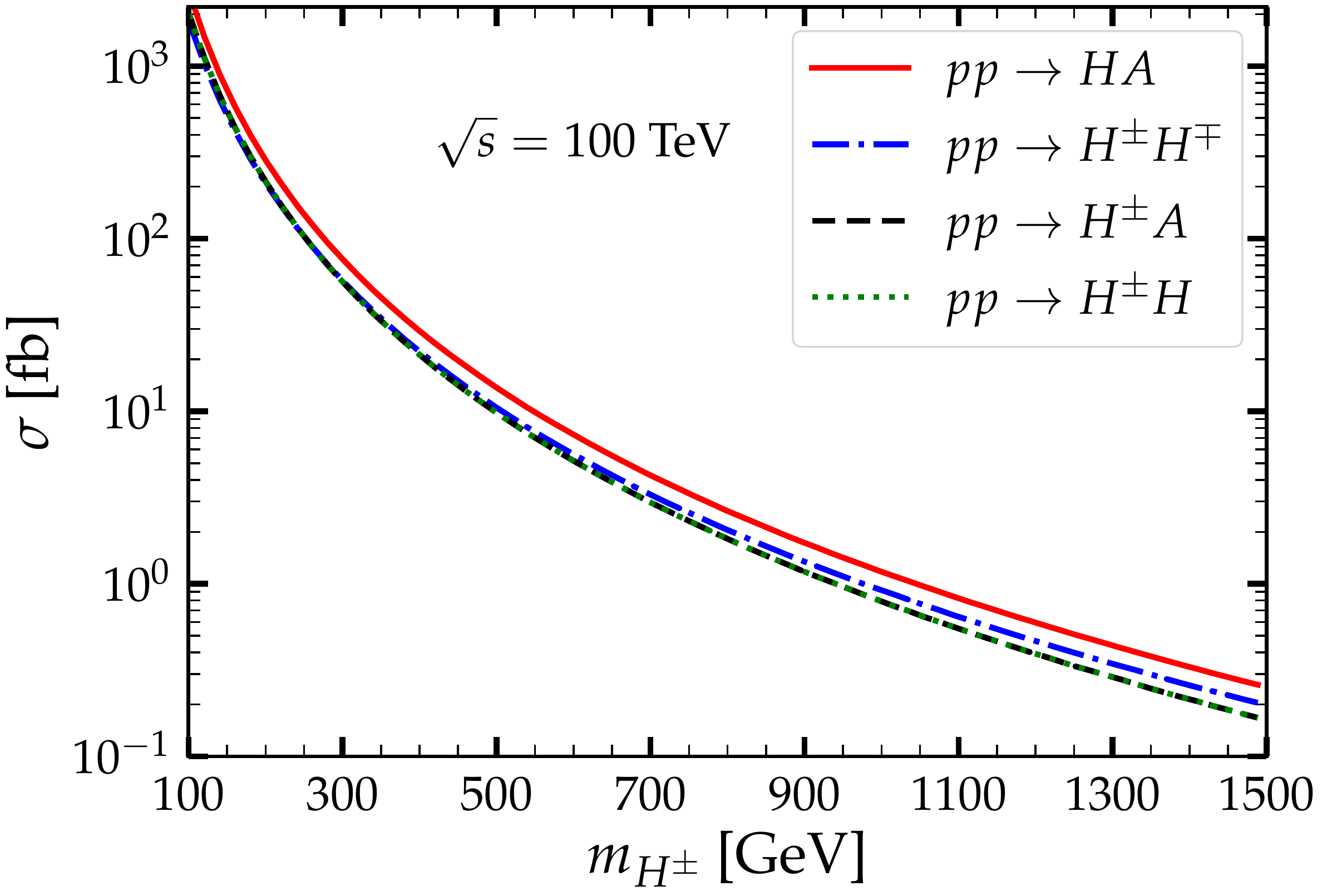}
	\includegraphics[width=0.49\linewidth]{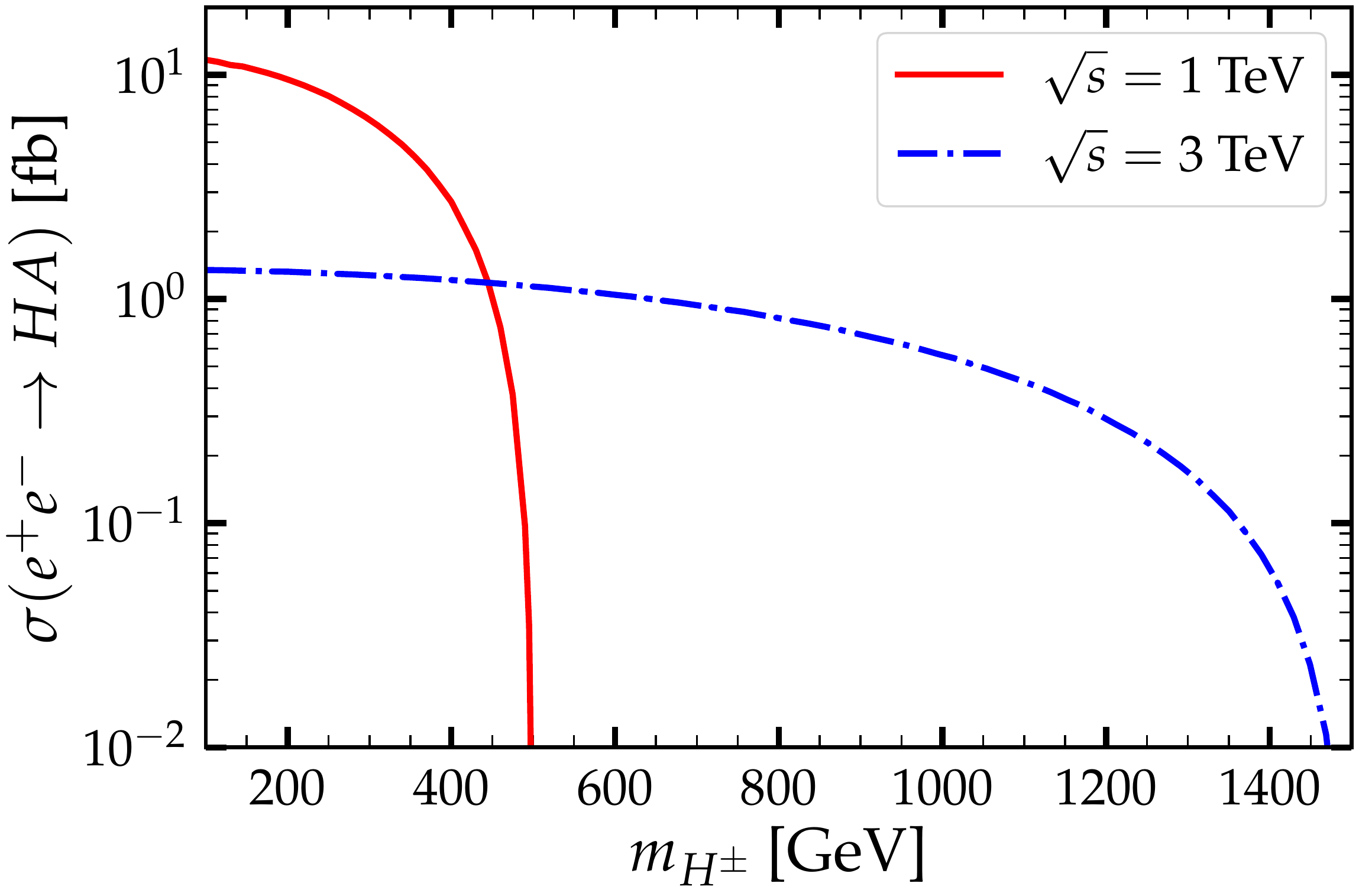}
	\caption{
           Cross section for pair and associated production of new scalars vs the charged scalar mass at $pp$~(upper panel) and $e^+e^-$~(bottom panel) colliders.
            We took center of mass energies $\sqrt{s}=14$ TeV~(left panel) and $\sqrt{s}=100$ TeV~(right panel) for $pp$ colliders
            and $\sqrt{s}=1$ TeV~(red line) and $\sqrt{s}=3$ TeV~(blue line) for the $e^+e^-$ collider.
          These results assume $m_{H^\pm}\approx m_H\approx m_A$, in agreement with the constraints discussed in previous sections.}
	\label{fig:production}
\end{figure}

Indeed, the decays of such Higgs bosons to the heavy neutrinos are not
suppressed by the light-heavy neutrino mixing.
This is because the pair production and associated production of
scalars $\Phi\Phi'$~(with $\Phi,\Phi'\in \{H,A,H^\pm\}$) via the
neutral or charged current Drell-Yan mechanism involving s-channel
$\gamma/Z, W^\pm$ exchange~(see Fig.~\ref{fig:feynman-production}) can
be large.
In Fig.~\ref{fig:production} we display these cross-sections both at
$pp$~($\sqrt{s}=14$ TeV, 100 TeV) as well as $e^+e^-$
colliders~($\sqrt{s}=$1 TeV, 3 TeV).
One sees that at such large center of mass energies the production
cross section is large enough that multi-TeV Higgs masses can be
explored.
\par
\begin{figure}[!h]
	\includegraphics[width=0.22\linewidth]{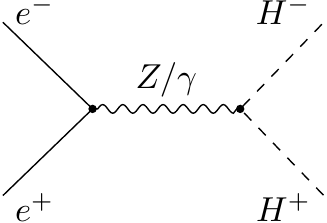}~~~~~~~~~~~~~
	\includegraphics[width=0.22\linewidth]{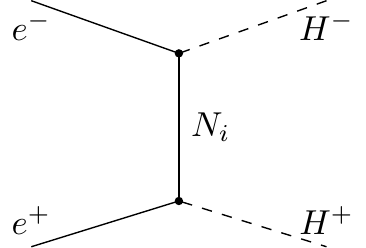}
	\caption{
          Feynman diagrams for the charged Higgs production in an
          $e^+e^-$ collider. Left panel: Drell-Yan production. Right
          panel:
           t-channel heavy neutrino mediated production.}
	\label{fig:feyn-prod-hphm}
\end{figure}

 Beyond Drell-Yan production, in the linear seesaw, one can have heavy-neutrino-mediated t-channel charged Higgs boson pair-production,
as shown in Fig.~\ref{fig:feyn-prod-hphm}.  This new process is a characteristic feature of $e^+e^-$ colliders, absent at $pp$ colliders.
The contribution from the t-channel exchange of heavy neutrinos is
proportional to $Y_S^4$.
Hence, for relatively large $Y_S$, this production cross section can
be large and can even surpass the Drell-Yan production.

\begin{figure}[h]
	\includegraphics[width=0.49\linewidth]{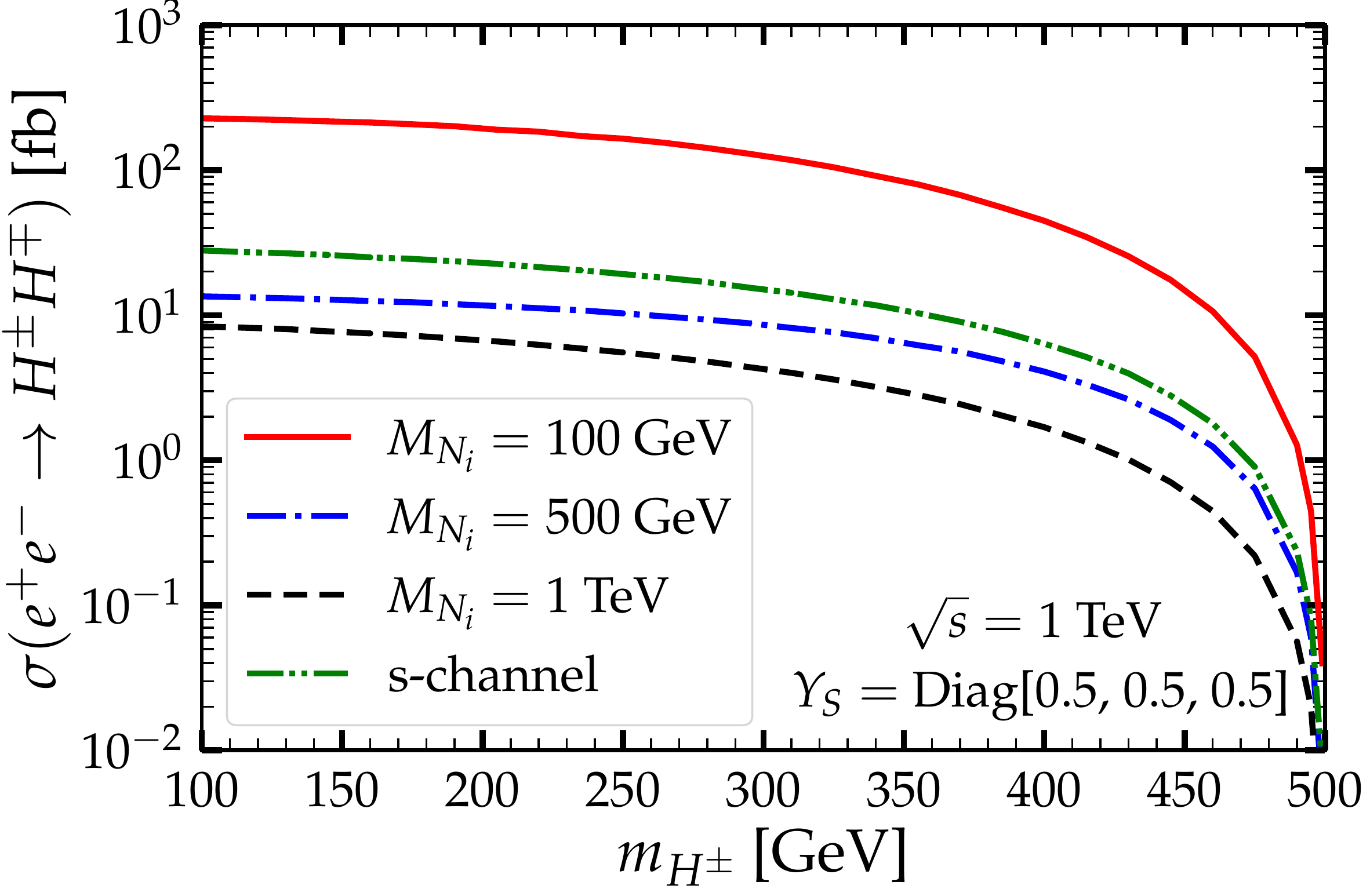}
	\includegraphics[width=0.49\linewidth]{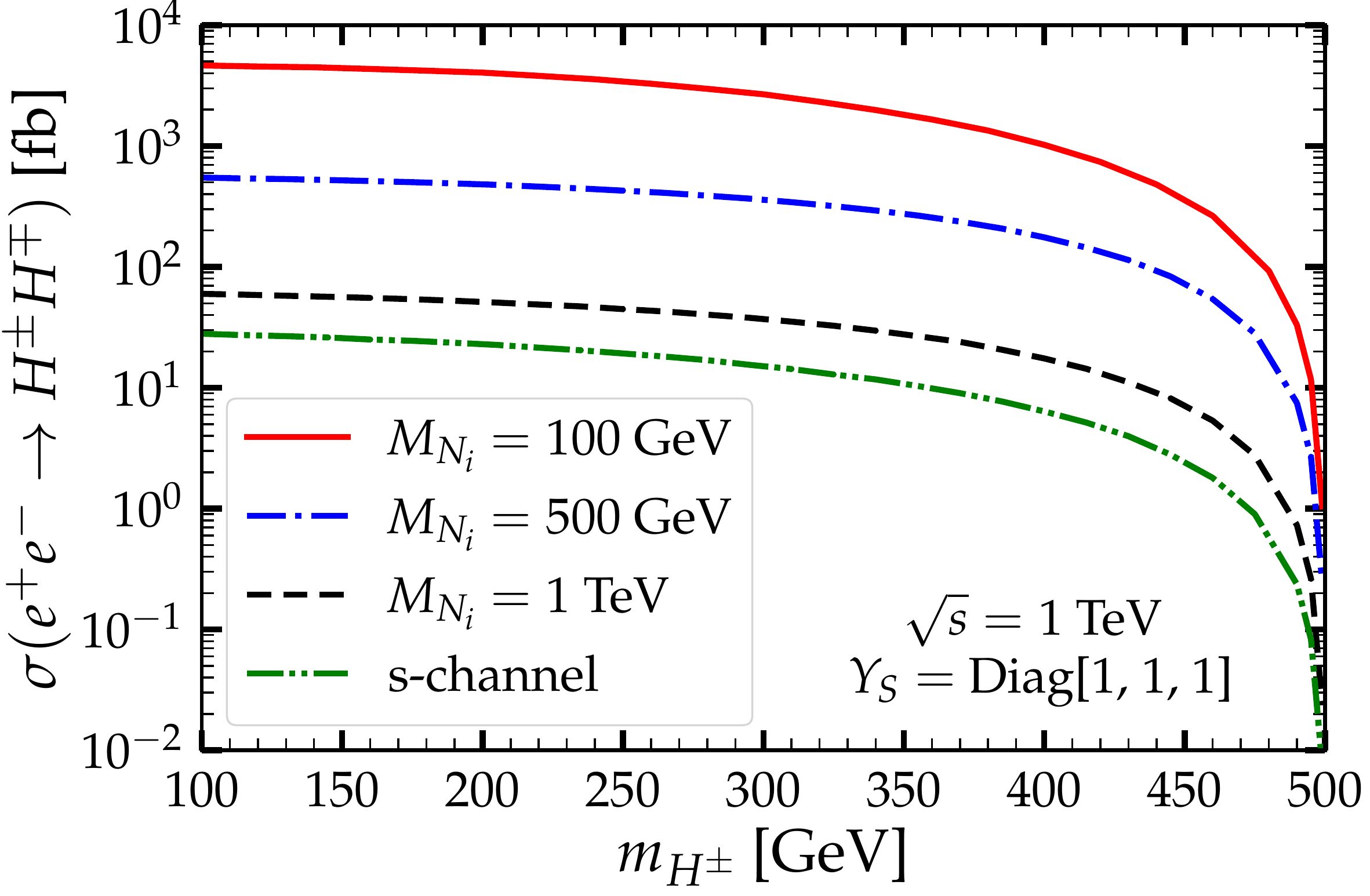}
	\includegraphics[width=0.49\linewidth]{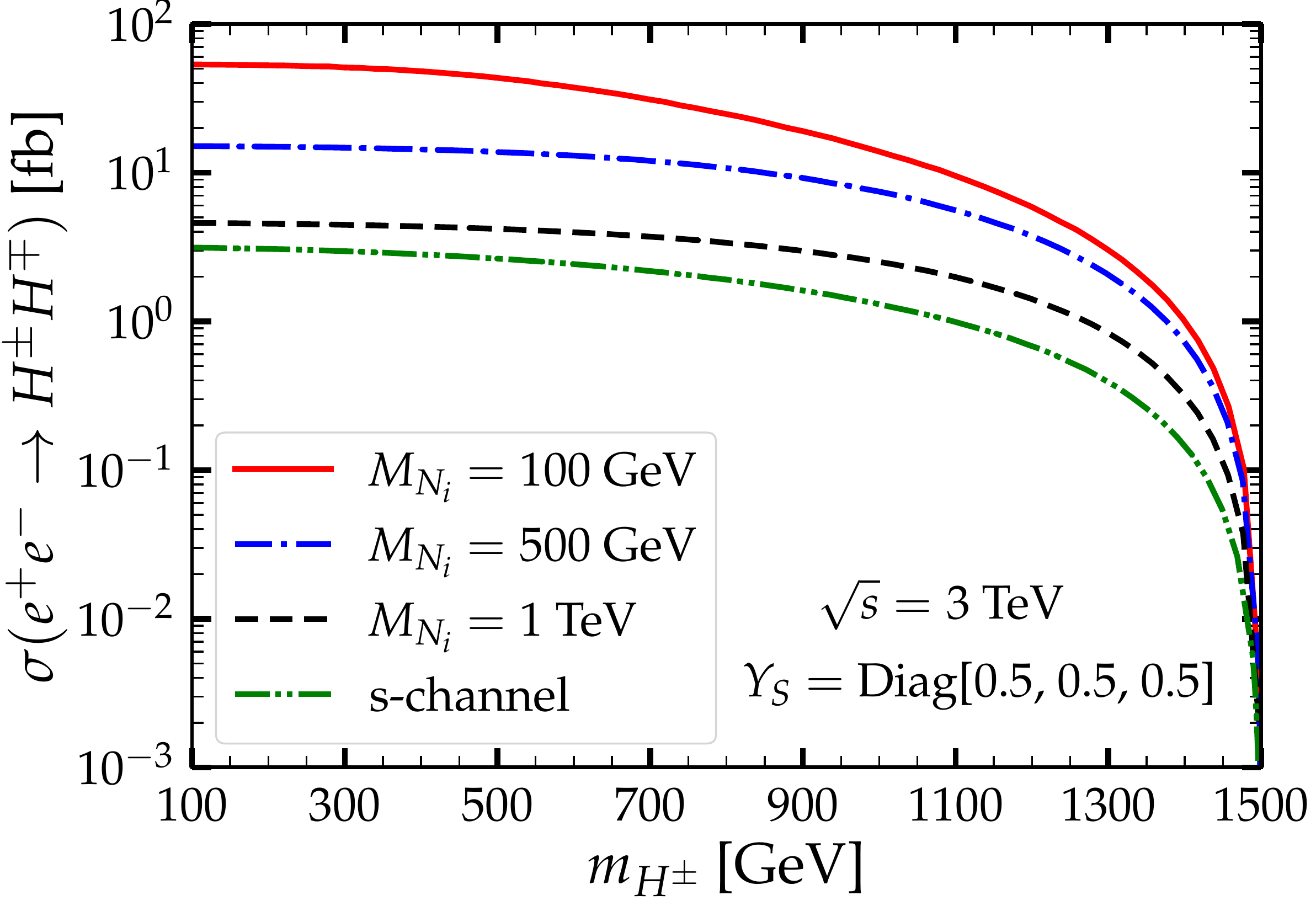}
	\includegraphics[width=0.49\linewidth]{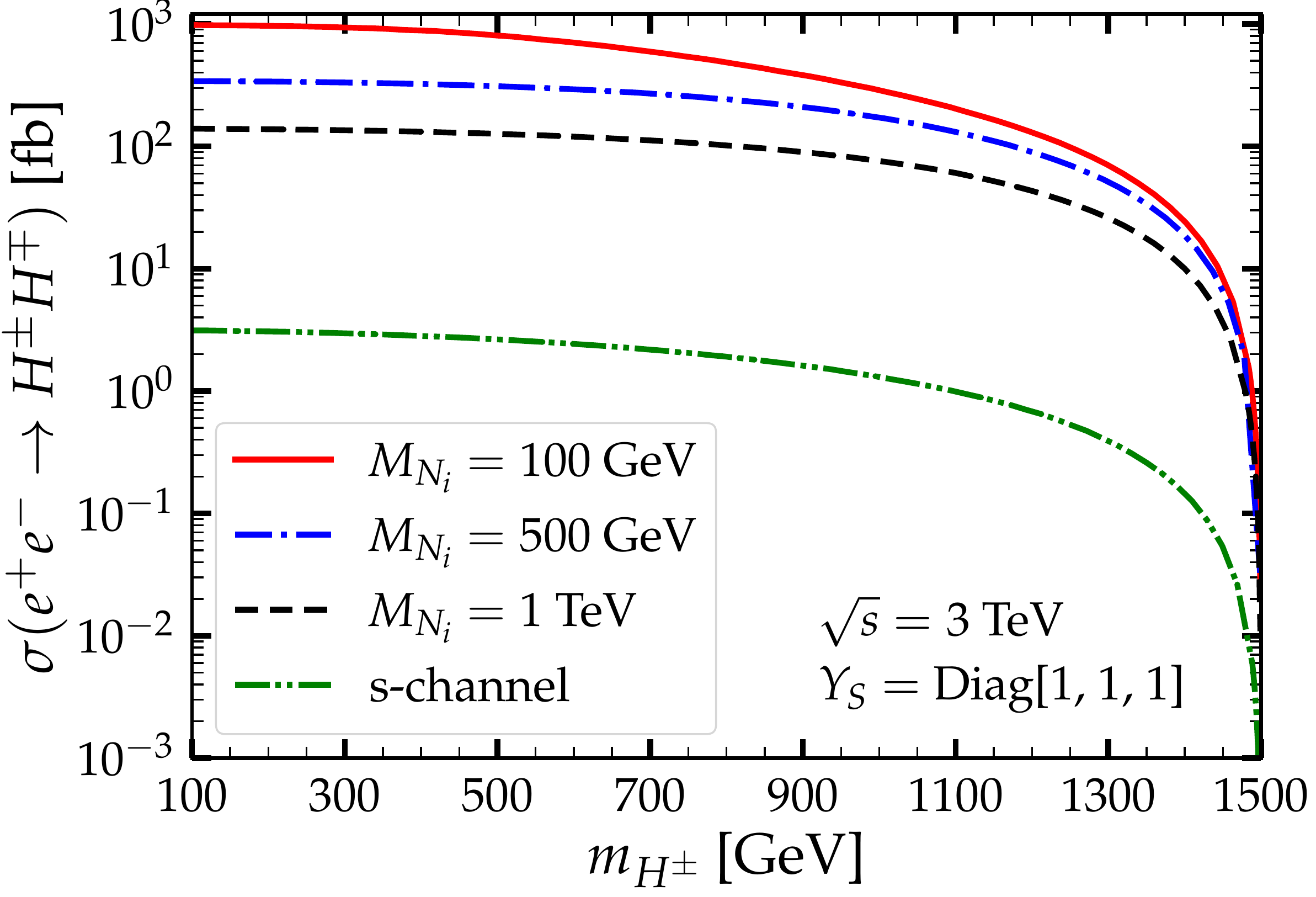}
	\caption{
          Cross-section for the charged scalar pair production in an $e^+e^-$ collider at center of mass energy $\sqrt{s}=1$ TeV (top) and $\sqrt{s}=3$ TeV (bottom)
          versus the charged scalar mass. The green line represents the s-channel contribution only, whereas the other three
          lines include both the s- and the t-channel.  They correspond to three mediator masses, $M_{N_i}=100$ GeV, 500
          GeV, and 1 TeV. Left and right panels assume $Y_S$ as Diag(0.5, 0.5, 0.5) and Diag(1, 1, 1), respectively.}
	\label{fig:HpHm-t}
\end{figure}

The analytical expression for this contribution to the production
cross-section is given as %
\small
\begin{align}
\frac{d\sigma}{d\cos\theta}=\frac{|Y_{S}^{1i}|^4\sin^2\theta\big(s-4m_{H^\pm}^2\big)^{\frac{3}{2}}}{128\pi\sqrt{s}\Big(s+2M_{N_i}^2-2m_{H^\pm}^2+\cos\theta\sqrt{s(s-4m_{H^\pm}^2)}\Big)^2}.
\label{eq:Sigma_eeHH}
\end{align}
\normalsize
In Fig.~\ref{fig:HpHm-t}, we compare the Drell-Yan contribution with the t-channel contribution for this production channel.
%
The left and right panels correspond to two different choices of Yukawa couplings $Y_S=$ Diag(0.5, 0.5, 0.5) and $Y_S=$ Diag(1, 1, 1).
The top and bottom panels correspond to two different choices of center of mass energies $\sqrt{s}=1$ TeV and $\sqrt{s}=3$ TeV, respectively.
In each panel, the double-dotted dashed green line gives the pure
Drell-Yan contribution, whereas the other three lines give the
combined contribution coming from Drell-Yan and t-channel heavy
neutrino mediation
for three benchmark values of the heavy neutrino mass
$M_{N_i}=100$~GeV, 500 GeV and 1 TeV.
One sees that there is a substantial enhancement in the cross-section
once the t-channel contribution is included along with the Drell-Yan
contribution.
\par Before closing this section, we would like to stress that once the new charged and neutral Higgs bosons are produced at a
$pp$ or $e^+e^-$ collider, the heavy seesaw mediator neutrinos can be
generated from their decays, such as $H^\pm\to\ell^\pm N_i$ or
$H/A\to\nu N$. Hence this will effectively enhance the heavy neutrino
production.
The pair-production cross-section of the charged scalar at a hadron
collider becomes smaller for large charged-Higgs masses, see
Fig.~\ref{fig:production}.
Moreover, the existence of multiple SM backgrounds reduces the
physics reach for charged Higgs boson discovery at a hadron collider.
In contrast, an $e^+e^-$ collider with enhanced charged-scalar
pair-production cross-section thanks to the new t-channel contribution
and with a considerably cleaner environment is more promising for
charged-scalar searches.

\section{Decay modes of heavy neutrinos and new scalars}
\label{sec:decay}

The decays of the new scalars are controlled by the underlying $\rm U(1)_L$ lepton symmetry.  In the limit of very small $v_\chi$, the neutral
scalars $H$ and $A$ are almost degenerate in mass. They can be lighter or heavier than the charged Higgs boson, i.e. $m_{A,H} \ge m_{H^\pm}$ for $\lambda_4\ge 0$ and $m_{A,H}\le m_{H^\pm}$ for $\lambda_4\le 0$.
The new scalar masses can also be smaller or larger than the mass of heavy neutrinos.
Let us now discuss the decay modes of the $H/A$, $H^\pm$ scalars as well as the heavy neutrino mediators.

\subsection{Decay modes of heavy neutrinos $N$}

The general structure of the weak interactions of the heavy neutrino
mass mediators $N_i$ in seesaw models is well-known~\cite{Schechter:1980gr,Schechter:1981cv}.
The heavy neutrinos can decay into various final states depending on
their mass $M_{N_i}$ via their mixing with light neutrinos present in
the SM charged and neutral currents.

For $M_{N_i}< m_W$ these are 3-body decays, either purely leptonic
$N_i\to\ell_1\ell_2\nu$, $3\nu$ or semileptonic, such as $\ell_1
u\bar{d}$ and $\nu_{\ell_1}q\bar{q}$.
The analytical expressions for these decay widths can be found in Ref.~\cite{Chun:2019nwi,Atre:2009rg}.
On the other hand, for relatively large $M_{N_i}$, two-body decay channels such as $\ell W$, $\nu_{\ell} Z$ and $\nu_{\ell} h$ start to dominate.
\begin{figure}[!htbp]
	\includegraphics[width=0.54\linewidth]{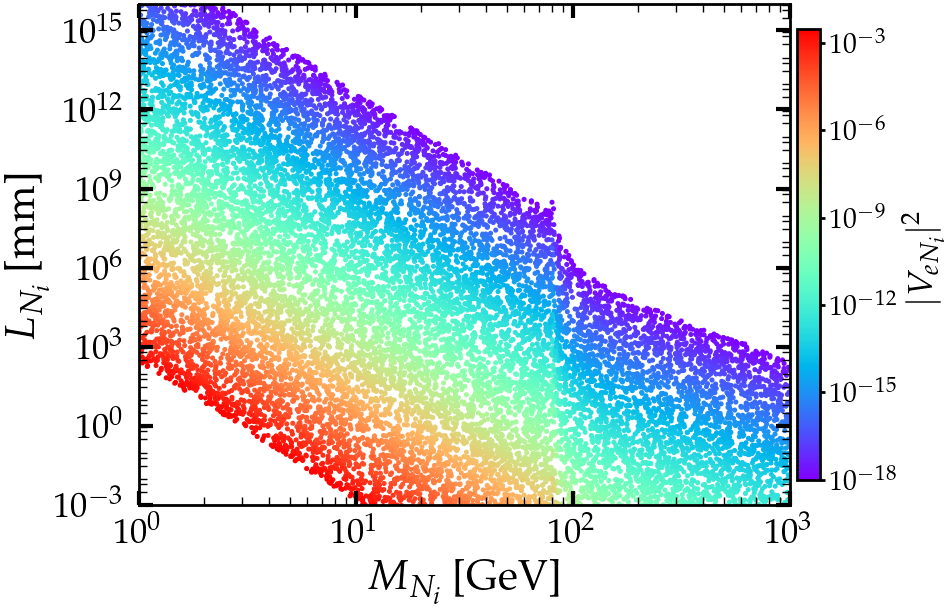}
	\includegraphics[width=0.45\linewidth]{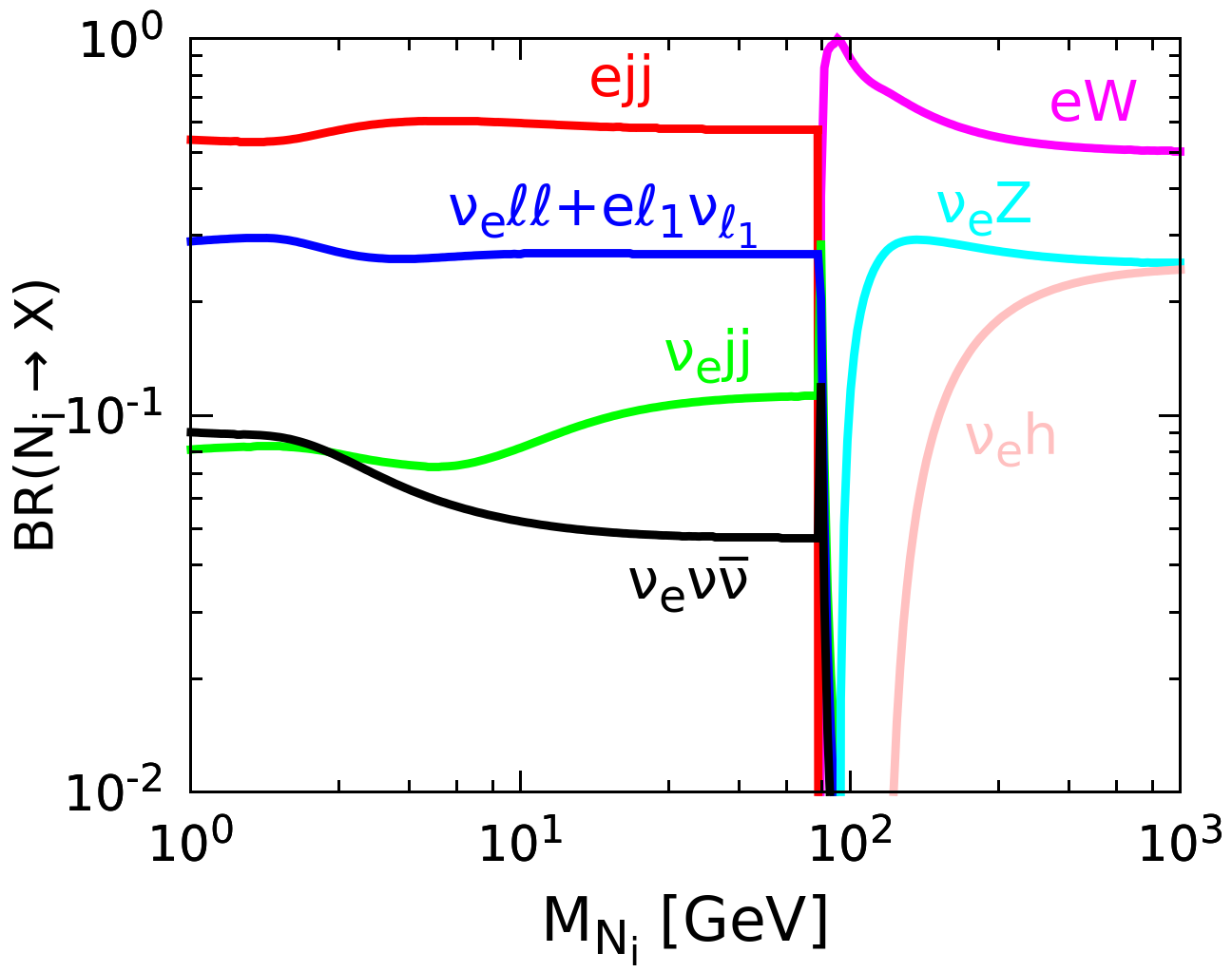}
	\caption{
          Left panel: The heavy neutrino decay length as a function of
          its mass for different values of the light-heavy mixing
          parameter $V_{eN_i}$.
          Right panel: Branching ratios of $N_i$ to various SM final
          states under the assumption $V_{eN_i} \neq 0$, $V_{\mu N_i}=
          0$ and $V_{\tau N_i}= 0$. }
        	\label{fig:BRN}
\end{figure}

In the left panel we show the heavy neutrino decay length as a
function of its mass and light-heavy neutrino mixing angle.
The white triangular region is excluded from electroweak precision data~(EWPD-e)~\cite{delAguila:2008pw,Antusch:2015mia}.
Note that the discontinuity in $L_N$ is due to the jump of the $\Gamma_N$ around $M_{N_i}\sim 80$ GeV which comes from the threshold
of gauge boson masses $M_{W,Z}$.
We see that for sufficiently small mixing and relatively small $M_{N_i}$, the heavy neutrinos are long-lived particles that can
travel macroscopic distances before they decay, giving rise to displaced vertex signatures.
However, for $M_{N_i} >50\,\text{GeV}$ and $|V_{eN_i}|^2>10^{-8}$, the heavy-neutrino decay-length is quite small, so we can take the
heavy-neutrino decay as prompt in most of the parameter space.

In the right panel of Fig.~\ref{fig:BRN}, we show the branching ratios (BR) of heavy neutrinos $N_i$ to various final states.
For simplicity we took $V_{eN_i} \neq 0$, $V_{\mu N_i}= 0$ and $V_{\tau N_i}= 0$, and assumed that the mixing angle is large enough
so that the decay is prompt.
With our assumption that the heavy neutrino only mixes with one lepton
generation, the branching ratio does not depend on neutrino mixing and
the heavy neutrino can decay to one, two or three lepton final states.
For relatively small $M_{N_i}$ three-body decays dominate,
specially $\ell jj$.
Once the heavy neutrino mass cross the $W,Z$ and $h$ mass thresholds, it starts to decay dominantly to two-body final states.
Indeed, Fig.~\ref{fig:BRN} shows that for large $M_{N_i}$ two-body decay such as $e W$, $\nu_e Z$ and $\nu_e h$ dominate.

\par Note that in our model, as long as the \lnv VEV $v_\chi$ is very
small, the Yukawa coupling $Y_S$ can be large and still be consistent
with neutrino mass.  One sees that, as long as
$M_{N_i}>m_{H^\pm},m_{H/A}$, the following decay channels will
dominate over decay channels coming from light-heavy neutrino mixing:
\begin{align} & \Gamma(N_i\to\ell^\pm H^\mp)\approx
\frac{|(Y_S)_{\ell
i}|^2\sin^2\beta}{32\pi}M_{N_i}\Big(1-\frac{m_{H^\pm}^2}{M_{N_i}^2}\Big)^2
,\\ & \Gamma(N_i\to\nu_\ell H)\approx
\frac{|(Y_S)_{\ell
i}|^2\cos^2\alpha}{64\pi}M_{N_i}\Big(1-\frac{m_{H}^2}{M_{N_i}^2}\Big)^2
,\\ & \Gamma(N_i\to\nu_\ell A)\approx
\frac{|(Y_S)_{\ell i}|^2
\sin^2\beta}{64\pi}M_{N_i}\Big(1-\frac{m_{A}^2}{M_{N_i}^2}\Big)^2, 
\end{align}
for large $Y_S$ and neglecting the lepton masses.
 \begin{figure}[!htbp]
 	\centering \includegraphics[width=0.49\linewidth]{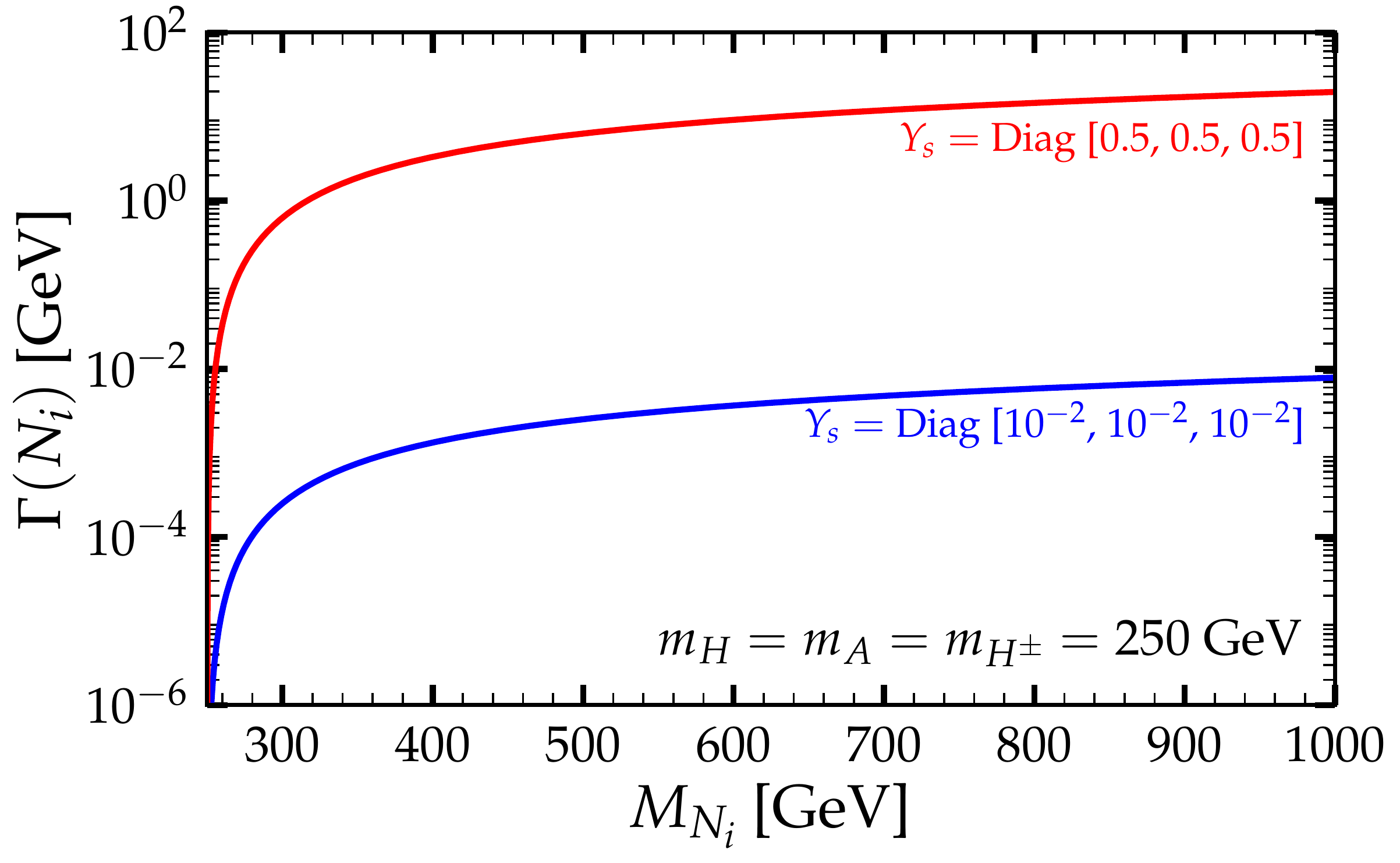}
 	\includegraphics[width=0.49\linewidth]{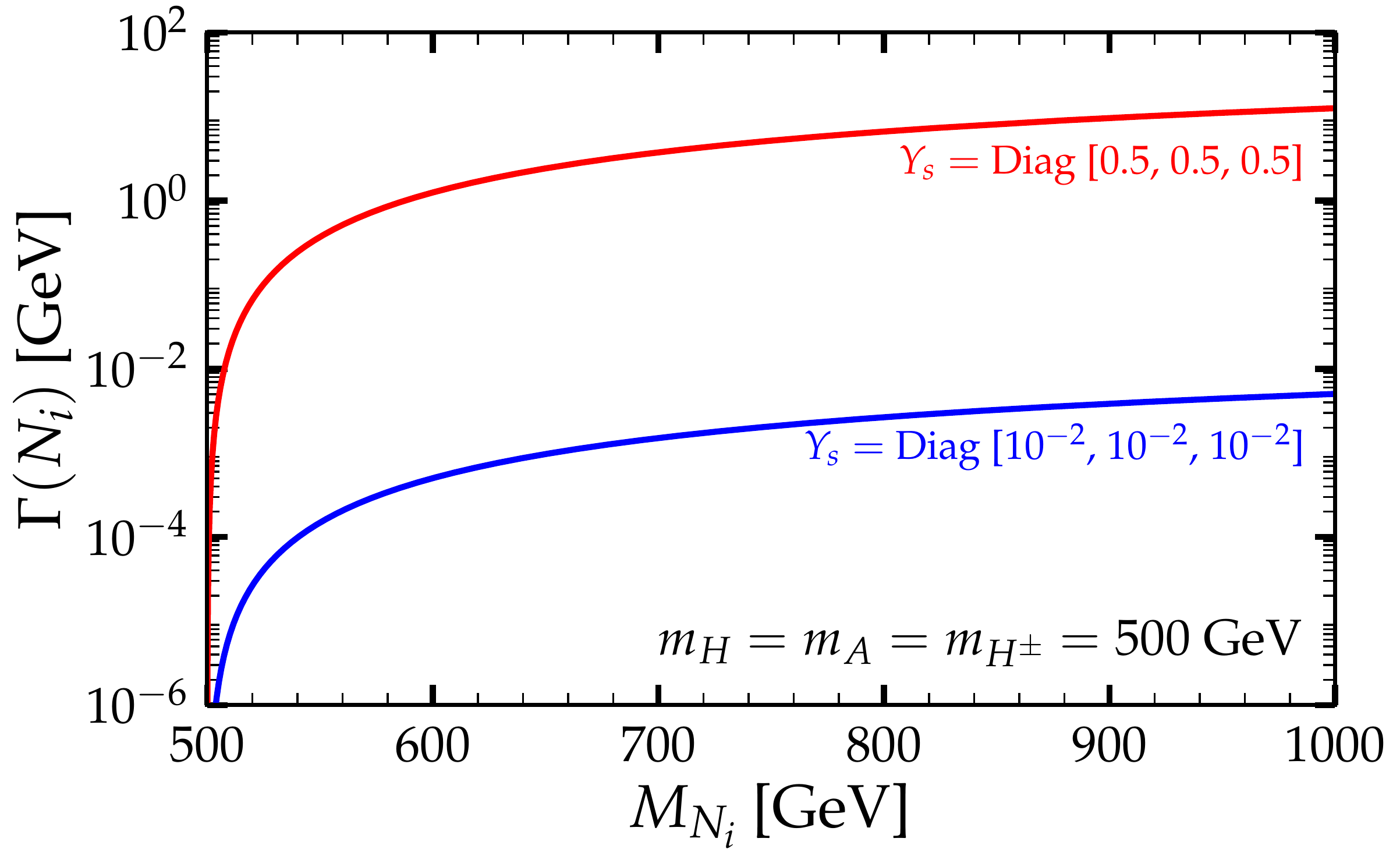}
 	\caption{
          Total $ N_i $ decay width versus its mass for different $Y_S $ and scalar mass values.
          The left and right panels are for $m_{H/A,H^{\pm}}=250$ GeV and $m_{H/A,H^{\pm}}=500$ GeV, respectively.
          The blue and red lines correspond to different Yukawa coupling values $Y_S$, as indicated. 
          }
 	\label{fig:Ndecay}
 \end{figure}
 
 In Fig.~\ref{fig:Ndecay}, we show the total decay width of $N_i$ for
 $M_{N_i}>m_{H^\pm},m_{H/A}$.
 One can see that, for larger $Y_S$ values, the decay width is much larger than that coming just from light-heavy mixing and, as expected, increases as $M_{N_i}$ increases.
 One sees that, as long as masses are far enough from the edge of phase space, 250 GeV and 500 GeV in our case, there is no substantial effect on the decay width.
 Moreover, even for relatively small $Y_S\sim\mathcal{O}(10^{-2})$, the decay width is always large enough such that $N_i$ is not long-lived. In this case, its branching ratios obey the following relations:
\begin{align} \text{BR}(N_i\to\ell^\pm H^\mp):\text{BR}(N_i\to\nu_\ell
H):\text{BR}(N_i\to\nu_\ell A)=2:1:1.
\end{align}
Note that for smaller mass of $N_i$, one can have decay modes such as
$\ell_i^\pm qq'$, $\ell_i^\pm\ell_j^\mp\nu$ and $\nu_\ell
\nu\bar{\nu}$ through the off-shell decay of $H^\pm,H/A$.
However in our range of interest for $v_\chi$, the contribution of new scalars to these 3-body decay modes will be negligible compared to the
contribution coming from SM gauge boson exchange involving light-heavy neutrino mixing. 
This feature is apparent in the small variation of the $N_i$ decay width with changing masses of the scalars.
This follows from the leptophilic nature of the new
scalars, for example, their couplings to quarks $q' q H^\pm $, $q q H$, $q q A$ are all suppressed as $\mathcal{O}(\frac{v_\chi}{v})$. 
We will discuss this further in the next section.
 
 \subsection{$H/A$ and $H^\pm$ decay modes}
 
\begin{figure}[!htbp]
	\centering \includegraphics[width=0.49\linewidth]{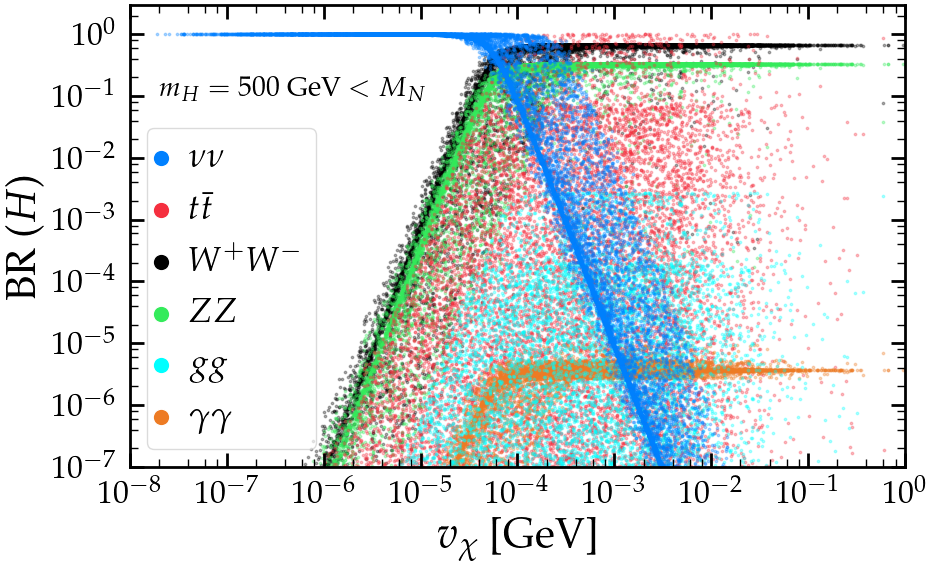}
	\includegraphics[width=0.49\linewidth]{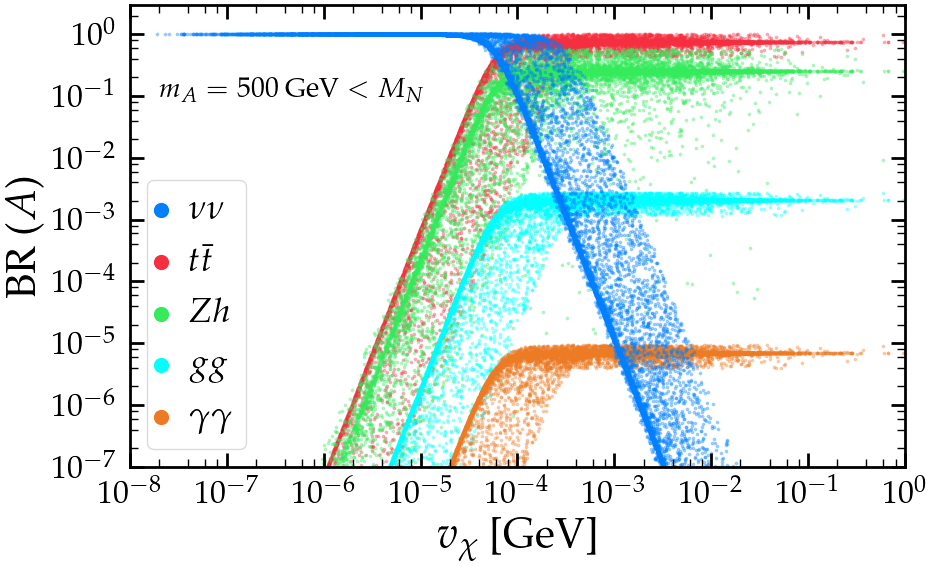}
	\includegraphics[width=0.49\linewidth]{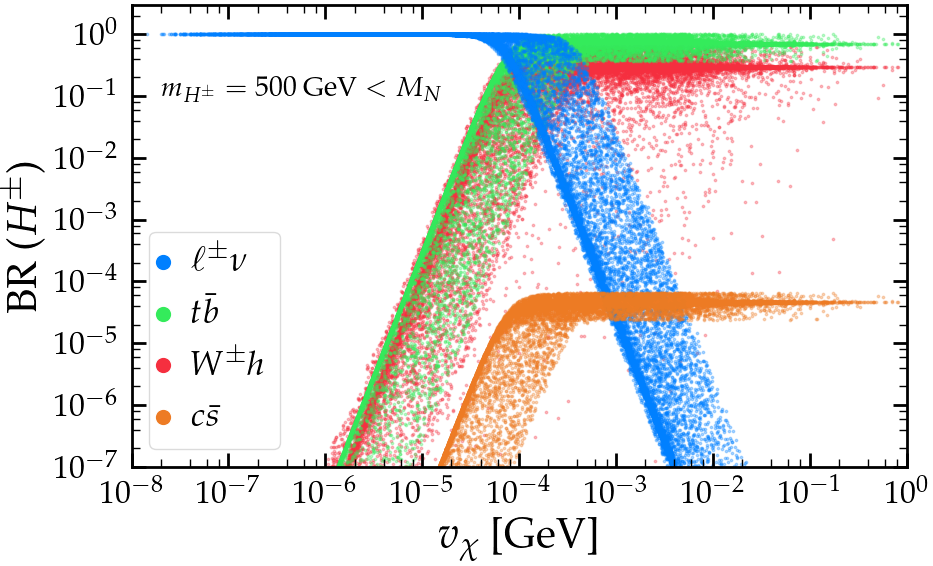}
	\caption{
          Neutral and charged Higgs boson decay branching ratios versus $v_\chi$ for $m_{H,A,H^\pm}=500$GeV $<M_{N_i}$  and $M_{N_i}$ randomly varied from $1$ TeV to $100$ TeV.}
	\label{fig:case1Hdecay}
\end{figure}

 We now turn to the decays of the new scalar particles, which are basically determined by the $\rm U(1)_L$ symmetry.
 The decay pattern of new scalars will also depend on whether these are heavier or lighter than heavy neutrinos $N_i$.  For example,
 Fig.~\ref{fig:case1Hdecay} shows the branching ratios of $H$, $A$ and $H^\pm$ to various decay channels as a function of $v_\chi$ assuming $m_{H,A,H^\pm}<M_{N_i}$.
Hence as long as $v_\chi$ is small, in this kinematical regime the
dominant decay modes are $H/A\to\nu \nu$ and $H^\pm\to\ell^\pm\nu$.
Since in our model the new neutral and charged scalars $H/A$ and $H^\pm$ are mainly composed by $\chi_L$ with just a tiny admixture of $\Phi$,
couplings such as $(H/A)WW$, $(H/A)ZZ$, $(H/A)\ell\ell$, $(H/A)qq$, $AhZ$, $H^\pm qq'$ and $H^\pm W^\mp h$ are all suppressed by $\mathcal{O}(v_\chi/v)$.
It follows that for small values of $ v_\chi $ the $H/A\to\nu\nu $ and
$ H^+\to l^+\nu $ decay channels dominate.
On the other hand, Fig.~\ref{fig:case1Hdecay} shows that other channels
such as $H/A\to qq$, $H^\pm\to qq'$ or $H/A$ or $H^\pm$ decaying to the SM gauge bosons
dominate over the $H/A\to\nu\nu $ or $H^+\to l^+\nu $ channels if
$v_\chi$ is relatively large. Similar results follow for the case of
$m_{A/H,H^\pm} >M_N$. In this case for small $v_\chi$, the dominant decay modes are $H/A\to\nu N$ and $H^\pm\to\ell^\pm N$.

Hence, in our range of interest for $v_\chi$, the relevant decay modes
are either $H/A\to\nu \nu$, $H^\pm\to\ell^\pm\nu$~($m_{A/H,H^\pm} <
M_N$) or $H/A\to\nu N$, $H^\pm\to\ell^\pm N$~($m_{A/H,H^\pm} > M_N$).
In the following subsection we discuss in detail the expressions of
these relevant decay widths and how they depend on the model
parameters.
 
  \subsubsection{$m_{H/A}, m_{H^\pm}<M_{N_i}$}

In this scenario for very small values of $v_\chi$,~or $\alpha\approx 0,\beta\approx\frac{\pi}{2}$, the charged
Higgs boson dominantly decays to light neutrinos and charged leptons via the Yukawa interaction.  The decay width is
\begin{align} \Gamma(H^\pm\to\ell^\pm\nu_i)\approx \frac{|(Y_S
M_N^{-1}Y_\nu^{\dagger})_{\ell i}|^2 v^2\sin^4\beta}{32\pi}m_{H^\pm}.
\end{align}
On the other hand neutral scalars $H/A$ for small values of $v_\chi$ dominantly decay to light neutrinos:
\begin{align} & \Gamma(H\to\nu_i\nu_j)\approx \frac{|(Y_S
M_N^{-1}Y_\nu^{\dagger})_{i j}|^2
v^2\cos^2\alpha\sin^2\beta}{64\pi}m_{H},\\ &
\Gamma(A\to\nu_i\nu_j)\approx \frac{|(Y_S M_N^{-1}Y_\nu^{\dagger})_{i
j}|^2 v^2\sin^4\beta}{64\pi}m_{A}.
\end{align}
Hence, in the limit of very small $v_\chi$, the decay width of $H^\pm,H/A$ will be strongly suppressed for small values of
$Y_S$ or for heavy neutrino mass and therefore, these particles can be long-lived.
On the other hand, for relatively large $v_\chi$, the Yukawa coupling $Y_\nu$ is small if $Y_S\ne 0$ implying that decay modes such as $H^\pm\to\ell^\pm\nu$ and $H/A\to\nu\nu$ will again be suppressed.
However, the scalars will not necessarily be long-lived as with relatively large $v_\chi$, the small mixing in the scalar sector becomes crucial.
In this scenario, charged and neutral Higgs will dominantly decay into quarks or gauge bosons.
The expressions for these decay widths are given in Appendix~\ref{app:decaywidth}.

Fig.~\ref{fig:case1Hdecay} shows the branching ratios  of $H$, $A$ and $H^\pm$ to various decay channels as a function of $v_\chi$ assuming
$m_{H,A,H^\pm}<M_{N_i}$. For these plots, we have scanned the parameter space according to Table.~\ref{tab:param}.
We have ensured vacuum stability and perturbativity of the $\lambda$
parameters.
We have also taken into account neutrino oscillation data in varying
the Yukawas $Y_\nu$, $Y_S$, imposing also the perturbative
requirements $\text{Tr}(Y_\nu^\dagger Y_\nu)<4\pi$,
$\text{Tr}(Y_S^\dagger Y_S)<4\pi$.
One can see that for small values of $v_\chi$, the $H\to\nu\nu $,
$A\to\nu\nu$ and $H^+\to l^+\nu$ decay channels dominate over all
other channels. This is easy to understand since in this case the
Yukawa coupling $Y_\nu$ can be relatively large.
For large values of $ v_\chi $, as the Yukawa coupling $Y_\nu$ needs to
be small, other channels involving scalar mixing (particularly the $
H\to VV $, $ A\to q\bar{q} $, $ A\to hZ $ and $ H^+\to t\bar{b} $,
$H^\pm\to W^\pm h$ channels) dominate over the $ H/A\to\nu\nu $ and $
H^+\to l^+\nu $ channels.

\subsubsection{$m_{H/A}, m_{H^\pm}>M_{N_i}$}

In the complementary case where $m_{H/A},m_{H^\pm}>M_{N_i}$, the charged Higgs bosons dominantly decay through the Yukawa interaction
into charged leptons $\ell^\pm$ and on-shell heavy neutrinos $N_i$ as
\begin{align} \Gamma(H^\pm\to \ell^\pm N_i)\approx
\frac{|(Y_S)_{\ell
i}|^2\sin^2\beta}{16\pi}m_{H^\pm}\Big(1-\frac{M_{N_i}^2}{m_{H^\pm}^2}\Big)^2,
\end{align}
whereas the neutral Higgs bosons dominantly decay into light neutrinos $\nu_\ell$ and on-shell heavy neutrinos $N_i$ as
\begin{align} &\Gamma(H\to\nu_\ell N_i)\approx
\frac{|(Y_S)_{\ell
i}|^2\cos^2\alpha}{32\pi}m_H\Big(1-\frac{M_{N_i}^2}{m_{H}^2}\Big)^2
\\ & \Gamma(A\to\nu_\ell N_i)\approx
\frac{|(Y_S)_{\ell i}|^2
\sin^2\beta}{32\pi}m_{A}\Big(1-\frac{M_{N_i}^2}{m_{A}^2}\Big)^2.
\end{align}
Note that in the limit of small $v_\chi$, the dominant decay modes in this case are proportional to $Y^2_S$.
Therefore, in contrast to the previous case, within the linear seesaw the decays are not suppressed by heavy-neutrino mass beyond the phase-space factor, as the Yukawa coupling $Y_S$ can be large even for weak-scale mediators consistent with neutrino mass.
It follows that the scalars will not be long-lived even for small $v_\chi$.
As in the previous case, for $Y_S$ small or $v_\chi$ relatively large, the small mixing in the scalar sector becomes important and the
charged and neutral Higgs scalars will dominantly decay into quarks or gauge bosons. The expressions for these decay widths are given in
Appendix~\ref{app:decaywidth}.

\section{Collider signatures}
\label{sec:collider-signatures}

\par Having discussed the various heavy-neutrino-mediator and new scalar production modes, we now discuss the associated collider signatures.
Once the heavy neutrinos or new scalars are produced, they will further decay to various possible decay channels depending on
kinematical considerations.
For example, if $M_{N_i}> m_{H^\pm},m_{H/A}$ and $Y_S$ is large, heavy neutrinos will dominantly decay to final states such as $\ell^\pm H^\mp$ and $\nu_\ell H/A$.
On the other hand, if $M_{N_i}<m_{H^\pm},m_{H/A}$, heavy neutrinos will decay to SM final states such as $\ell W$, $\nu Z$ and $\nu h$
through light-heavy neutrino mixing.
Let us now discuss the possible signatures associated to various production processes:

\subsubsection*{\underline{$\mathbf{e^+e^-\to N_i N_i}$}}

\par
\begin{figure}[h]
	\includegraphics[width=0.65\linewidth]{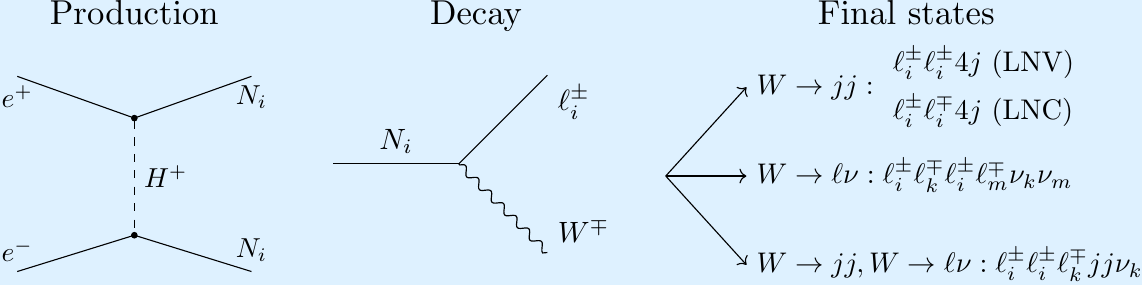}
    	\includegraphics[width=0.65\linewidth]{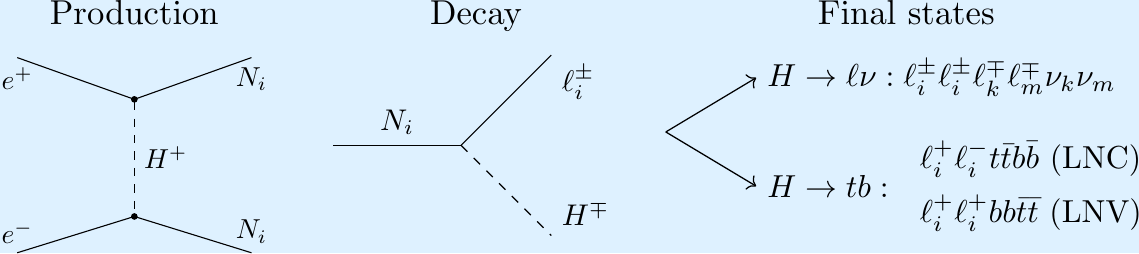}
	\caption{
          Illustrative representation of heavy-neutrino
          pair-production and decay channels.  The upper row is for
          the case of $M_{N_i}<m_{H^\pm}$ and the bottom one is for
          $M_{N_i}>m_{H^\pm}$.}
	\label{fig:eetoNNtoFS}
\end{figure}

 
In Fig.~\ref{fig:eetoNNtoFS}, we show the various possible final states coming from the pair production of heavy neutrinos at $e^+e^-$ collider.
The first row of Fig.~\ref{fig:eetoNNtoFS} corresponds to the case of $M_{N_i}<m_{H^\pm}$, whereas the second row is for the case of $m_{N_i}>m_{H^\pm}$.
When $M_{N_i}<m_{H^\pm}$, $N_i$ will dominantly decay to SM final states through light-heavy neutrino mixing
and we depict the charged-current-mediated decay mode $N_i\to\ell W$ with $W$ boson decaying leptonically or hadronically.
\begin{figure}[h]
	\includegraphics[width=0.49\linewidth]{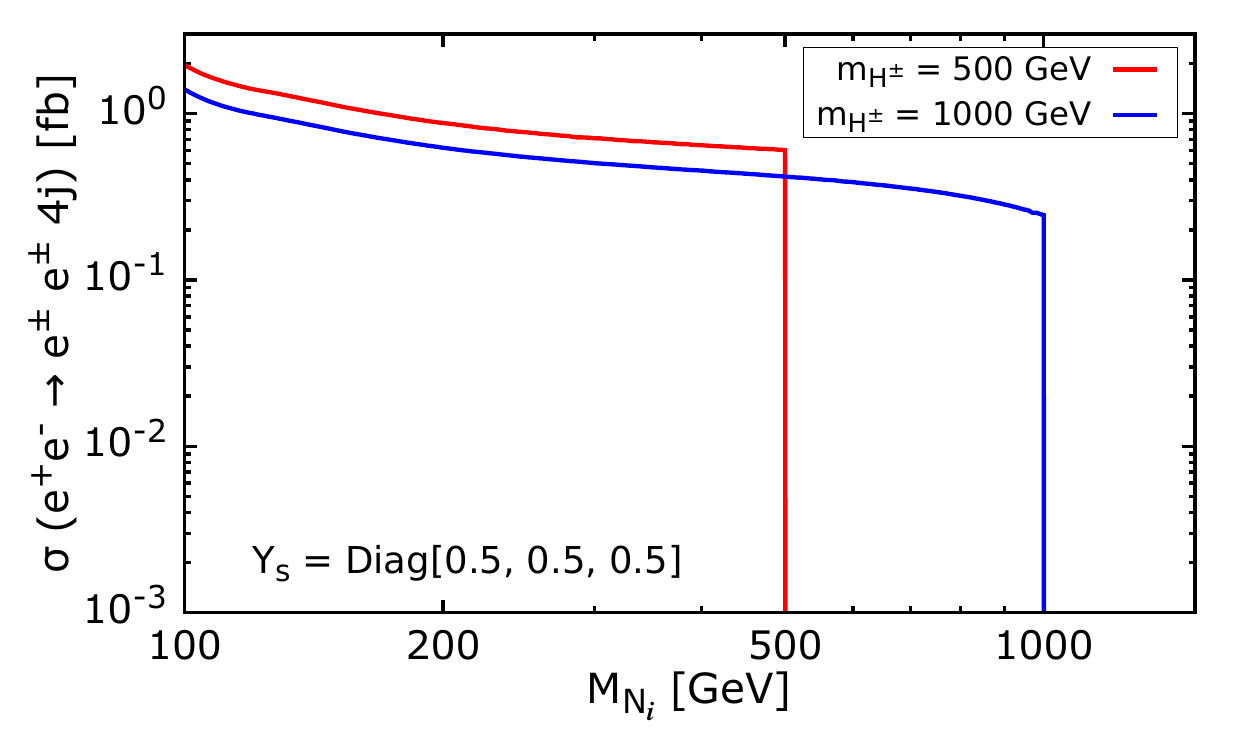}
    	\includegraphics[width=0.49\linewidth]{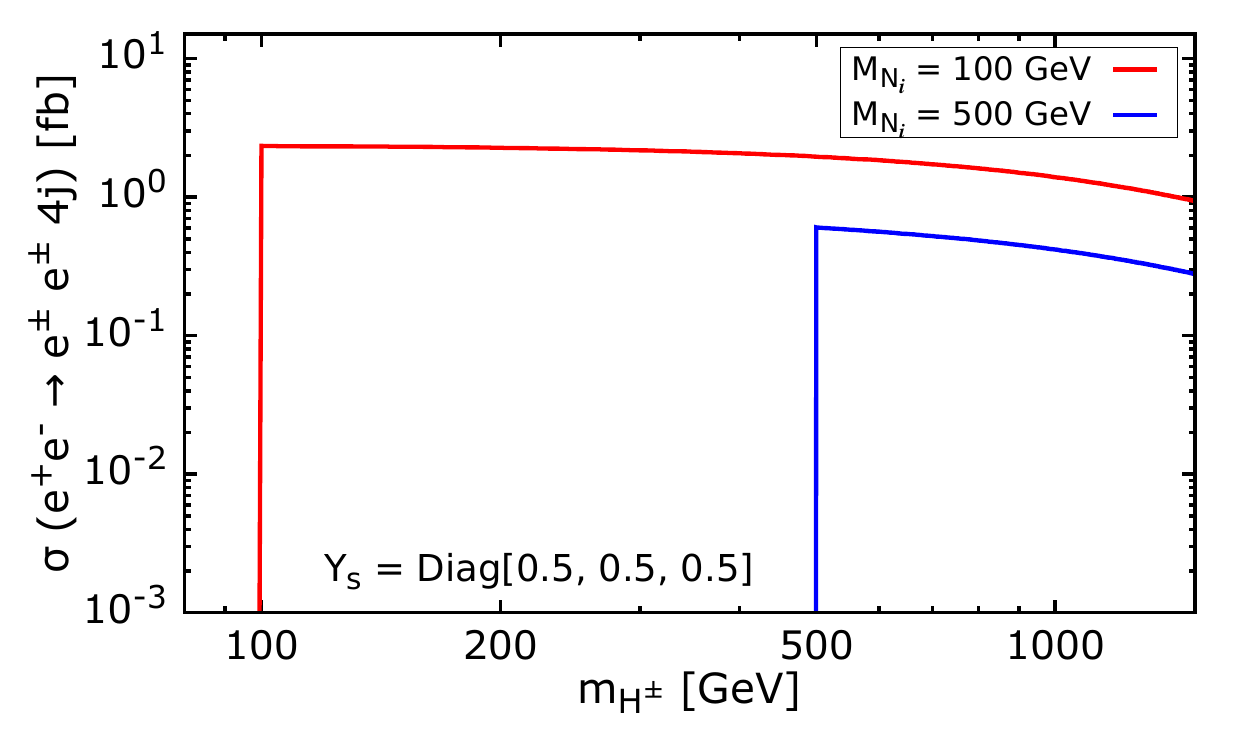}
	\caption{
          Cross-section for the process $e^+e^-\to N_iN_i\to e^\pm
          e^\pm 4j$ versus the mass of heavy neutrinos $M_{N_i}$ (left
          panel), and versus the mass of charged Higgs $m_{H^+}$
          (right panel). }
	\label{fig:CSeetoee4j}
\end{figure}

If both $W$ boson decays hadronically, we have lepton number violating~(LNV) and lepton number conserving~(LNC) final states
$\ell_i^\pm\ell_i^\pm 4j$ and $\ell_i^\pm\ell_i^\mp 4j$, respectively.
When the decay width $\Gamma_N$ is comparable or smaller than the mass splitting within the quasi-Dirac pair $\Delta M\sim m_\nu$, the cross
section for both the LNV and LNC final states can be of the same order, see Appendix~\ref{app:LNV_vs_LNC} for a detailed discussion.
The LNV same-sign dilepton final state is quite interesting as it may help probing the Majorana nature of neutrinos~\cite{Schechter:1981bd}.
Moreover, the LNV signal $\ell_i^\pm\ell_i^\pm 4j$ is almost free of SM backgrounds.

In Fig.~\ref{fig:CSeetoee4j}, we show the cross section for the process $e^+e^-\to N_i N_i\to e^\pm e^\pm 4j$ at $\sqrt{s}=3$ TeV
under the assumption $|V_{eN_i}|\neq 0$, $|V_{\mu N_i}|=|V_{\tau N_i}|=0$.
The left and right panel stands for the cross-section with respect to $M_{N_i}$ and $m_{H^\pm}$ where we fix the Yukawa coupling
$Y_S=\text{Diag}(0.5,0.5,0.5)$.
Different lines in each panel correspond to different values of charged Higgs or heavy neutrino mass.
\par If both the $W$ bosons decay leptonically, we will have multileptonic final states associated with missing transverse energy such as
$\ell_i^\pm \ell_i^\pm \ell_k^\mp \ell_m^\mp +\slashed{E}_T$.
On the other hand, if one of the $W$ bosons decays leptonically and other one decays hadronically, we can have semileptonic final states
in association with missing transverse energy such as $\ell_i^\pm\ell_i^\pm\ell_k^\mp jj+\slashed{E}_T$.
Such final states will have large enough cross-sections, as the leptonic branching ratio $\text{BR}(W\to\ell\nu)$ is almost $10\%$
while the hadronic branching ratio is $\text{BR}(W\to jj)\approx 67\%$.
However, unlike the LNV final states, for these final states there
will be SM backgrounds, requiring a dedicated analysis in order to
reduce them.
\par The lower diagrams in Fig.~\ref{fig:eetoNNtoFS} show the other
final states possible for the case of $M_{N_i}>m_{H^\pm}, m_{H/A}$.
In this case $N_i$ will dominantly decay to $\ell^\pm H^\mp$ or $\nu
H/A$.  Here we only show the possible final states coming from the
decay chain $N_i\to\ell^\pm H^\mp$ and $H^\pm\to\ell^\pm\nu$ or
$H^\pm\to tb$.
As seen in Fig.~\ref{fig:case1Hdecay}, when $v_\chi$ is very small the
dominant decay mode is $H^\pm\to\ell^\pm\nu$, while for relatively
large $v_\chi$, the dominant mode is $H^\pm\to tb$.
Although in the small $v_\chi$ limit $H^\pm\to\ell^\pm\nu$ is the
dominant decay mode, the decay width can be strongly suppressed for
small values of $Y_S$ or for very heavy neutrino masses.
In this case $H^\pm$ can be long-lived and hence the secondary charged
tracks of the long-lived charged Higgs boson can be tagged at the ILC.
However, if $Y_S$ is relatively large or $M_{N_i}$ is not so large,
then $H^\pm$ will not be long-lived and $H^\pm$ to $\ell^\pm\nu$ can give
rise to multilepton final states containing missing energy,
i.e. $\ell_i^\pm\ell_i^\pm\ell_k^\mp\ell_m^\mp+\slashed{E}_T$.
The cross-section for this final state is of the same order as for the
LNV same-sign dilepton final state shown in Fig.~\ref{fig:CSeetoee4j}.
On the other hand, for relatively large $v_\chi$, the decay chain
$H^\pm\to tb$ will give rise to both LNV and LNC final states such as
$\ell_i^+\ell_i^+ bb\bar{t}\bar{t}$ and
$\ell_i^+\ell_i^-tt\bar{b}\bar{b}$, respectively. However, these are
suppressed in the small $v_\chi$ limit.
%

\subsubsection*{\underline{$\mathbf{e^-\gamma\to N_i H^-}$}}
\begin{figure}[h]
	\includegraphics[width=0.65\linewidth]{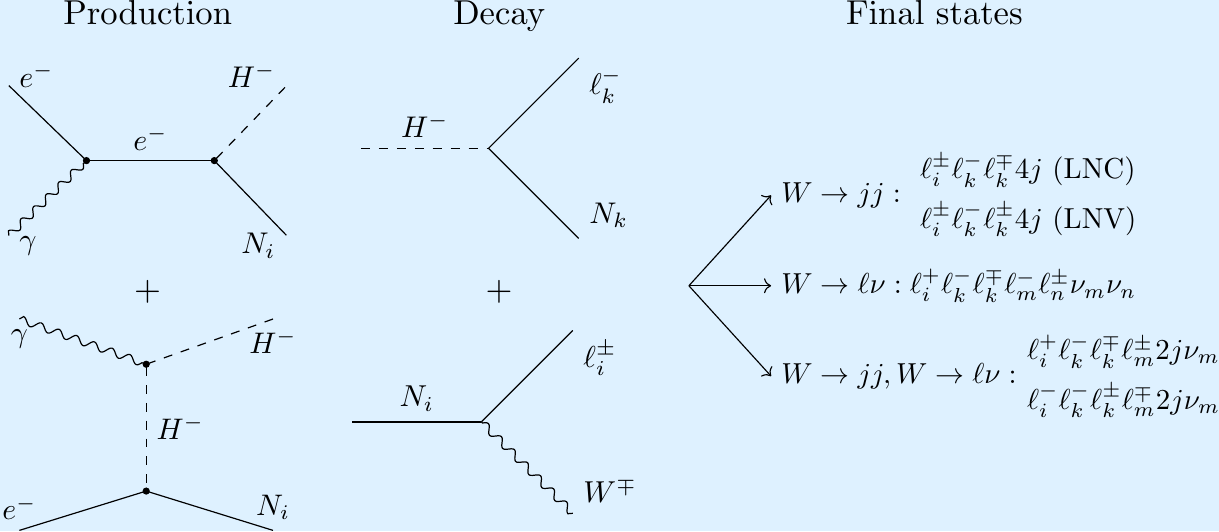}
	\caption{
          Illustrative representation of the heavy neutrino production
          in association with charged Higgs and the expected decay
          channels for the case $M_{N_i}<m_{H^\pm}$.}
	\label{fig:yetoHmNtoFS}
\end{figure}
\begin{figure}[h]
	\includegraphics[width=0.49\linewidth]{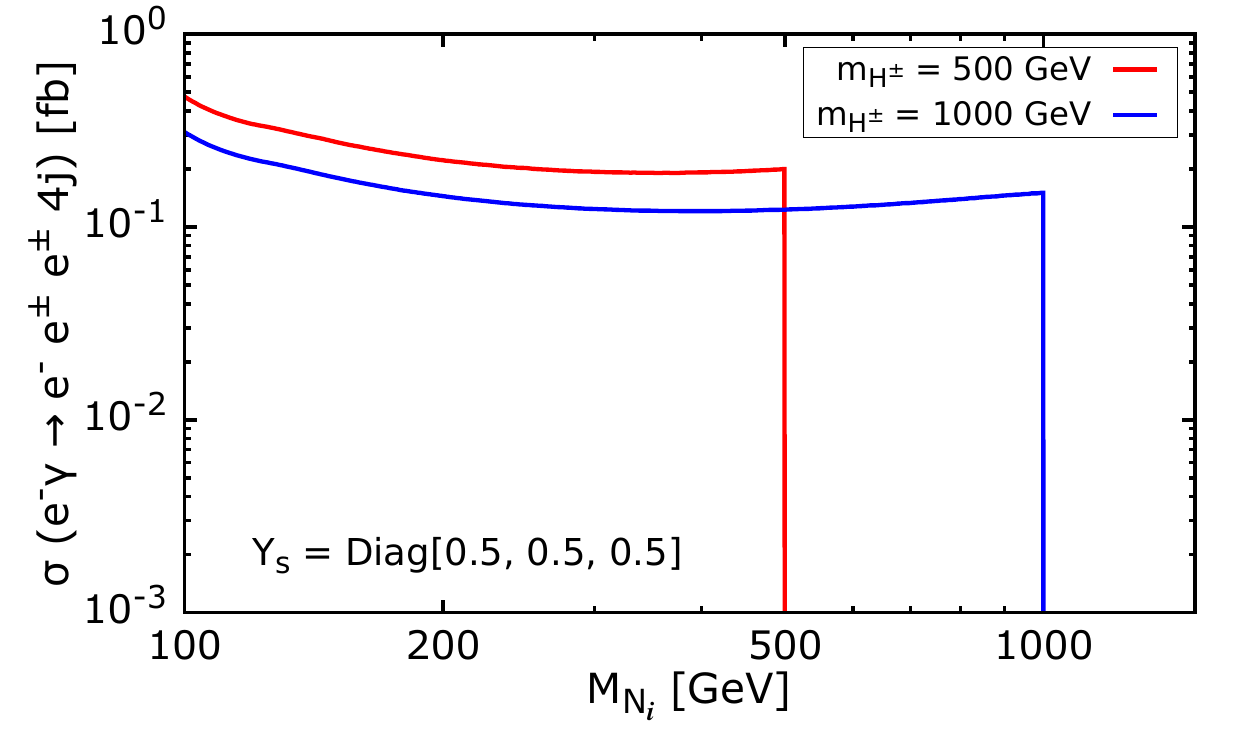}
    	\includegraphics[width=0.49\linewidth]{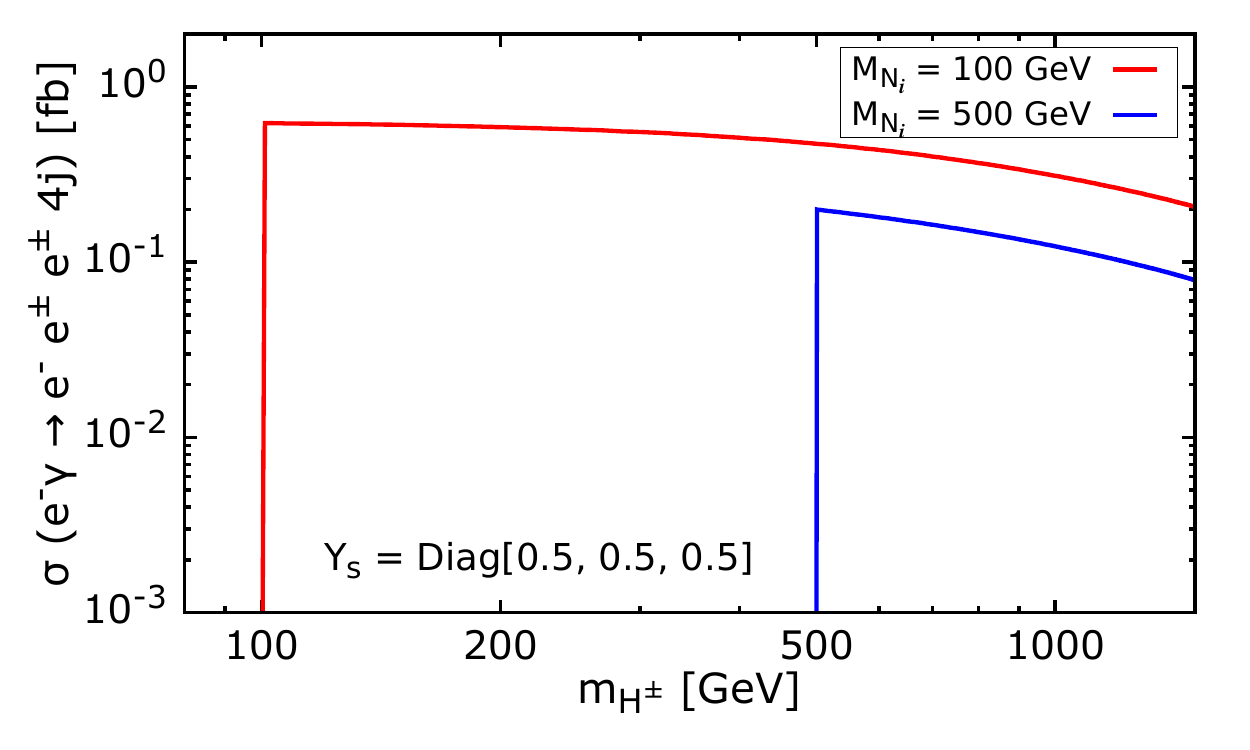}
	\caption{
          Cross-section for the process $e^-\gamma\to H^-N_i\to
          e^-e^\pm e^\pm 4j$ versus the heavy-neutrino mediator mass
          $M_{N_i}$ (left panel), and versus the charged Higgs mass
          $m_{H^+}$ (right panel).}
	\label{fig:CSeatoeee4j}
\end{figure}
\par

In Fig.~\ref{fig:yetoHmNtoFS} we show the possible interesting final states coming from the process $e^-\gamma\to N_i H^-$ when $m_{H^\pm}>M_{N_i}$.
A possible decay chain is $H^\pm\to\ell_i^\pm N_i$, with $N_i\to\ell_i^\pm W^\mp$, and either leptonic or hadronic decay of the $W$ boson.
If both $W$ bosons decay hadronically, one can have LNV or LNC trilepton final state such as $\ell_i^\pm\ell_k^-\ell_k^\pm 4j$ and
$\ell_i^\pm\ell_k^-\ell_k^\mp 4j$, respectively.
Again, the LNV signal $\ell_i^\pm\ell_k^-\ell_k^\pm 4j$ seems more interesting, as it is free from the SM background and might help 
probing the Majorana nature of neutrinos~\cite{Schechter:1981bd}.
In Fig.~\ref{fig:CSeatoeee4j} we show the cross section for the LNV final state $e^\pm e^- e^\pm 4j$ at $\sqrt{s}=3$~TeV assuming $|V_{eN_i}|\neq 0$, $|V_{\mu N_i}|=|V_{\tau N_i}|=0$.
The left and right panels show the cross-section versus $M_{N_i}$ and $m_{H^\pm}$ fixing the Yukawa coupling as $Y_S =$Diag(0.5, 0.5, 0.5).
Different lines in each panel correspond to different values of the charged Higgs or heavy neutrino masses.
Similar to the previous case, if both the $W$ bosons decay leptonically, or if one $W$ decays hadronically and other one
leptonically, we will have pure multileptonic or semi-leptonic final states, respectively, accompanied by missing energy.
\begin{figure}[h]
	\includegraphics[width=0.56\linewidth]{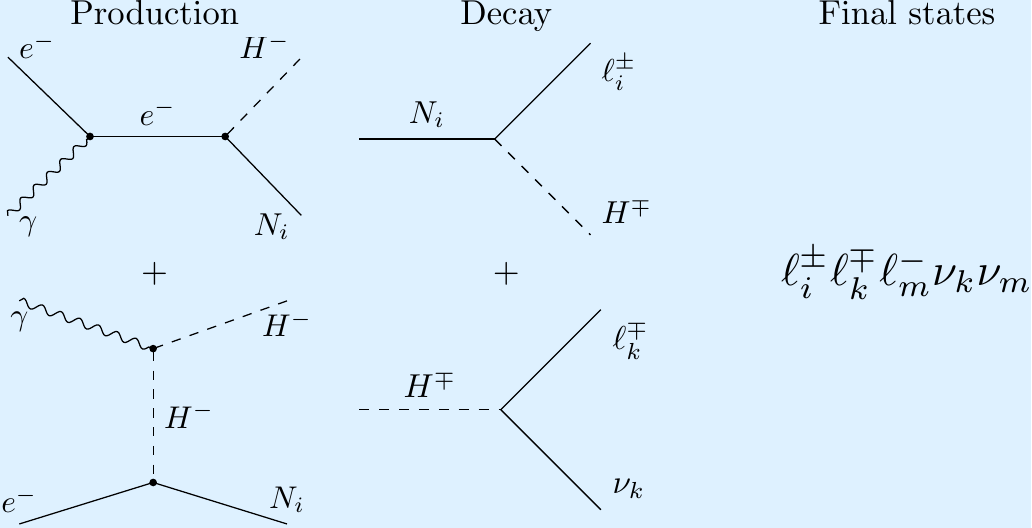}
	\caption{
          Illustrative representation of the heavy neutrino production
          in association with charged Higgs and their various decay
          channels for the case of $M_{N_i}>m_{H^\pm}$.}
	\label{fig:eatoFSC2}
\end{figure}
\par Fig.~\ref{fig:eatoFSC2} shows the possible final states arising from the process $e^-\gamma\to N_iH^-$ for the case when $M_{N_i}>m_{H^\pm}$ and $Y_S$ is large.
This case seems not too interesting phenomenologically, as the final states have large missing energy.
Other decay modes of $H^\pm$ for the $M_{N_i}>m_{H^\pm}$ case are small for small $v_\chi$, hence are not discussed here.

\subsubsection*{\underline{$\mathbf{e^+ e^-/pp \to H^+ H^-}$}}
\begin{figure}[!h]
	\includegraphics[width=0.63\linewidth]{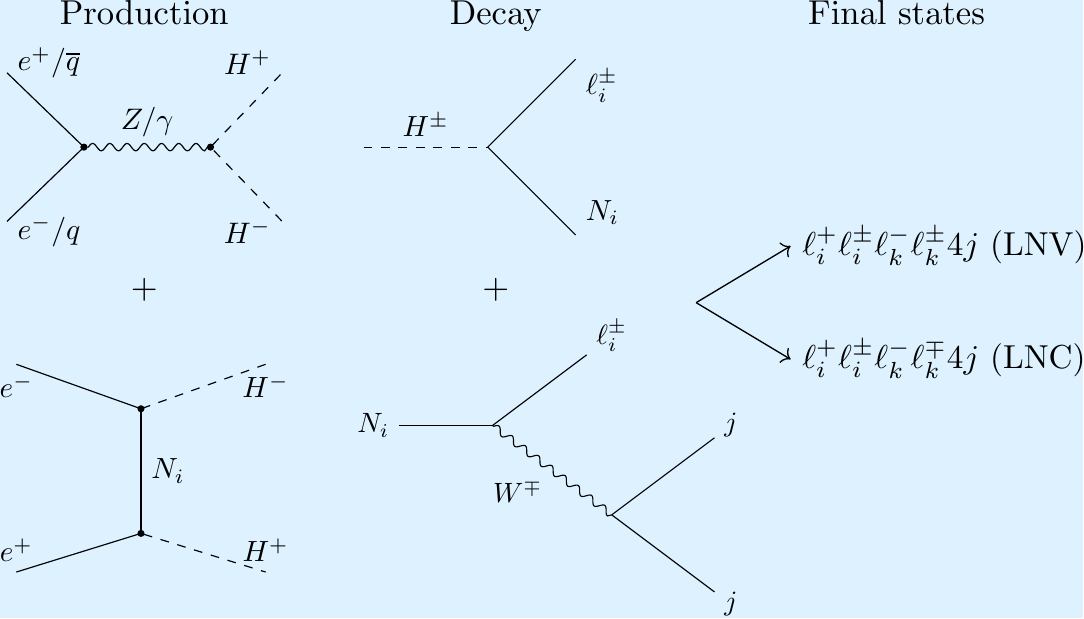}
	\caption{
          Illustrative representation of the pair of charged Higgs
          production and the expected decay channels for the case
          $M_{N_i}<m_{H^\pm}$.}
	\label{fig:eetoHHtoFS}
\end{figure}
\begin{figure}[!h]
	\includegraphics[width=0.49\linewidth]{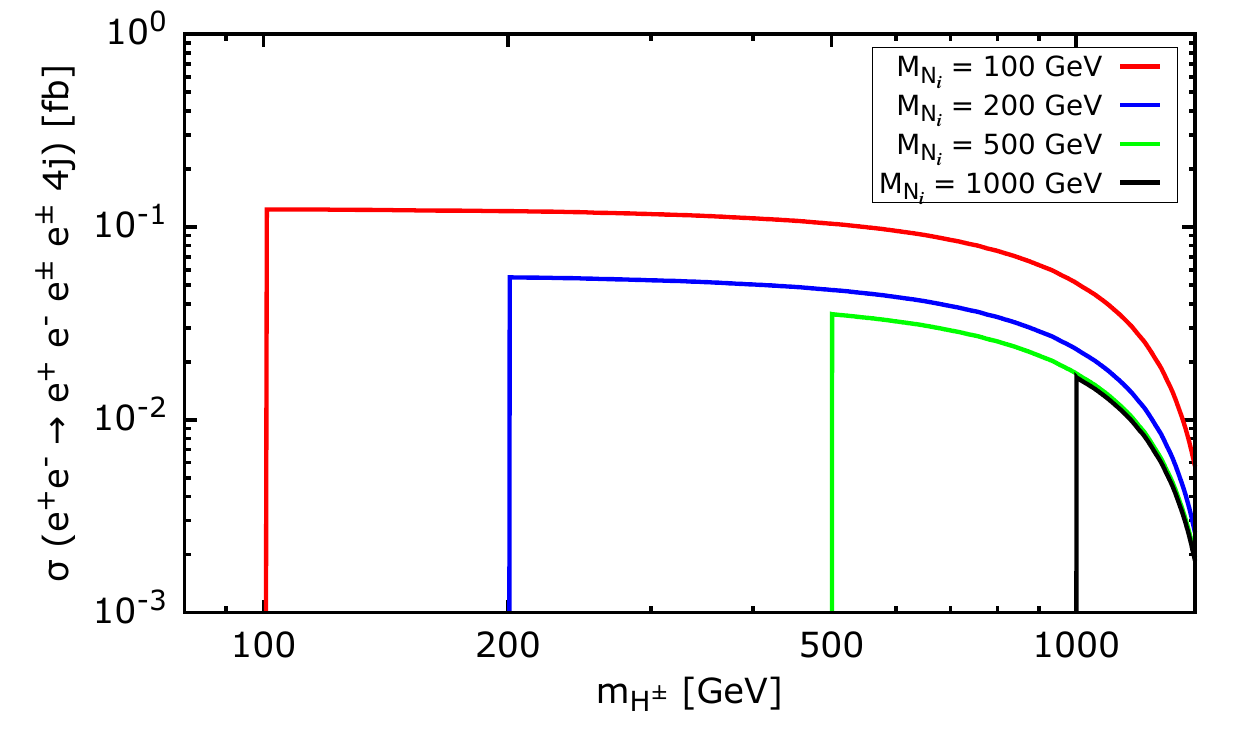}
	\includegraphics[width=0.49\linewidth]{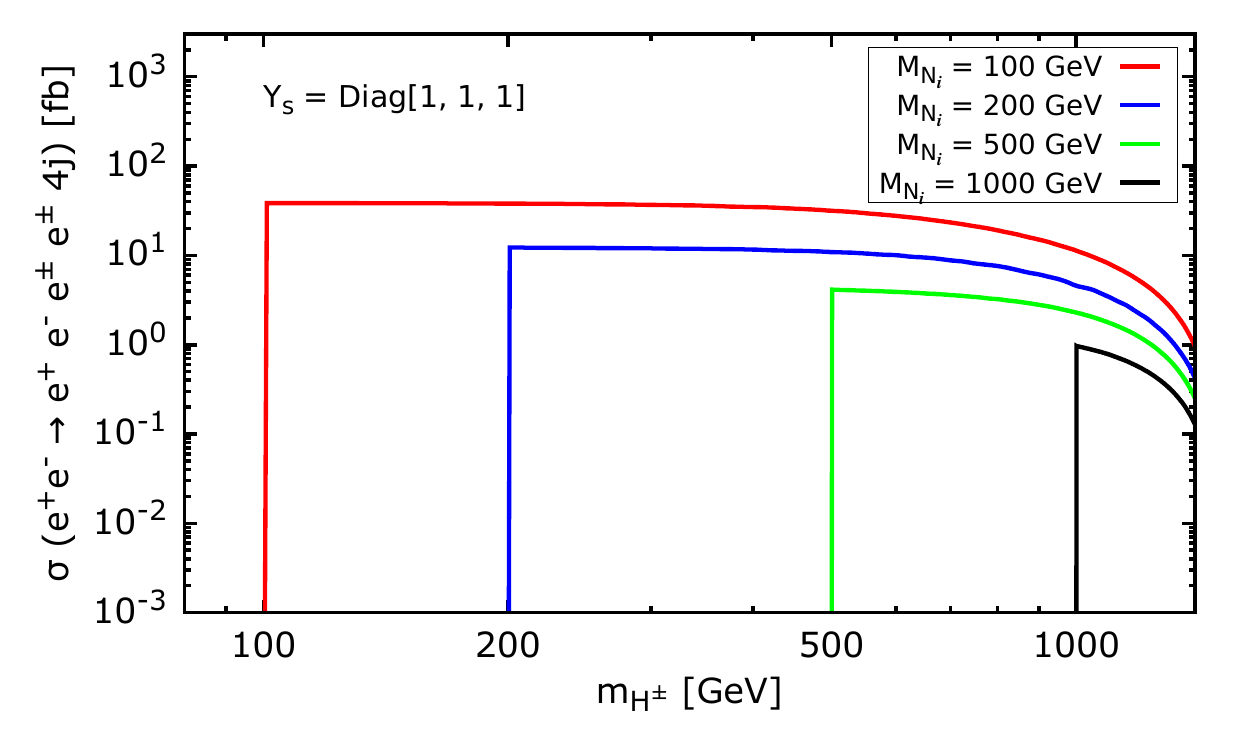}
	\includegraphics[width=0.49\linewidth]{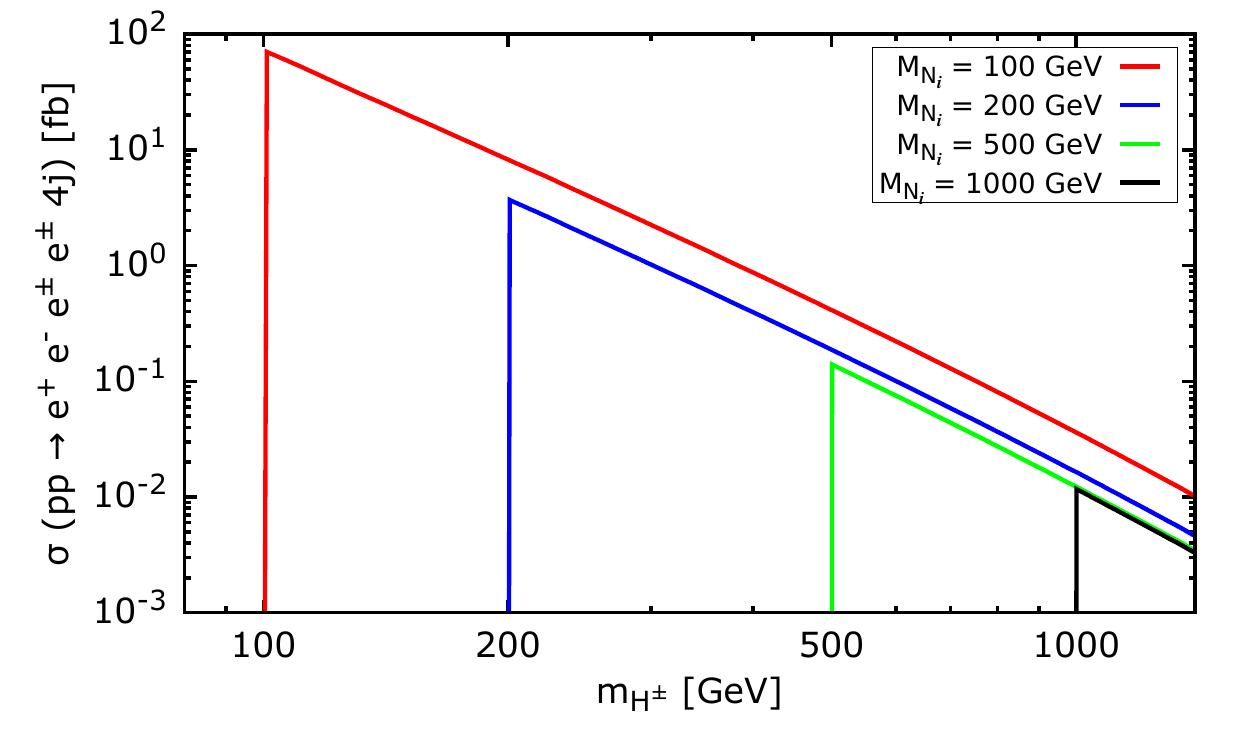}
	\caption{
          Cross-section for the process $XX\to H^+H^-\to e^+N_i
          e^-N_i\to e^+e^-e^\pm e^\mp 4j$ versus the charged scalar
          mass.  For the upper left panel we have taken only the
          s-channel contribution to the process $e^+e^-\to H^+H^-$
          whereas for the upper right panel we took both the s- and
          the t-channel contributions for $Y_s=$Diag(1, 1, 1).  We
          took the center of mass energies $\sqrt{s}=$ 3 TeV and
          $\sqrt{s}=$ 100 TeV for the $e^+e^-$ collider and $pp$
          collider, respectively.}
	\label{fig:CSXXtollll4j}
\end{figure}
\par
In Fig.~\ref{fig:eetoHHtoFS}, we show interesting final states coming from the production process $e^+e^-/pp\to H^+ H^-$ when $m_{H^\pm}>M_{N_i}$.
Note that at a $pp$ collider production comes only through the Drell-Yan mechanism, whereas for large $Y_S$ the production at lepton colliders is dominated by the t-channel heavy
neutrino mediation.  From the decay chain $H^\pm\to\ell_i^\pm N_i$, $N_i\to\ell_i^\pm W^\mp$, $W\to jj$, we find two very interesting
semileptonic final sates: $\ell_i^+\ell_i^\pm\ell_k^-\ell_k^\pm 4j$~(LNV) and $\ell_i^+\ell_i^\pm\ell_k^-\ell_k^\mp 4j$~(LNC).
In Fig.~\ref{fig:CSXXtollll4j} we show the cross section for the LNV final state $XX\to e^+e^\pm e^- e^\pm 4j$ under the assumption $|V_{eN_i}|\neq 0$, $|V_{\mu N_i}|=|V_{\tau N_i}|=0$.
Different lines in each panel correspond to different values of the heavy neutrino masses.
The upper panel is for $\sqrt{s}=3$ TeV $e^+e^-$ collider where in the left panel we only take into account the contribution coming from the
Drell-Yan mechanism while in the right panel we include both the Drell-Yan and the t-channel contribution with Yukawa coupling $Y_S=$Diag(1, 1, 1).
Comparing the left and right panels of Fig.~\ref{fig:CSXXtollll4j} one sees that there is a huge enhancement in the production rate once we
take into account the t-channel contribution.  In the bottom panel of Fig.~\ref{fig:CSXXtollll4j} we show the LNV cross-section at a proton-proton collider with $\sqrt{s}=100$~TeV.
Note that the specific Yukawa coupling values do not matter for the case of hadron collider, as the cross-section comes just from the Drell-Yan
contribution.  Therefore, the pair production cross-section of charged scalars at hadron collider becomes smaller for large charged Higgs masses.
\begin{figure}[h]
	\includegraphics[width=0.6\linewidth]{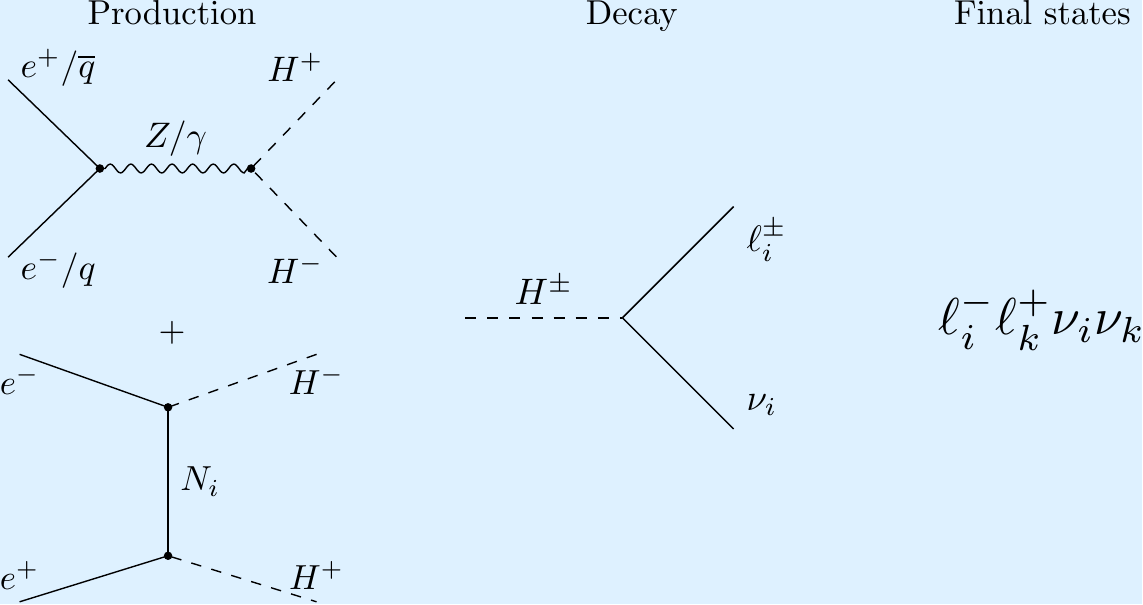}
	\caption{
          Illustrative representation of the charged-Higgs
          pair-production and their various decay channels for the
          case of $M_{N_i}>m_{H^\pm}$.}
	\label{fig:epemtoFSC2}
\end{figure}
\par Fig.~\ref{fig:epemtoFSC2} shows the possible final states coming
from $pp\to H^+H^-$ or $e^+e^-\to H^+H^-$ for $M_{N_i}>m_{H^\pm}$
case.  For small $v_\chi$ values, $H^\pm$ dominantly decays to
$\ell^\pm\nu$, which leads to the large missing energies in the final
states. 
$H^\pm$ can decay to heavy quarks, $tb$, but this channel is
$\mathcal{O}(v_\chi/v)$ suppressed. Altogether, the phenomenology for
the $M_{N_i}>m_{H^\pm}$ case seems not too interesting. 
\par Note that in all of our above discussion of collider signatures, we have considered the heavy neutrinos decay through the charged current interaction.
  One can also consider scenarios where both heavy neutrinos decay through neutral current interaction, such as $N\to\nu Z$ or a scenario where one of them decays
  as $N\to\ell W$ and other one decay as $N\to\nu Z$.
  Taking into account the $Z$ boson leptonic decay mode, and the leptonic/hadronic decay modes of the $W$ boson,
  one finds many interesting final states with high lepton multiplicity, as listed in Appendix~\ref{app:CS}.
\section{Conclusions}
\label{sec:conclusions}
In summary we studied the linear seesaw mechanism where neutrino masses arise from a second scalar doublet $\chi_L$ with lepton number $L[\chi_L]=-2$.
The model has three pairs of fermion singlets $\nu_i^c$, $S_i$ with lepton number $L[\nu_i^c]=-1$ and $L[S_i]=1$ as neutrino mass mediators, see Fig.~\ref{fig:neutrino}.
The smallness of neutrino mass is dynamically explained by the tiny induced VEV of the new Higgs doublet $\chi_L$. 
The scenario provides an attractive benchmark for collider physics, in which the new scalar Higgs bosons $H^\pm, H/A$ as well as the heavy neutrinos can be kinematically accessible. 
\par
We have discussed existing restrictions from neutrino masses and electroweak precision data, that leads to a compressed scalar boson spectrum, see Fig.~\ref{fig:compress}.
One sees from Fig.~\ref{fig:Rgamgam} that LHC searches for rare $\gamma\gamma$ Higgs boson decays do not place any important restrictions on the charged Higgs mass in Fig.~\ref{fig:htogamgam}.
Moreover, we have investigated the novel features associated to cLFV, see Figs.~\ref{fig:LFV} and \ref{fig:LFVplots}. 
Apart from the charged current contribution involving heavy or light neutrinos, there are also charged Higgs contributions to the cLFV rates coming from the Yukawa interactions in the right panel of Fig.~\ref{fig:LFV0}. 
\par Note that the scale of the new fermions mediating neutrino mass generation can lie at the TeV scale and may be accessible at various collider setups.  
As a consequence, our linear seesaw mechanism offers a very intriguing alternative to the classic high-scale type-I seesaw mechanism. 
Indeed, the fact that the Yukawa couplings $Y_\nu$ and $Y_S$ can be sizeable makes our model testable at collider experiments.  
We find that unlike the simplest type-I seesaw mechanism, our proposal offers new unsuppressed heavy-neutrino production mechanisms at various colliders. 
For example, as illustrated in Figs.~\ref{fig:NN-production}, \ref{fig:eetoNN} and Figs.~\ref{fig:HpN-production}, \ref{fig:eatohmN}, \ref{fig:eyNH}, the heavy neutrinos can be produced at $e^+e^-$ and $e^-\gamma$ colliders through charged-Higgs
boson exchange.
These production rates are proportional to $Y_S^4$, and hence can be sizeable while consistent with smallness of neutrino masses, since the latter can be ascribed to the small value of $v_\chi$.
\par We looked at the production of new scalars in $pp$ as well as the $e^-e^+$ colliders as shown in Figs.~ \ref{fig:feynman-production}, \ref{fig:production}, \ref{fig:feyn-prod-hphm} and \ref{fig:HpHm-t}.
Specifically we found that $e^+e^-\to NN$ and $e^-\gamma\to NH^-$ followed by $H^\pm\to\ell^\pm N$ decay provides a promising way to produce heavy neutrinos.  
Moreover, besides the usual Drell-Yan mechanism in $e^+e^-$ collisions, one can have heavy-neutrino-mediated charged-Higgs boson pair production $e^+e^-\to H^\pm H^\mp$ as shown in Figs.~\ref{fig:feyn-prod-hphm} and \ref{fig:HpHm-t}.
The latter can dominate over Drell-Yan production for large Yukawa coupling $Y_S$. 
\par Once the charged-Higgs bosons are produced, the heavy seesaw mediator neutrinos will be produced in their decays, $H^\pm\to\ell^\pm N$, effectively enhancing the heavy neutrino production rates. 
Moreover, assuming $m_{H^\pm,H/A}>M_{N_i}$, the produced heavy neutrinos dominantly decay to SM final states such as $\ell W$, $\nu Z$ and $\nu h$ through the light-heavy neutrino mixing.
We find that the decay chain $N_i\to\ell_j^\pm W^\mp$ with $W^\mp\to jj$ leads to interesting collider signatures. 
\par Fig.~\ref{fig:BRN} shows the decay length and branching fractions of the heavy neutrinos due to the light-heavy neutrino mixing, while Fig.~\ref{fig:Ndecay} displays the decay widths of the heavy neutrinos when $M_{N_i}>m_{H/A/H^\pm}$. 
In this case, the scalars decay to various SM final states as shown in Fig.~\ref{fig:case1Hdecay}. 
For the case $m_{H/A/H^\pm}>M_{N_i}$, the scalars dominantly decay into heavy neutrinos and SM leptons. 
The production and decay of the heavy neutrinos at $e^+ e^-$ collider as illustrated in Fig.~\ref{fig:eetoNNtoFS}, leads to lepton number conserving as well as violating final states.
The cross-section for LNV/LNC final states $e^\pm e^\pm4j/e^\pm e^\mp4j$ at $e^+e^-$ colliders is shown in Fig.~\ref{fig:CSeetoee4j}.
\par Moreover, we also discussed the decays of charged-Higgs and heavy-neutrinos at $e^-\gamma$ colliders, as shown in Fig.~\ref{fig:yetoHmNtoFS}. 
In Fig.~\ref{fig:CSeatoeee4j} we display the cross-section for the resulting three-lepton-four-jet final states $e^-e^\pm e^\pm4j/e^-e^\pm e^\mp4j$. 
Likewise the production and decay of charged scalars can also lead to four-lepton-four-jet final states
$e^+e^-e^\pm e^\pm4j/e^+e^-e^\pm e^\mp4j$ as depicted in Fig.~\ref{fig:eetoHHtoFS}. 
In Fig.~\ref{fig:CSXXtollll4j} we show the associated cross-sections at $pp$ and $e^-e^+$ colliders for the production of various LNV/LNC high multiplicity final states $e^+e^-e^\pm e^\pm4j/e^+e^-e^\pm e^\mp4j$.
\par We also stress that some of the above signatures involve lepton number violating final states, hence their possible detection would provide an indirect test of the Majorana nature of neutrinos, complementary to that provided by neutrinoless double beta decay searches.
\par Last, but not least, note that the discussion presented in our paper can be easily extended to the proposed muon collider~\cite{Li:2023tbx,Delahaye:2019omf}.
  In this case too one expects a plethora of interesting signatures arising from our leptophilic Higgs portal.
 
 \begin{acknowledgments}
  The work of A.B. is supported by Funda\c{c}\~ao para a Ci\^encia e a Tecnologia (FCT, Portugal) through the PhD grant UI/BD/154391/2023 and through the projects CFTP-FCT Unit UIDB/00777/2020 and UIDP/00777/2020, CERN/FIS-PAR/0019/2021, which are partially funded through POCTI (FEDER), COMPETE, QREN and EU.
  The work of P.B. is supported by the CSIR JRF-NET fellowship.
  The work of S.M. is supported by KIAS Individual Grants (PG086001) at Korea Institute for Advanced Study.
  The work of RS is supported by the Government of India, SERB Startup Grant SRG/2020/002303.
  The work of J.V. is supported by the Spanish grants PID2020-113775GB-I00~(AEI/10.13039/501100011033) and Prometeo CIPROM/2021/054 (Generalitat Valenciana).
\end{acknowledgments}

\appendix
\label{Appendix}
\section{Scalar Decay Width Expressions} 
\label{app:decaywidth}

Here we discuss the various possible decay modes of the new scalars present in our linear seesaw model, including decay channels to quarks, gauge, as well as Higgs bosons.

\subsection{$H_i\to f\bar{f}'$}

The decay width of the new scalars $ H $, $ A $ and $ H^\pm $ to fermions are given in equations \eqref{eq:Hff},\eqref{eq:Aff} and \eqref{eq:DW_Hpffbar}.
\begin{equation}
\Gamma(H\to f \bar{f}')=\frac{N_c m_{H}}{8\pi}\Bigl\{\bigl[1-\left(x_1+x_2\right)^2\bigr]|C_{Hff'}|^2\Bigr\}\lambda^{1/2}\left(1,x_1^2,x_2^2\right),
\label{eq:Hff}
\end{equation}
\begin{equation}
\Gamma(A\to f \bar{f}')=\frac{N_c m_{A}}{8\pi}\Bigl\{\bigl[1-\left(x_1-x_2\right)^2\bigr]|C_{Aff'}|^2\Bigr\}\lambda^{1/2}\left(1,x_1^2,x_2^2\right),
\label{eq:Aff}
\end{equation}
\begin{equation}
    \label{eq:DW_Hpffbar}
    \Gamma(H^\pm \to f\bar{f}')=\frac{N_c m_{H^\pm}}{8\pi}\Bigl\{[1-(x_1+x_2)^2]|C_S|^2+[1-(x_1-x_2)^2]|C_P|^2\Bigr\}\lambda^{1/2}(1,x_1^2,x_2^2)
\end{equation}
where $x_1=m_f/m_{H_i}$, $x_2=m_{\bar{f}'}/m_{H_i}$, $ C_{Hff'}=\sin\alpha\frac{m_f}{v} $, $ C_{Aff'}=\cos\beta\frac{m_f}{v} $ and
\begin{equation}
\lambda(1,x,y)=\left(1-x-y\right)^2-4xy,
\label{eq:lambda}
\end{equation}  
For the physical electrically chareged Higgs boson $H^\pm \to u_i\bar{d_j}$ we have defined
$$
C_S=\frac{1}{\sqrt{2}v}(-m_{d_j}+m_{u_i})V_{\text{CKM}}^{ij}\cos\beta$$
$$C_P=\frac{1}{\sqrt{2}v}(m_{d_j}+m_{u_i})V_{\text{CKM}}^{ij}\cos\beta
$$
here $i,j=1,2,3$.

For decays into quarks, the QCD radiative corrections are included as:
\begin{equation}
\Gamma = \Gamma_0\left[1+5.67\frac{\alpha_s}{\pi}+(35.94-1.36n_f)\left(\frac{\alpha_s}{\pi}\right)^2\right]
\label{eq:gamma_hff_nlo}
\end{equation} 
where $n_f$ is the number of quark flavours with $m_q<m_{H_i}$. 
%

\subsection{$H_i\to VV$}

Only the new scalar $ H $ has a sizeable decay width to two gauge bosons. The expression for this decay width is shown in equation \eqref{eq:HVV}.
\begin{equation}
\Gamma(H\to VV)=\delta_V\frac{|C_{H_{i}VV}|^2\, m_{H}^3}{128\pi\, m_V^4}\left(1-4k+12k^2\right)\sqrt{1-4k}.
\label{eq:HVV}
\end{equation}
where $k=\frac{m_V^2}{m_{H}^2}$, $\delta_V=2\,(1)$ for $V=W\,(Z)$,
\begin{gather*}
	C_{HW^+W^-}=\frac{1}{2} i g^2 v (\cos\alpha \cos\beta + 
	\sin\alpha \sin\beta)\\
	C_{HZZ}=\frac{1}{2} i v (\cos\alpha \cos\beta + 
	\sin\alpha \sin\beta) (g c_w + g_Y s_w)^2
\end{gather*}
Here $ g=e/c_w $ and $ g_Y=e/s_w $
%

\subsection{$H_i\to VH_j$} 

The decay of the new scalars to one SM gauge boson and the Higgs boson are given in equations \eqref{eq:hvh} and \eqref{eq:DW_HpWph}.
\begin{equation}
\Gamma(A\to h Z)=\frac{|C_{AhZ}|^2\, m_V^2}{16\pi^2\, m_{A}}\lambda\left(1,\frac{m_{A}^2}{m_{V}^2},\frac{m_{H_j}^2}{m_{V}^2}\right)\lambda^{1/2}\left(1,\frac{m_V^2}{m_{A}^2},\frac{m_{H_j}^2}{m_{A}^2}\right),
\label{eq:hvh}
\end{equation}
\begin{equation}
    \label{eq:DW_HpWph}
    \Gamma(H^\pm \to W^\pm h)=\frac{|C_{H^\pm W^\pm h}|^2 m_{W^\pm}^2}{16\pi^2 m_{H^\pm}}\lambda\left(1,\frac{m_{H^\pm}^2}{m_{W^\pm}^2},\frac{m_{h}^2}{m_{W^\pm}^2}\right)\lambda^{1/2}\left(1,\frac{m_{W^\pm}^2}{m_{H^\pm}^2},\frac{m_{h}^2}{m_{H^\pm}^2}\right)
\end{equation}
Here 
\begin{equation*}
	C_{AhZ}=\frac{1}{2} (\cos\beta \cos\alpha + \sin\alpha \sin\beta) (-g c_w -g_Y s_w)
\end{equation*}
\begin{equation*}
C_{H^\pm W^\pm h}=-\frac{i}{2}g(\sin\alpha\sin\beta+\cos\alpha\cos\beta)\\
\end{equation*}
%

\subsection{$H_i\to \gamma\gamma$} 

The loop-induced decay width of the new neutral scalars $ H $ and $ A $ to photons are given in equations \eqref{eq:Hgamgam} and \eqref{eq:Agamgam}.
\begin{equation}
\Gamma(H\to\gamma\gamma)=\frac{\alpha^2M_{H}^3}{256\pi^3 v^2}\left|S^\gamma(M_{H})\right|^2,
\label{eq:Hgamgam}
\end{equation}
\begin{equation}
\Gamma(A\to\gamma\gamma)=\frac{\alpha^2M_{A}^3}{256\pi^3 v^2}\left|P^\gamma(M_{A})\right|^2,
\label{eq:Agamgam}
\end{equation}
where the loop factors are
\begin{equation}
S^\gamma(M_{H})=2\sum_f N_c Q_f^2 C_{Hf\bar{f}} \frac{v}{m_f}F_s(\tau_f)-C_{HW^+W^-}\frac{v}{2m_W^2}F_1(\tau_W)-C_{HH^+H^-}\frac{v}{2m_{H^+}^2}F_0(\tau_{H^+})
\end{equation}
and
\begin{equation}
P^\gamma(M_{A})=2\sum_f N_c Q_f^2 C^P_{Af\bar{f}} \frac{v}{m_f}F_p(\tau_f)
\end{equation}
Here $N_c=3\,(1)$ for quarks (leptons), and $Q_f$ is the electric charge and 
\begin{equation*}
C_{HH^+H^-}=-i v (\cos\alpha \cos\beta^3 \lambda_3 + \sin\alpha \sin\beta^3 \lambda_3 + \cos\beta^2 \sin\alpha \sin\beta (2 \lambda_1 - \lambda_4) + \cos\alpha \cos\beta \sin\beta^2 (2 \lambda_2 - \lambda_4))
\end{equation*}
The functions $ F_s(\tau) $, $ F_p(\tau) $, $ F_0(\tau) $ and $ F_1(\tau) $ are given as in equation \eqref{eq:Fs}.
\begin{equation}
\begin{aligned}
F_s(\tau)&=\tau^{-1}\Bigl[1+\left(1-\tau^{-1}\right)f(\tau)\Bigr]\qquad F_p(\tau)=\frac{f(\tau)}{\tau}\\
F_0(\tau)&=\tau^{-1}\left[\tau^{-1}f(\tau)-1\right]\qquad\qquad\;\; F_1(\tau)=2+3\tau^{-1}+3\tau^{-1}(2-\tau^{-1})f(\tau)
\end{aligned}.
\label{eq:Fs}
\end{equation}
where $\tau_f=m_{H_i}^2/4m_f^2$, $\tau_W=m_{H_i}^2/4m_W^2$, $\tau_{H^+}=m_{H_i}^2/4m_{H^+}^2$ and 
\begin{equation}
f(\tau)=\left\{
\begin{array}{lr}
\arcsin^2\left(\sqrt{\tau}\right) & \tau \leq 1 \\
\displaystyle -\frac{1}{4}\left[\ln\left(\frac{\sqrt{\tau}+\sqrt{\tau-1}}{\sqrt{\tau}-\sqrt{\tau-1}}\right)-\mathrm{i}\pi\right]^2 & \tau > 1.
\end{array}
\right.
\label{eq:ftau}
\end{equation}
%

\subsection{$H_i\to gg$} 

The loop-induced decay width of the new neutral scalars $ H $ and $ A $ to gluons are given in equations \eqref{eq:Hgg} and \eqref{eq:Agg}.
\begin{equation}
\Gamma(H\to gg)=\frac{\alpha_s^2M_{H}^3}{32\pi^3 v^2}\left|S^g(M_{H})\right|^2,
\label{eq:Hgg}
\end{equation}
\begin{equation}
\Gamma(A\to gg)=\frac{\alpha_s^2M_{A}^3}{32\pi^3 v^2}\left|P^g(M_{A})\right|^2,
\label{eq:Agg}
\end{equation}
with the loop factors
\begin{equation}
S^g(M_{H})=\sum_q C_{Hq\bar{q}} \frac{v}{m_q}F_s(\tau_q), \qquad
P^g(M_{A})=\sum_q C_{Aq\bar{q}} \frac{v}{m_q}F_p(\tau_q)
\end{equation}

\section{LNV versus LNC in Linear seesaw model} 
\label{app:LNV_vs_LNC}

\subsection{Linear seesaw with quasi-Dirac heavy neutrinos}  

Here we discuss the prospects of having large rates for LNV processes. 
Let's start by recalling that the rates for LNV and LNC processes are predicted to be the same when mediated by heavy Majorana neutrinos.
Within the linear seesaw with softly broken lepton number, LNV processes should be severely suppressed. 
This follows from the fact that the quasi-Dirac heavy neutrinos can be thought of as two fermions of opposite CP-phase and small Majorana mass-splitting~($\Delta M\sim m_\nu$).
This leads to a cancellation between the individual contributions to LNV processes involving virtual heavy-neutrino propagation.

However, at collider energies heavy neutrino mediators can be produced on-shell, so that oscillations can occur between the members of the quasi-Dirac pair.
This can prevent the cancellation suppressing LNV signals. This applies to all low-scale seesaw setups, such as the inverse seesaw or our linear seesaw model.  
\begin{figure}[!htbp]
	\includegraphics[width=0.45\linewidth]{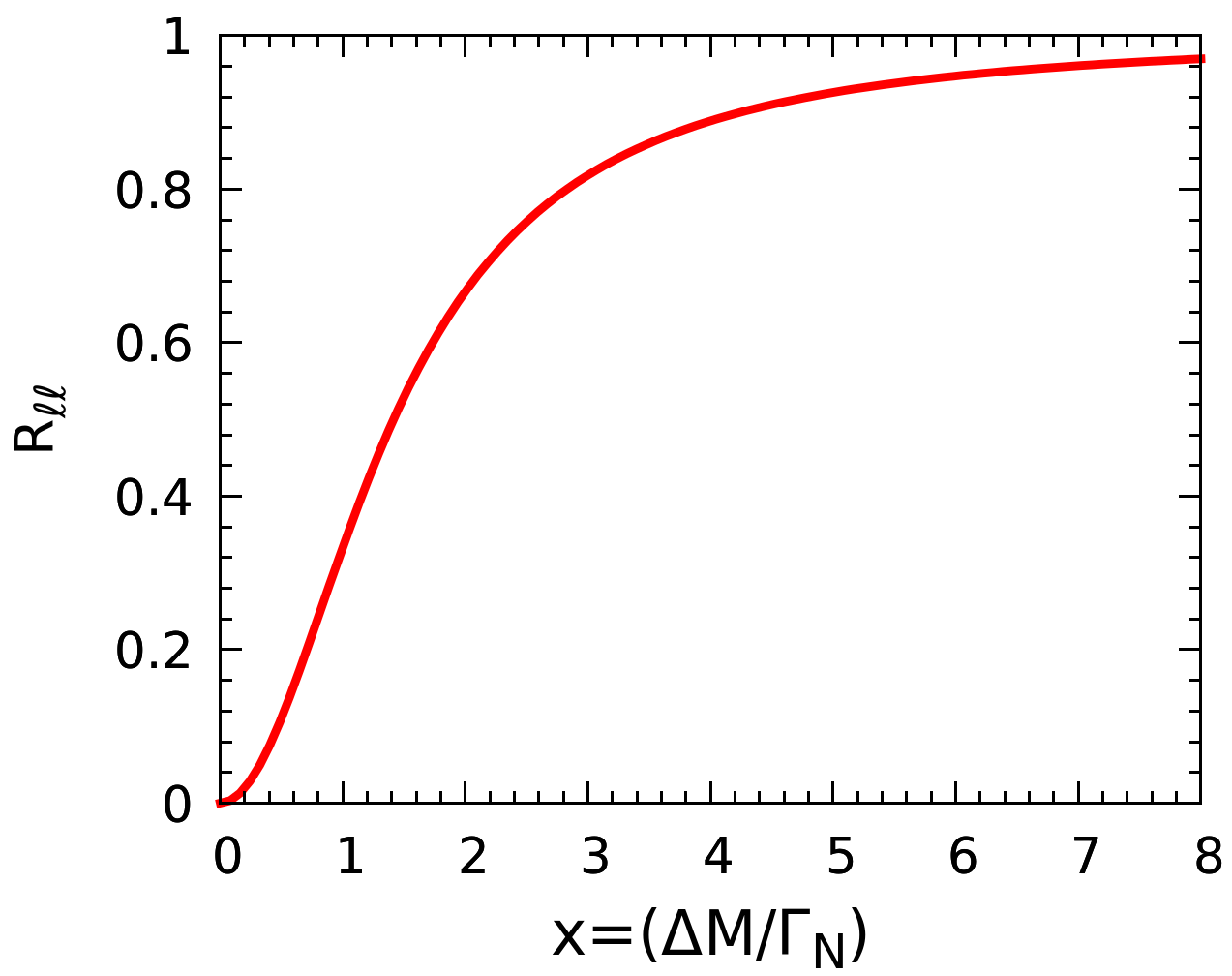}~~~~
	\includegraphics[width=0.45\linewidth]{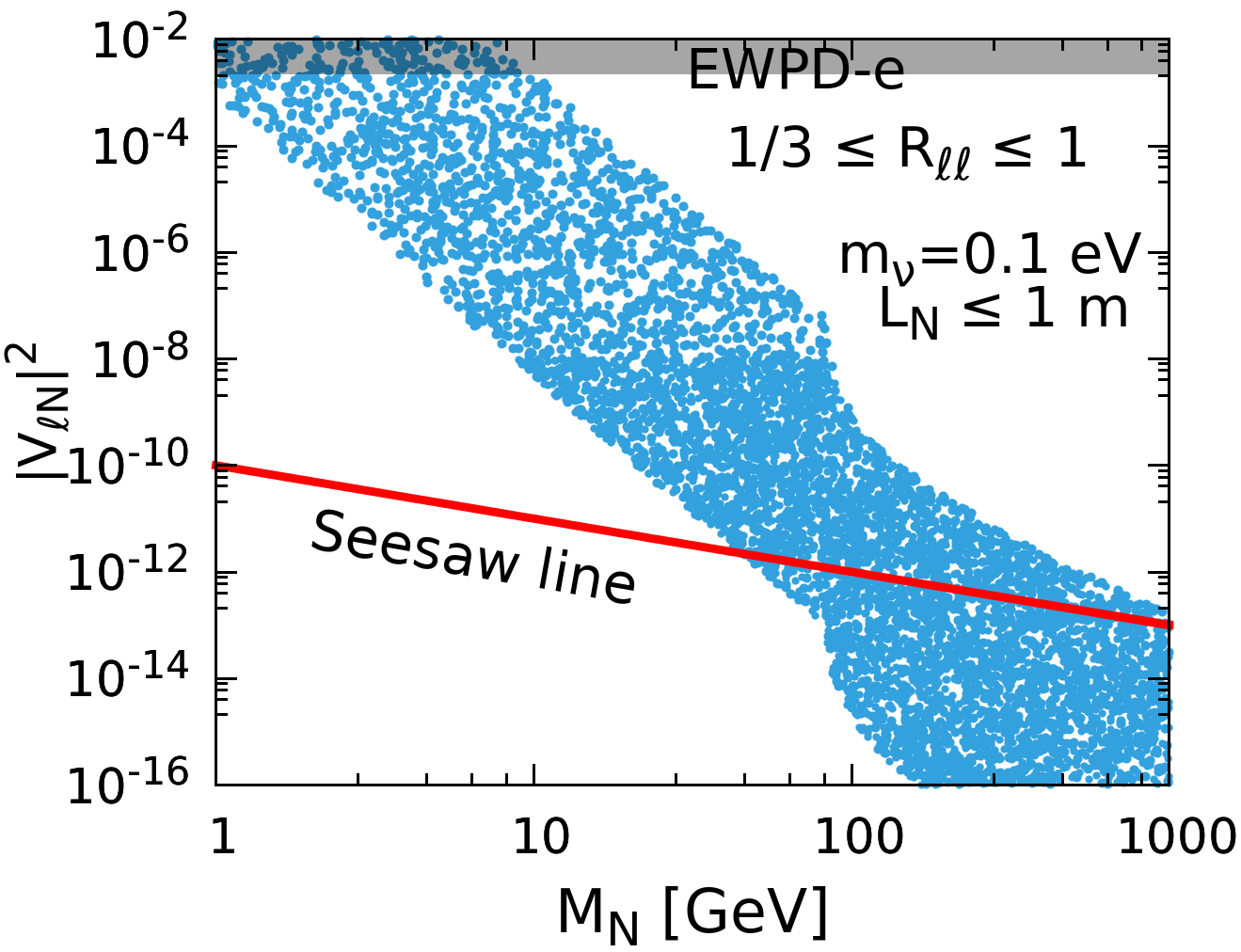}
	\caption{
          Left panel: LNV to LNC ratio $R_{\ell\ell}$ versus the quantity $x=\frac{\Delta M}{\Gamma_N}$.
          Right panel: parameter region in the plane $|V_{\ell N}|^2-M_N$ for $m_\nu\sim 0.1$~eV yielding $R_{\ell\ell}$ in the range $1/3\leq R_{\ell\ell}\leq 1$.
          We also required that the heavy neutrino decay length $L_{N}<1$~meter, so that $N$ decays inside the linear collider detector.
          The gray band is excluded by electroweak precision data, and the red line represents a naive seesaw expectation $|V_{\ell N}|^2\sim m_\nu/M_N$. }
	\label{fig:Rll}
\end{figure}
Note that the oscillations between the members of quasi-Dirac pair are determined by the mass-splitting $\Delta M$ and if $\Delta M\ll \Gamma_N$,
they will decay before they have time to oscillate. 
In this case, the destructive interference between heavy neutrinos will hold and LNV processes are suppressed.
On the other hand, if $\Delta M \gg \Gamma_N$, as the heavy neutrinos oscillate many times before they decay, the oscillations are averaged out and the LNV rates
are expected to be similar to those of LNC processes. 
A more dedicated discussion was given in Ref.~\cite{Anamiati:2016uxp,Antusch:2022ceb,Antusch:2020pnn},
where it was found that the ratio of the rates for opposite and same-sign dilepton events is effectively characterised by the following expression~ 
\begin{align}
R_{\ell\ell}=\frac{\Delta M^2}{2\Gamma_N^2+\Delta M^2}=\frac{x^2}{2+ x^2},  \text{   with   }x=\frac{\Delta M}{\Gamma_N}.
\end{align}
Note that $R_{\ell\ell}\to 1$ as $x\to \infty$~(limiting Majorana case) and $R_{\ell\ell}\to 0$ as $x\to 0$~(limiting Dirac case).
In models with quasi-Dirac neutrinos, the ratio $R_{\ell\ell}$ can take any value between 0 and 1.  
In the left panel of Fig.~\ref{fig:Rll} we show $R_{\ell\ell}$ versus $x=\frac{\Delta M}{\Gamma_N}$ for a benchmark value of the heavy neutrino mass, $M_N=100$~GeV.  
We see that when the mass splitting $\Delta M$ is larger than a few times the width $\Gamma_N$, $R_{\ell\ell}$ approaches rapidly the Majorana limit $R_{\ell\ell}=1$.  
Note that this result is independent of the absolute heavy-neutrino mass scale. 
In our case the mass splitting is $\Delta M\sim m_\nu$ and hence is very tiny.
On the other hand the decay width $\Gamma_N$ is controlled by the mixing parameter $V_{\ell N}$~(when $M_N<m_{H^\pm,H/A}$)  
or by the Yukawa coupling $Y_S$~($M_N > m_{H^\pm,H/A}$). 
In either case the decay width $\Gamma_N$ is large for relatively large mixing $V_{\ell N}$ or large Yukawa coupling $Y_S$, see Fig.~\ref{fig:BRN} and Fig.~\ref{fig:Ndecay}.
Of course, for large decay width and small mass-splitting, the quantity $x$ is small and hence $R_{\ell\ell}$ is small. 
However the decay width can be small for small mixing $V_{\ell N}$, (at least when $M_N<m_{H^\pm,H/A}$), 
comparable or even smaller than the mass splitting $\Delta M$, so that $x$ becomes large and $R_{\ell\ell}$ becomes close to 1.  
This is precisely what is shown in the right panel of Fig.~\ref{fig:Rll}. 
Assuming a light neutrino mass $m_\nu\sim 0.1$~eV we show the value of the mixing parameter $|V_{\ell N}|^2$ and mass $M_N$ 
required to achieve LNV to LNC ratio $R_{\ell\ell}\geq 1/3$. 
We use the criterion $R_{\ell\ell}=1/3$~($\Delta M\approx \Gamma_N$) to distinguish between suppressed and unsuppressed LNV rates. 
For smaller light-neutrino-mass values the required value of the mixing angle $|V_{\ell N}|$ will be even smaller for a given mass $M_N$. 
Note that, although the mixing angle required to achieve maximum $R_{\ell\ell}$ is relatively small, the decay width is large enough to have heavy-neutrino decays inside the detector. 
\par All in all, the above discussion shows that within our linear seesaw setup there is a wide parameter region with $\Delta M\approx \Gamma_N$,
so that the ratio $R_{\ell\ell}$ can have any value within the range $[0,1]$. 
This indicates the intrinsic presence of detectable \lnv rates at colliders, suggesting the that they may play a complementary role in probing the Majorana nature of neutrinos.

\subsection{A generalized scheme} 

So far we have always assumed a genuine linear seesaw scenario where the heavy neutral leptons form quasi-Dirac pairs. 
Another way to achieve a large $R_{\ell\ell}\approx 1$ is to consider a generalized scheme containing a large Majorana mass entry breaking lepton number by two units and characterized by a large violation scale.
Indeed, we can introduce such a $\Delta L=2$ term $M$ for the right-handed neutrinos $\nu^c$, in co-existence with the $\Delta L=1$ linear seesaw terms in Eq.~\ref{eq:Yukawa}, that break lepton number by one unit, i.e.  
\begin{equation}
  \label{eq:Yukawa2}
  - \mathcal{L}_{\rm Yuk}= Y_{\nu}^{ij} L_i^T C \nu^c_j \Phi 
  + M_R^{ij} \nu^c_i C S_j + M^{ij} \nu^c_i C \nu^c_j + Y_{S}^{ij} L_i^T C  S_j \chi_L+ \text{h.c.}
\end{equation}
The resulting neutral lepton mass matrix in the basis $(\nu_L, \nu^c, S)$ is given as 
\begin{align}
\mathcal{M}_{\nu}=
 \begin{pmatrix}
  0 & m_D  & M_L  \\
  m_D^T & M & M_R \\
  M_L^T &  M_R^T  &  0 \\
 \end{pmatrix}~.
 \label{eq:neutrino-mass-matrix2}
\end{align}
Under the assumption $M > M_R \gg m_D \gg M_L$ we integrate out the heavy fields $\nu^c$ and $S$ and obtain the complete diagonalization of the extended seesaw mechanism.
This results in the following physical masses for the neutral leptons, 
\begin{align}
& m_{\nu}\approx m_D(M_L M_R^{-1})^T+(M_L M_R^{-1})m_D^T - (M_L M_R^{-1})M(M_L M_R^{-1})^T \\
& M_{\nu^c}=\frac{1}{2}\Big(M+\sqrt{M^2+4M_R^2}\Big) \text{ and  } M_S=\frac{1}{2}\Big(M-\sqrt{M^2+4 M_R^2}\Big).
\end{align}
One sees that the light neutrinos acquire mass from the same source $M_L \sim v_\chi$ but now there is a quadratic term in addition to those characteristic of the linear seesaw mechanism.
This new term is reminiscent from the conventional $\Delta L=2$ seesaw mechanism. 
It is clear from the above mass matrices that the mass splitting of the heavy neutral leptons is now large, so they are closer to being pure Majorana.
Thus, the addition of the large Majorana mass term for the singlet lepton $\nu^c$ makes it easier to induce large values for the ratio $R_{\ell\ell}$ characterizing LNV processes.

\section{Other collider signatures}
\label{app:CS}

\begin{figure}[!h]
	\includegraphics[width=0.815\linewidth]{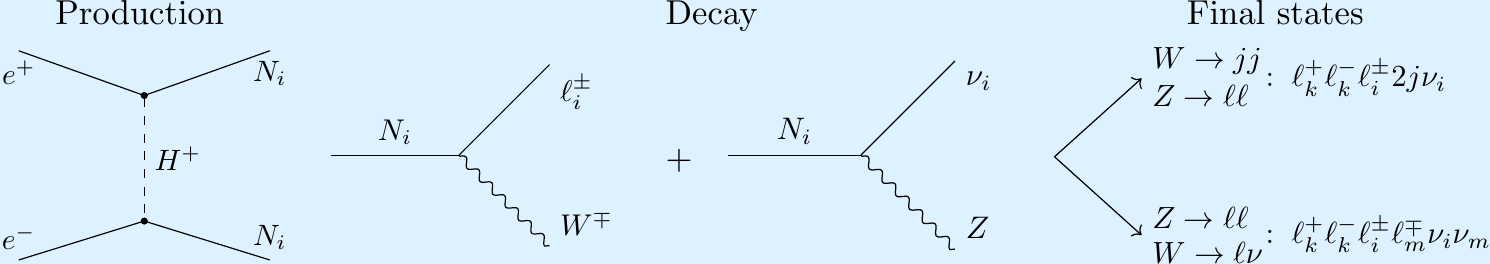}
	\caption{ Illustrative heavy-neutrino pair production and their decays via the charged and neutral currents for the case of $M_{N_i}<m_{H^\pm}$.}
	\label{fig:eeNNtoFSmix}
\end{figure}
\begin{figure}[!h]
	\includegraphics[width=0.815\linewidth]{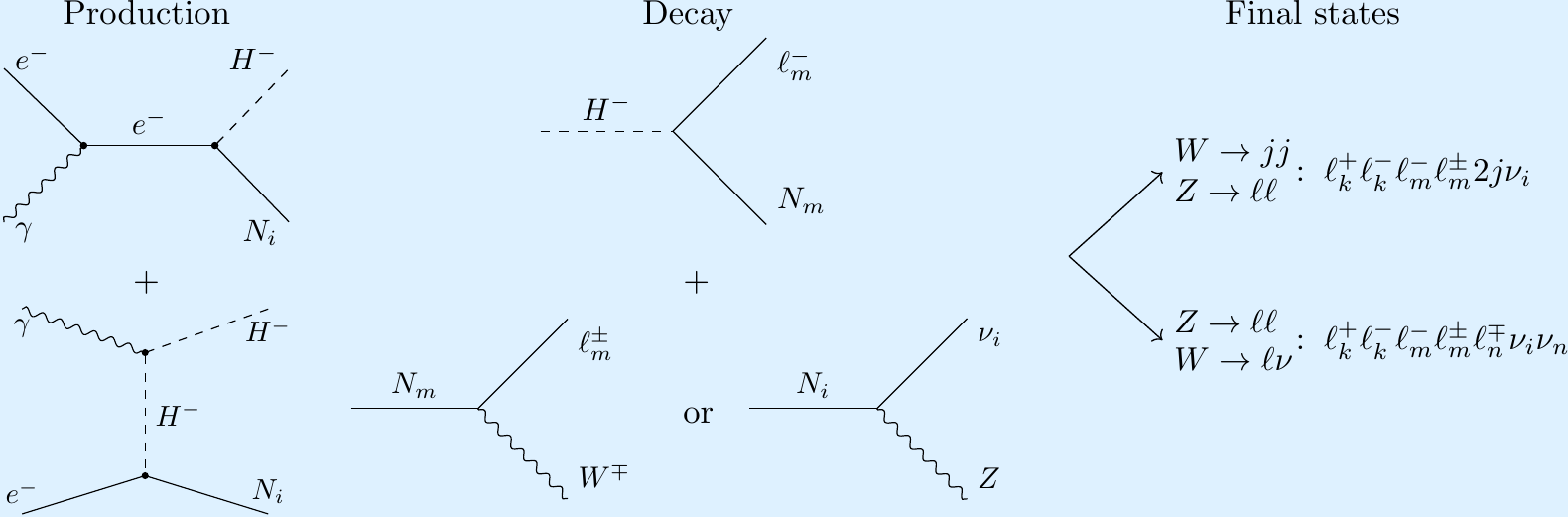}			
	\caption{
		Illustrative SM final states arising from the decay of the heavy neutrinos via the charged and neutral currents
		after being produced in association with charged Higgs, for the case of $M_{N_i}<m_{H^\pm}$.}
	\label{fig:eaNHtoFSmix}
\end{figure}
\begin{figure}[!h]
	\includegraphics[width=0.815\linewidth]{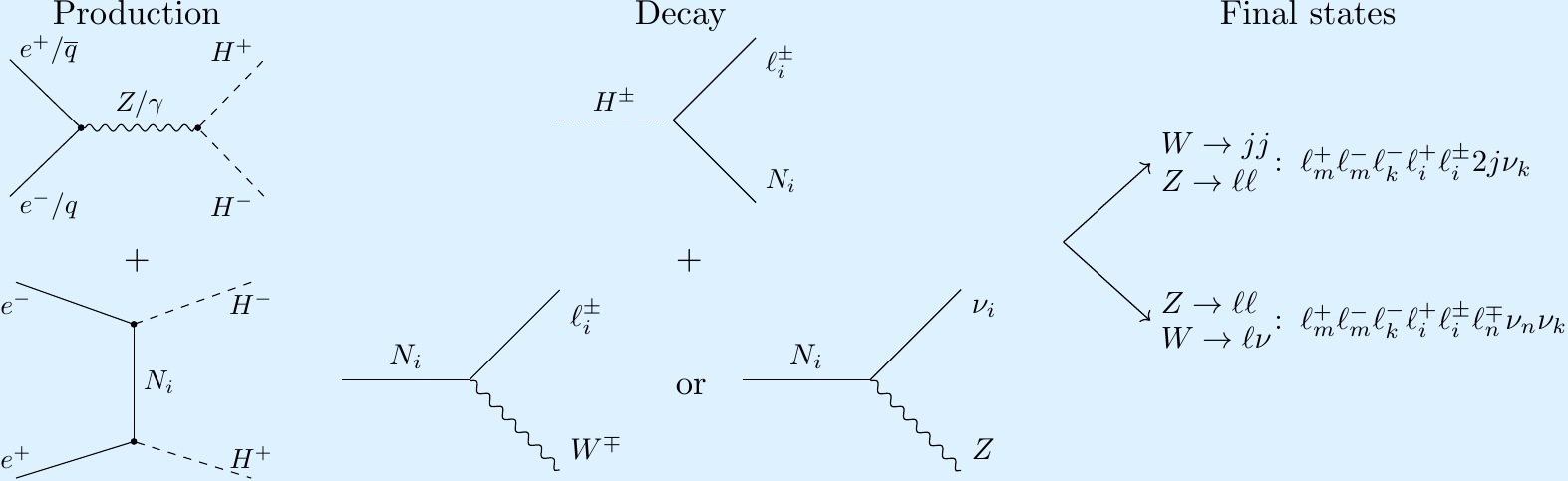}			
	\caption{
		Illustrative SM final states arising from the decay of the heavy neutrinos via the charged and neutral currents
		after being produced as the decay products of charged Higgs, for the case of $M_{N_i}<m_{H^\pm}$.}
	\label{fig:eeHHtoFSmix}
\end{figure}
Our study can be expanded to include also the case where heavy neutrino mediators decay through the neutral current interaction.
  In Figs.~\ref{fig:eeNNtoFSmix}, \ref{fig:eaNHtoFSmix} and \ref{fig:eeHHtoFSmix}, we show possible final sates that result when
  all heavy neutrinos decay through the neutral current, such as $N\to\nu Z$, or when one of them decays as $N\to\ell W$ and other one decay as $N\to\nu Z$.
  We have assumed the leptonic mode of the $Z$ boson and leptonic/hadronic decay mode of the $W$ boson.


\bibliographystyle{utphys}
\bibliography{bibliography} 
\end{document}